\documentclass[12pt]{article}
\usepackage{amssymb,amsmath}
\usepackage{epsfig,psfrag}

\catcode`\@=11
\def \tr {\mathop{\rm tr}\nolimits}

\def \e  {\mathop{\rm e}\nolimits}
\newcommand\lr[1]{{\left({#1}\right)}}
\newcommand \widebar [1] {\overline{#1}}

\newcommand \ket [1] {|{#1}\rangle}
\newcommand \bra [1] {\langle {#1}|}
\newcommand\re[1]{(\ref{#1})}
\def \qqquad {\qquad\quad}
\def \qqqquad {\qquad\qquad}

\newcommand{\pa}{\partial}
\newcommand{\be}{\begin{equation}}
\newcommand{\ee}{\end{equation}}
\newcommand{\bea}{\begin{eqnarray}}
\newcommand{\eaa}{\end{eqnarray}}
\newcommand{\nn}{\nonumber}

\renewcommand{\a}{\alpha}

\newcommand{\da}{{\dot\alpha}}
\newcommand{\db}{{\dot\beta}}
\newcommand{\bl}{{\tilde\lambda}}

\renewcommand{\b}{\beta}

\newcommand{\q}{\theta}

\newcommand{\ep}{\epsilon}
\newcommand{\m}{\mu}

\newcommand{\cN}{{\cal N}}

\newcommand{\p}[1]{(\ref{#1})}
\newcommand{\bt}[1]{{\bar t}}

\newcommand \vev [1] {\langle{#1}\rangle}
\newcommand \ran [1] {|{#1}\rangle}
\newcommand \lan [1] {\langle{#1}|}
\def\numberbysection{\@addtoreset{equation}{section}
                     \def\theequation{\thesection.\arabic{equation}}}
\numberbysection

\begin{document}

\thispagestyle{empty}
\null\vskip-12pt \hfill  LAPTH-1257/08 \\
\null\vskip-12pt \hfill LPT--Orsay--08--60
\vskip2.2truecm
\begin{center}
\vskip 0.2truecm {\Large\bf
{\Large Dual superconformal symmetry of  scattering\\[3mm] amplitudes in $\cN=4$ super-Yang-Mills theory}
}\\
\vskip 1truecm
{\bf J.M. Drummond$^{*}$, J. Henn$^{*}$, G.P. Korchemsky$^{**}$ and E. Sokatchev$^{*}$ \\
}

\vskip 0.4truecm
$^{*}$ {\it
LAPTH\footnote{Laboratoire d'Annecy-le-Vieux de Physique Th\'{e}orique, UMR 5108}, Universit\'{e} de Savoie, CNRS\\
B.P. 110,  F-74941 Annecy-le-Vieux Cedex, France\\
\vskip .2truecm $^{**}$ {\it
Laboratoire de Physique Th\'eorique%
\footnote{Unit\'e Mixte de Recherche du CNRS (UMR 8627)},
Universit\'e de Paris XI, \\
F-91405 Orsay Cedex, France
                       }
  } \\
\end{center}

\vskip 1truecm 
\centerline{\bf Abstract} 
We argue that the scattering amplitudes in the maximally supersymmetric $\mathcal{N}=4$
super-Yang-Mills theory possess a new symmetry which extends the previously discovered dual
conformal symmetry. To reveal this property we formulate  the scattering amplitudes  as functions in
the appropriate dual superspace. Rewritten in this form, all tree-level MHV and next-to-MHV
amplitudes exhibit manifest dual superconformal symmetry. We propose a new, compact and Lorentz
covariant formula for the tree-level NMHV amplitudes for arbitrary numbers and types of external
particles.  The dual conformal symmetry is broken at loop level by infrared divergences. However, we
provide evidence that the anomalous contribution to the MHV and NMHV superamplitudes is the same
and, therefore, their ratio is a dual conformal invariant function. We identify this function by an
explicit calculation of the six-particle amplitudes at one loop. We conjecture that these properties
hold for all, MHV and non-MHV, superamplitudes in $\mathcal{N}=4$ SYM both at weak and at strong
coupling.
\medskip

 \noindent

\newpage
\setcounter{page}{1}\setcounter{footnote}{0} {\tableofcontents}
\newpage

\section{Introduction}

The scattering amplitudes in maximally supersymmetric ${\cal N}=4$ super-Yang-Mills theory (SYM)
have a number of remarkable properties both at weak and at strong coupling. Defined as matrix
elements of the $S-$matrix between asymptotic on-shell states, they inherit the symmetries of the
underlying gauge theory. In addition, trying to understand the properties of the scattering
amplitudes, one can discover new dynamical symmetries of the  ${\cal N}=4$ theory. A well-known
example is the twistor formulation of the scattering amplitudes by Witten \cite{Witten:2003nn}.

In this paper we argue that the planar scattering amplitudes in ${\cal N}=4$ SYM theory have a
hidden symmetry that we call  {\em dual superconformal symmetry}. It appears on top of all
well-known symmetries of the scattering amplitudes (supersymmetry, conformal symmetry, etc.) and is
not related (not in an obvious way, at least) to an invariance of the Lagrangian of the theory.
Quite remarkably, the same symmetry also emerges, in the planar limit,  from fermionic T-duality of the sigma model on an
AdS${}_5\times$S${}^5$ background in the AdS/CFT description
of scattering amplitudes~\cite{BM}.

Hints of such a symmetry first came from the classification of the loop integrals entering the
perturbative expansion of the planar four-gluon maximally helicity violating (MHV) scattering
amplitude~\cite{abdk}. The latter was calculated up to three loops in terms of a restricted class of
planar integrals by Bern, Dixon and Smirnov (BDS) in \cite{bds}. In \cite{dhss} it was observed that
these integrals all have a very special property. If one changes variables from momenta $p_i^\mu$ to
`dual coordinates' $x_i^\mu$ via
\be
p_i^\mu = x_i^\mu - x_{i+1}^\mu\,, \label{dualx}
\ee
then the integrals exhibit a formal conformal covariance in the dual $x-$space. This dual conformal
symmetry is formal because the integrals are in fact infrared divergent and the introduction of
dimensional regularisation breaks the symmetry. Nevertheless, even broken, the dual conformal
symmetry still imposes constraints on the on-shell scattering amplitudes. Their precise formulation
required further developments, both at strong and weak coupling. At strong coupling, Alday and
Maldacena applied the AdS/CFT correspondence to the study of scattering amplitudes in
$\mathcal{N}=4$ super-Yang-Mills \cite{am1}. According to their proposal, the planar $n-$gluon
amplitudes with various helicity configurations are related at strong coupling to the area of a
minimal surface in AdS${}_5$ space. This surface is attached to a closed polygon contour $C_n$ made
of $n$ light-like segments $[x_i,x_{i+1}]$, defined by the on-shell gluon momenta according to
\re{dualx}. In the case of the four-particle amplitude the area could be calculated explicitly and
was shown to agree with the ABDK/BDS conjecture about the all-order structure of MHV amplitudes
\cite{abdk,bds}. An important observation made in \cite{am1} was that the calculation of the
amplitude is mathematically identical to that of a Wilson loop $W(C_n)$ at strong coupling
\cite{malda98}.

These findings together with previous results on the relation between infrared singularities of
scattering amplitudes and Wilson loops \cite{GK} subsequently lead to a formulation of the duality
between scattering amplitudes and light-like Wilson loops at weak coupling (see Fig.~1). Firstly, in
\cite{dks} it was found that at one-loop the four-point MHV amplitude matched the corresponding
Wilson loop $W(C_4)$. This was then generalised to $n$ points at one loop \cite{bht}. The duality
was shown to hold beyond one loop at four \cite{dhks1}, five \cite{dhks2} and six
\cite{dhks4,bdkrsvv} points. As discussed in \cite{dhks3}, the confirmation of the duality at six
points and two loops necessarily implies the breakdown of the BDS conjecture beyond five points.
Indeed, evidence that the conjecture should fail for some number of gluons had been found in
\cite{am2}. The analysis of the Regge limits also shows the BDS conjecture had indeed to fail at six
points and two loops \cite{bls}. This was later confirmed by an explicit two-loop six-gluon
calculation \cite{bdkrsvv}. Not only was it shown that the BDS conjecture fails at two loops, but it
fails exactly so that the duality with the Wilson loop is maintained \cite{dhks4,bdkrsvv}.

\begin{figure}[h]
%
\psfrag{p1}[cc][cc]{$p_1$}\psfrag{p2}[cc][cc]{$p_2$}\psfrag{p3}[cc][cc]{$p_3$}\psfrag{pn}[cc][cc]{$p_n$}
\psfrag{x1}[cc][cc]{$x_1$}\psfrag{x2}[cc][cc]{$x_2$}\psfrag{x3}[cc][cc]{$x_3$}\psfrag{xn}[cc][cc]{$x_n$}
\psfrag{xn1}[cc][cc]{$x_{n-1}$}
\centerline{{\epsfysize8cm \epsfbox{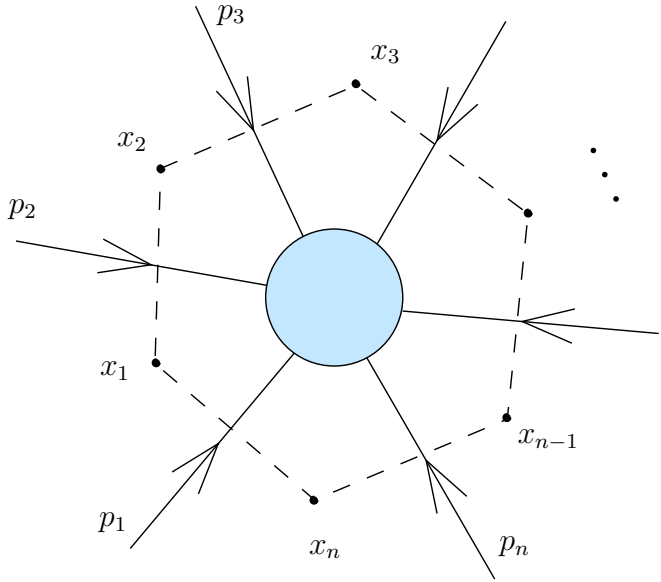}}} \caption{Diagrammatic representation of the
scattering amplitude/Wilson loop duality. The dashed line depicts the contour $C_n$.}
\end{figure}

An important feature of this duality is that the Wilson loop in $\mathcal{N}=4$ SYM has a natural
conformal symmetry. Applied to the dual scattering amplitudes, this implies an unexpected `dual'
conformal symmetry, acting on the particle momenta via their relation to the dual coordinates $x_i$
\p{dualx}. The conformal symmetry of the Wilson loop is formulated as an anomalous conformal Ward
identity (the anomaly is due to the ultraviolet divergences of the Wilson loop) \cite{dhks1,dhks2}.
This symmetry is sufficient to fix the form of the finite part of the four- and five-cusp Wilson
loops, both at weak~\cite{dhks1,dhks2} and at strong coupling~\cite{am2,Komargodski:2008wa}. For
six-cusp Wilson loops and beyond, the conformal symmetry leaves some freedom. So, the fact that the
MHV amplitude continues to agree with the Wilson loop at this level shows that there is more to the
amplitude/Wilson loop duality than just dual conformal symmetry.

The above developments referred to MHV amplitudes. A remarkably simple features of MHV amplitudes is
the fact that supersymmetry allows to reduce the problem of calculating all $n$-particle MHV
amplitudes to computing a single scalar function of the Mandelstam variables. Amplitudes with other
helicity configurations, such as next-to-MHV (NMHV), next-to-next-to-MHV (N${}^2$MHV) and so on, are
known to have a more complicated
structure~\cite{bddk94-2,Bern:2004ky,Bern:2004bt,Britto:2004nc,Britto:2004tx}. In this paper we
provide evidence that amplitudes with arbitrary helicity configurations enjoy a new, bigger symmetry
which can be thought of as a supersymmetric generalization of dual conformal symmetry.

In the planar limit, a generic $n-$particle scattering amplitude in the ${\cal N}=4$ SYM theory with an
$SU(N)$ gauge group has the form
\begin{align}\label{A-planar}
\mathcal{A}_n(\{p_i,h_i,a_i\}) =(2\pi)^4\delta^{(4)}\big(\sum_{i=1}^n p_i\big)\!\! \sum_{\sigma\in
S_n/Z_n} 2^{n/2}g^{n-2} \tr [t^{a_{\sigma(1)}}\ldots t^{a_{\sigma(n)}} ]
A_n\left(\sigma(1^{h_1},\ldots, n^{h_n})\right) \,,
\end{align}
where each scattered particle (scalar,  gluino with helicity $\pm 1/2$ or gluon with helicity
$\pm 1$) is characterized by its on-shell momentum $p_i^\mu$ ($p_i^2=0$), helicity $h_i$ and color
index $a_i$. Here the sum runs over all possible non-cyclic permutations $\sigma$ of the set
$\{1,\ldots,n\}$ and the color trace involves the generators $t^a$ of $SU(N)$ in the fundamental
representation normalized as $\tr(t^a t^b) =\frac12\delta^{ab}$. All particles are treated as
incoming, so that the momentum conservation takes the form $\sum_{i=1}^n p_i=0$.

The color-ordered partial amplitudes  $A_n\left(\sigma(1^{h_1},\ldots, n^{h_n})\right)$ only depend
on the momenta and helicities of the particles and admit a perturbative expansion in powers of `t
Hooft coupling $a=g^2N/(8\pi^2)$. The best studied so far are the gluon scattering amplitudes. In
some cases, amplitudes with external particles other than gluons can be obtained from them with the
help of supersymmetry. In particular, the supersymmetric Ward identities~\cite{Ward-SUSY} imply that
the partial amplitudes $A_n$ vanish to all orders in the coupling when at least $n-1$ gluons have
the same helicity, \footnote{This is true for $n\ge4$. However, for $n=3$ one can construct e.g.
amplitudes $A_{3}(1^-,2^+,3^+)\neq0$ provided the on-shell momenta are complex
\cite{Witten:2003nn}.}
\be
A_n(1^\pm,2^+,\ldots,n^+)=0\,.
\ee
In the same way, for maximally helicity violating amplitudes (MHV) amplitudes, i.e. amplitudes with two gluons of negative helicity, say the $j-$th and the $k-$th, all perturbative corrections can be factorized into
a universal scalar factor $M_n^{\rm (MHV)}$ independent of $j$ and $k$
\be\label{MHV}
A^{\rm MHV} _n(1^+...\, j^-...\,  k^-...\,  n^+) = A^{\rm MHV} _{n;0}  + a A^{\rm MHV} _{n;1} +
O(a^2) = A^{\rm MHV} _{n;0}  \, M_n^{\rm MHV} \,.
\ee
Here $M_n^{\rm MHV}=M_n^{\rm MHV}(\{s_{ij}\},a)$ is a function of the Mandelstam kinematical invariants
and of the `t Hooft coupling. The tree amplitude is given, in the spinor helicity formalism
$p^{\dot\a\a} =(\sigma_\mu)^{\dot \a\a} p^\mu =\lambda^\a\bl^{\dot \a} $, by the Parke-Taylor
formula \cite{PT,BG}
\be\label{PT}
A^{\rm MHV} _{n;0}  = i \frac{\vev{j\, k}^4}{\vev{12} \vev{23}\ldots \vev{n1}}\,,
\ee
where $\vev{jk} =\epsilon_{\alpha\beta} \lambda_j^\alpha \lambda_k^\beta$ and $\lambda_j^\alpha$ (or
$\bl_j^{\dot \alpha}$ ) is a two-component Weyl commuting spinor with helicity
$-1/2$ (or $+1/2$). Two special features of the MHV amplitude \re{MHV}  are that, firstly, the tree-level factor is
holomorphic in the $\lambda-$spinors and, secondly, the same function of spinors \re{PT} appears to
all orders in `t Hooft coupling. This allows one to reduce the problem of calculating the MHV
amplitude to all loops to determining the scalar function $M_n^{\rm MHV}$. We shall return to this
function shortly.

The above-mentioned simple properties of the MHV amplitudes are {already} lost for next-to-MHV gluon
amplitudes $A_n$ (with $n\ge 6)$ which have three gluons of negative helicity. In that case, the
tree-level amplitude $A^{\rm NMHV} _{n;0}(\lambda,\bl)$ depends on  both $\lambda-$ and
$\bl-$spinors and, most importantly, the perturbative corrections produce new helicity  structures
$A_{n;1} ^{\rm NMHV,(\ell)}(\lambda,\bl)$ different from the tree-level
one~\cite{bddk94-2,Bern:2004ky,Bern:2004bt,Britto:2004nc,Britto:2004tx},
\be\label{NMHV}
A^{\rm NMHV} _n(1^+...\, i^-...\, j^-...\,  k^-...\,  n^+) = A^{\rm NMHV} _{n;0}(\lambda,\bl) + a
\sum_\ell A_{n;1} ^{\rm  NMHV,(\ell)}(\lambda,\bl)\, M^{\rm NMHV,\,(\ell)}_{n;1}  + O(a^2)  \,.
\ee
Another complication with the NMHV amplitude is due to the fact that  the coefficients $A_{n;1} ^{\rm
 NMHV, (\ell)}$ and the corresponding scalar Feynman integrals $M^{\rm  NMHV,(\ell)}_{n;1}$ also
depend on the positions of the gluons $i$, $j$ and $k$ carrying negative helicities and, therefore,
they have to be calculated separately for each configuration of helicities. A general expression for
the  one-loop NMHV gluon amplitude was given in \cite{Bern:2004bt} and it was later generalized to
amplitudes with external particles different from gluons in \cite{Risager:2005ke}. The situation
becomes even more complex  for the N${}^2$MHV, $\ldots$ amplitudes. In this case only {certain parts
of the} one-loop amplitudes have been  constructed using the quadruple cut method
\cite{Britto:2004nc}, and at  tree level only the restricted class of the so-called split-helicity
amplitudes were found \cite{Britto:2005dg}.

Bearing in mind the simplicity of the Parke-Taylor formula \re{PT} for MHV amplitudes, one may wonder
whether the complexity of the NMHV amplitudes is genuine or whether there exists some symmetry which
relates different NMHV amplitudes to each other and which allows us to write them all in a
compact form. In this paper we show that such a symmetry does exist. As a hint, we return to the
MHV amplitudes and recall~\cite{Nair} that the various tree-level MHV amplitudes can be
combined into a single MHV superamplitude by introducing Grassmann variables $\eta_i^A$ (with
$A=1,\ldots,4$), one for each external particle. Since the perturbative corrections to the MHV amplitude
are factorized into a universal scalar factor \re{MHV}, we can write the all-loop generalization of
the MHV superamplitude as
\be\label{A-SMHV}
{\cal A}_{n}^{\rm MHV} (p_1,\eta_1; \ldots; p_n,\eta_n)= i(2\pi)^4  \frac{ \delta^{(4)}
\left(\sum_{j=1}^n p_j\right)\,\delta^{(8)}\lr{\sum_{i=1}^n \lambda_{i}^\alpha \eta_i^A}} {\vev{12}
\vev{23}\ldots \vev{n1}}\ M_n^{\rm MHV} \,,
\ee
where $\delta^{(8)}\lr{\sum_{i=1}^n \lambda_{i\,\alpha}
\eta_i^A}=\prod_{\alpha=1,2}\prod_{A=1}^{4}\lambda_{i\,\alpha} \eta_i^A$ is a  Grassmann
delta function. The all-loop MHV amplitudes appear as coefficients in the expansion of ${\cal
A}_{n}^{\rm MHV}$ in powers of $\eta_i$ in such a way that $(\eta_i)^h\equiv \prod_{k=1}^h
\eta_i^{A_k}$ is associated with an external particle with momentum $p_i$ and helicity $1-h/2$. In
particular, the gluon MHV amplitude \re{MHV} arises as
\begin{align}\label{gen-MHV}
{\cal A}_{n}^{\rm MHV} =(2\pi)^4 \delta^{(4)} \big(\sum_{i=1}^n p_i\big)\sum_{1\le j<k\le n}
\lr{\eta_j}^4  \lr{\eta_k}^4 A^{\rm MHV} _n(1^+...\, j^-...\,  k^-...\,  n^+) + \ldots\,,
\end{align}
where the dots denote terms describing MHV amplitudes with {some external particles different from
gluons}. As was already mentioned, the function $M_n^{\rm MHV}$ exhibits a very interesting symmetry
when expressed in terms of the dual coordinates defined in \re{dualx}. Explicit perturbative
calculations show~\cite{abdk,bds,Bern:2006ew} that  $M_n^{\rm MHV}$  is given by a sum of Feynman
integrals which are formally (in $D=4$ dimensions) covariant under the $SO(2,4)$ transformations of
the dual coordinates $x_i^\mu$. Moreover, the conjectured duality between the MHV scattering
amplitudes and light-like Wison loops leads to the following relation~\cite{dks,bht},
\be\label{duality}
\ln M_n^{\rm MHV} = \ln W_n+ \text{const} + O(\epsilon,1/N^2)\,,
\ee
where $W_{n}=W(C_n)$ is the expectation value of a Wilson loop evaluated over the light-like polygonal
contour $C_n$ in Minkowski space-time, with vertices located at the points $x_i^\mu$ (with
$i=1,\ldots,n$),
\be
\qquad W_n =\frac1{N}\vev{0|\tr P\exp\lr{ig\oint_{C_n} dx^\mu A_{\mu}} |0}\,.
\ee
The divergences of $M_n^{\rm MHV}$ and $W_n$ in \re{duality} are regularized using dimensional
regularization with $D=4-2\epsilon$. {As mentioned earlier, the natural (anomalous) conformal symmetry of the Wilson loop induces a `dual' conformal symmetry of the MHV amplitude.}

In the present paper we show that the MHV superamplitude \re{A-SMHV} possesses an even bigger, dual superconformal symmetry. This symmetry acts on the dual coordinates $x_i^\mu$ and their superpartners $\q^A_{i\, \a}$ which are defined
in close analogy with
\re{dualx}  in terms of odd variables $\eta$ as follows,
 \begin{equation}\label{defrel2}
   \lambda_{i}^\a\, \eta^A_i =\q^{A\,\a}_{i} - \q^{A\,\a}_{i+1}\,.
\end{equation}
We demonstrate that when expressed in terms of the dual supercoordinates $(x_i,\, \lambda_{i}^\a,\,
\q^{A\, \a}_{i} )$ the MHV superamplitude \re{A-SMHV} transforms covariantly under $\cN=4$
superconformal ($SU(2,2|4)$) transformations. Most importantly, dual superconformal symmetry also
allows us to understand much better the complicated structure of the one-loop NMHV amplitudes. We
argue that, in a close analogy with the MHV amplitudes, all NMHV amplitudes can be combined into a
single superamplitude ${\cal A}_{n}^{\rm NMHV}$ {which is a homogeneous polynomial of degree 12 in
the Grassmann variables $\eta^A_i$}. Similarly to \re{gen-MHV}, e.g., the gluon NMHV amplitudes
arise as the coefficients in front of $\lr{\eta_i}^4  \lr{\eta_j}^4  \lr{\eta_k}^4$,
\begin{align}\label{gen-NMHV-glue}
{\cal A}_{n}^{\rm NMHV} =(2\pi)^4 \delta^{(4)} \big(\sum_{i=1}^n p_i\big) \sum_{i,j,k} \lr{\eta_i}^4
\lr{\eta_j}^4  \lr{\eta_k}^4 A^{\rm NMHV} _n(1^+...\, i^-...\, j^-...\,  k^-...\, n^+) + \ldots\ .
\end{align}
Quite remarkably, ${\cal A}_{n}^{\rm NMHV}$  transforms covariantly under the dual
superconformal transformations and has the same conformal weights as the MHV superamplitude
${\cal A}_{n}^{\rm MHV}$.  As a result, the `ratio' of the two superamplitudes is given by a linear
combination of {\it superinvariants} of the form (to one-loop order)
\be\label{gen-NMHV}
{\cal A}_{n}^{\rm NMHV} =  {\cal A}_{n}^{\rm MHV}\times\lr{\frac1n\sum_{p,q,r=1}^n c_{pqr} \,
 \delta^{(4)}\,\lr{\Xi_{pqr}} \left[ 1 + a V_{pqr} + O(\epsilon)\right] + O(a^2)}\,.
\ee
Here $\delta^{(4)}\,\lr{\Xi_{pqr}} \equiv \prod_{A=1}^4 \Xi_{pqr}^A$ are Grassmann delta functions.
The integers $p\neq q\neq r$  label three points in the dual superspace $(x_i,\lambda_{i}^\a,\q^{A\,
\a}_{i})\ $. From the coordinates of these three points one makes the dual superconformal covariants
$\Xi_{pqr}(x,\lambda,\q)$, linear in the odd variables $\q\ $. Both $\Xi_{pqr}$ and the c-valued
coefficients $c_{pqr}(x, \lambda)$ transform covariantly under dual superconformal transformations
in such a way that the product $c_{pqr} \, \delta^{(4)}\,\lr{\Xi_{pqr}}$ remains invariant.
Equivalently, the dual superconformal invariants can be rewritten as functions of the on-shell
superspace coordinates $(\lambda_i,\bl_i,\eta_i)$ as follows,
\begin{equation}\label{eqdusco}
    c_{pqr} = c_{pqr}(\lambda,\bl)\,, \qqqquad \Xi_{pqr} = \Xi_{pqr}(\lambda,\bl,\eta)\,,
\end{equation}
with the corresponding induced action of the dual superconformal algebra,
in accord with the defining relations \p{dualx} and \p{defrel2}.

The dependence on the coupling constant enters the right-hand side of \re{gen-NMHV} through the
perturbative corrections to the MHV superamplitude \re{MHV} and through the scalar factor involving
the functions $V_{pqr}(x_i)$. According to the duality relation \re{duality}, the former are
determined by the light-like Wilson loop $W_n$. Unlike $W_n$, the  functions
$V_{pqr}(x_i)$ remain finite as $\epsilon\to 0$ and, most importantly, they are exactly invariant under dual
conformal transformations of $x_i^\mu$. This means that the dual conformal symmetry of the
superamplitudes ${\cal A}_{n}^{\rm MHV}$ and ${\cal A}_{n}^{\rm NMHV} $ is broken by the infrared
divergences in such a way that the factor in parentheses in the right-hand side of \re{gen-NMHV}, to which
we shall refer as the `ratio' of the two superamplitudes, remains dual conformal as $\epsilon\to 0$.

We demonstrate this property by an explicit one-loop calculation for $n=6$ and we conjecture that it
should be true to all loops and for {\it all} superamplitudes in the $\mathcal{N}=4$ SYM theory,
\begin{align}\notag
{\cal A}_{n} &\equiv {\cal A}_{n}^{\rm MHV}+{\cal A}_{n}^{\rm NMHV}+{\cal A}_{n}^{\rm N^2MHV}+\ldots
+{\cal A}_{n}^{\rm N^{n-4}MHV}
\\[2mm]
& =  {\cal A}_{n}^{\rm MHV} \left[ R_n(\eta_i,
\lambda_i,\bl_i)+O(\epsilon)\right]\,.
\end{align}
In the first of these relations the sum runs over all N$...$NMHV amplitudes and ${\cal A}_{n}^{\rm
N^{n-4}MHV}={\cal A}_{n}^{\rm \widebar{MHV}}$ is the googly, or anti-MHV amplitude. The ratio
function
\be
R_n(\eta_i, \lambda_i,\bl_i)= 1 + R_n^{\rm NMHV}+R_n^{\rm N^2MHV}+\ldots +R_{n}^{\rm
N^{n-4}MHV}
\ee
is finite as $\epsilon\to 0$. Most importantly, it satisfies the dual conformal Ward identities
\be
 {K}^{\mu} R_n(\eta_i, \lambda_i,\bl_i) = {D}\, R_n(\eta_i,
\lambda_i,\bl_i) = 0\,,
\ee
with the dilatation $D$ and conformal boost $K^\mu$ operators defined appropriately in
the on-shell superspace $(\eta_i, \lambda_i,\bl_i)$.

In the simplest case $n=6$, the general expression for the NMHV superamplitude \re{gen-NMHV}
simplifies due to the fact that all possible supercovariants $\Xi_{pqr}$ can be expressed in terms
of $\Xi_{146}$ and five other supercovariants obtained from  $\Xi_{146}$ by consecutive cyclic shift
of the indices $i\to i+1$, with the periodicity condition $i+6\equiv i$. This leads to the following
remarkably simple one-loop expression for the $n=6$ NMHV superamplitude
 \be\label{n6-ansatz2}
{\cal A}^{\rm  NMHV }_{6} ={\cal A}^{\rm  MHV }_{6}
  \left[ \frac12 c_{146}\, \delta^{(4)}\lr{\Xi_{146}}\left(1+ a V_{146}\right) + \text{(cyclic)}+O(a^2)\right]
  \,,
\ee
where the MHV superamplitude ${\cal A}^{\rm  MHV }_{6}$ is given by \re{A-SMHV} and \re{duality}.
Here $\Xi_{146}$ can be expressed in terms of three Grassmann $\eta-$variables
\be
 \Xi_{146}= \vev{61}\vev{45}\big({\eta_4 [56]+\eta_5[64]+\eta_6[45]}\big)\,,
\ee
while the function $c_{146}$ is given by
\be
c_{146} = -  \frac{\vev{34}\vev{56}}{x_{46}^2\bra{1}\,x_{16}x_{63}|3]
\bra{1}\,x_{16}x_{64}|4]\bra{1}\,x_{14}x_{45}|5] \bra{1}\,x_{14}x_{46}|6]}  \,,
\ee
where the standard notation for contracting spinors $\lambda,\bl$ and vectors $x$ is used (see
Appendix A for more details). Finally, the scalar function $V_{146}$ is given by a linear
combination of scalar (1-mass, 2-mass-hard and 2-mass-easy in the nomenclature of \cite{bddk94}) box
integrals
\begin{align} 
V_{146}
& = - \ln u_1 \ln u_{2} +\frac12\sum_{k=1}^3  \bigg[{ \ln u_k \ln u_{k+1} + {\rm Li}_2(1-u_k)
}\bigg],
\end{align}
where the periodicity condition $u_{i+3}=u_i$ is implied. This function depends on the  conformally invariant cross-ratios $u_1$, $u_2$ and $u_3$ made from the  dual coordinates,
\be
u_1=\frac{x_{13}^2x_{46}^2}{x_{14}^2x_{36}^2}\,,\qqquad
u_2=\frac{x_{24}^2x_{15}^2}{x_{25}^2x_{14}^2}\,,\qqquad
u_3=\frac{x_{35}^2x_{26}^2}{x_{36}^2x_{25}^2}\,.
\ee
Expanding the right-hand side of  the relation \re{n6-ansatz2} in powers of $\eta$'s and comparing
the result with \re{gen-NMHV-glue}, it is straightforward to extract the expressions for the three different
six-gluon NMHV one-loop amplitudes $A^{+++---}$, $A^{++-+---}$ and $A^{+-+-+-}$ and to verify that
they agree with the known results \cite{bddk94-2}. In the same manner, one can also verify
that the superamplitude \re{n6-ansatz2} correctly reproduces the one-loop expressions for various NMHV
amplitudes involving scalars and gluinos~\cite{Risager:2005ke}.

The paper\footnote{The results of this paper were reported at the workshops ``Wonders of gauge
theory", Paris/Saclay, June 2008 \cite{PS} and ``Gauge theory and string theory", Zurich, July
2008.}  is organized as follows. In Sect.~2 we discuss the dual conformal properties of gluon
amplitudes. After recalling the formulation of MHV amplitudes in terms of commuting spinors
$\lambda_\a$ and $\bl_{\da}$, we introduce dual coordinates $x_{\a\da}$ and the notions of on-shell
(coordinates $\lambda,\bl$), full (coordinates $x,\lambda,\bl$) and dual (coordinates $x,\lambda$)
spaces. We determine the action of dual conformal symmetry (inversion) in these spaces, in
particular, we derive the transformations of the spinor variables $\lambda$ and $\bl$ from the known
ones of $x$. We then show that all split-helicity tree-level gluon amplitudes (MHV as well as
non-MHV) are manifestly dual conformal covariant. We conclude the section by recalling the anomalous
dual conformal behavior of the loop corrections to the MHV gluon amplitudes.

In Sect.~3 we introduce on-shell superspace, parametrized by the spinor variables $\lambda_\a$ and $
\bl_{\da}$, as well as by Grassmann variables $\eta^A$. We recall the content of the on-shell
$\cN=4$ gluon supermultiplet and give its realization in this superspace. Implementing the
conditions for on-shell supersymmetry, we rederive Nair's description of MHV tree-level
superamplitudes. We then discuss the general structure of $n-$particle superamplitudes of the
N${}^k$MHV type (with $k=0,1,\ldots, n-4$) formulated in the on-shell superspace. We write them in a
factorized form, with the MHV superamplitude as a prefactor, followed by a homogeneous polynomial of
degree $4k$ in the $\eta-$variables.

In Sect.~4, by analogy with the bosonic case, we go from on-shell superspace to full superspace with
coordinates $x_{\a\da},\lambda_\a,\bl_{\da},\q^A_\a,\eta^A$ and then to chiral dual superspace with
coordinates $x_{\a\da},\lambda_\a,\q^A_\a$. We describe the realization of the $\cN=4$ dual
superconformal algebra $su(2,2|4)$ in these spaces. This algebra has a central charge which is
identified with helicity. We show that the MHV tree-level superamplitudes become manifestly dual
superconformal covariant, if rewritten in the chiral dual superspace.

In Sect.~5 we apply this formalism to the simplest example of a NMHV superamplitude, the
six-particle case. We propose a compact description of the $n=6$ tree-level superamplitude as a
product of the MHV superamplitude, followed by a set of three-point superconformal invariants. The
generalization to one loop  is achieved by turning the coefficients of the superinvariants into
exactly dual conformal functions given in terms of one-loop box integrals. The dual conformal
anomaly is then confined to the MHV prefactor.  By expanding the superamplitude in the
$\eta-$variables, we obtain explicit expressions for various gluon amplitudes, and show that they
agree with the known results from the literature.

In Sect.~6 we generalize analysis to $n$-particle NMHV superamplitudes. We explain how to
systematically construct the three-point superconformal invariants, which match the coefficients of
the 3-mass-box integrals in the gluon amplitude. We show that their twistor coplanarity is an
immediate corollary of dual supersymmetry, combined with the obvious property of
`super-coplanarity'.  We then make a proposal for the complete one-loop NMHV superamplitude. As a
byproduct, we obtain a new, very compact representation of the tree-level NMHV superamplitude,
written down in a manifestly dual superconformal form.

Sect.~7 contains concluding remarks and outlook. Various technical details are collected in three appendices.

\section{Dual conformal symmetry of  gluon amplitudes}\label{DCSGA}

Since $\cN=4$ SYM is a (super)conformal theory, we can expect that the scattering amplitudes bear
some traces of this symmetry. This is indeed true, as shown by Witten for tree-level MHV amplitudes
in \cite{Witten:2003nn}. However, the action of the conformal group, being linear in configuration
space, is rather complicated in momentum space. For example, the conformal boosts are generated by a
second-order differential operator (see \re{witt} below). After a twistor transform this action
again becomes linear and the restrictions it imposes on the amplitude can be made manifest in the
twistor {representation}. However, the inverse twistor transform is difficult to perform explicitly,
therefore it is  not easy to exploit this type of conformal symmetry.

The main subject of the present paper is a completely different kind of conformal symmetry of the
scattering amplitudes, with linear action on the {\it momenta} of the particles. Its origin cannot
be traced back to the Lagrangian of the theory. We may say that this is a {\it dynamical symmetry}.
It becomes manifest after we introduce a {\it dual space} in Sect.~\ref{DS}. Its coordinates
$x_{\a\da}$ are expressed in terms of the particle momenta $p_{\a\da}=\lambda_\a\bl_{\da}$, where
$\lambda_\a$ and $\bl_{\da}$ are commuting spinor variables discussed in Sect.~\ref{MHVGA}. In
Sect.~\ref{DCS} we define a conformal group $SO(2,4)$ with the standard {\it linear action} on the
dual space coordinates $x_{\a\da}$ and derive the corresponding dual conformal transformations of
the spinor variables $\lambda_\a$ and $\bl_{\da}$. In Sect.~\ref{Tpma} we establish the
transformation properties of the tree-level MHV gluon amplitudes under dual conformal
transformations. In Sect.~\ref{split} we generalize the analysis to all split-helicity gluon
amplitudes at tree level. In Sect.~\ref{sect27} we give the form of the dual conformal boost
generator. Finally, in Sect.~\ref{DCPFM} we discuss the loop corrections to the MHV amplitude. They
produce infrared divergences which break the dual conformal symmetry. This breakdown is  controlled
by an anomalous dual conformal Ward identity.

\subsection{MHV gluon tree-level amplitudes}\label{MHVGA}

The main objects of study in this paper are scattering amplitudes of massless particles in gauge theories.
Specifically, we are interested in $\cN=4$ SYM theory, which involves bosons  (gluons and
scalars), as well as fermions  (gluinos). Each of these particles is characterized by its on-shell
momentum $p_i^\mu$ ($i=1\ldots n$) and helicity $h_i = \pm1$ (gluons), $\pm 1/2$ (gluinos), $0$
(scalars). We shall treat all particles as incoming, so that the total momentum conservation
condition reads
\begin{equation}\label{momcon}
    \sum_{i=1}^n\ (p_i)^{\da\a} = 0\,.
\end{equation}
To solve the on-shell condition for $p_i$, it is convenient to introduce {\it commuting spinor}
variables\footnote{In this paper we use two-component Weyl
spinor notation for Lorentz vectors and spinors, see Appendix \ref{A} for details.}
\begin{equation}\label{2'}
   p_i^2 ={1\over 2} (p_i)^{\dot\alpha\alpha}(p_i)_{\alpha\dot\alpha}= 0 \qquad
   \Longrightarrow \qquad  (p_i)^{\da\a} = \bl_{i}^{\da}\, \lambda_{i}^{\a} \,,
\end{equation}
where $\lambda_{i}^{\a}$ ($\a=1,2$) and $\bl_{i}^{\da}$ ($\da=\dot 1, \dot 2$) are two-component
spinors. We will refer to the space with coordinates $(\lambda_i,\tilde\lambda_i)$ as the `on-shell'
space. {If we wish to have real momenta $p_i^\mu$ in a space-time with Minkowskian signature, these
spinors transform under the Lorentz group $SL(2, \mathbb{C})$, and $\bl = \pm \lambda^*$.  However,
in various applications of the spinor formalism (for instance, in the generalized unitarity cuts
approach of \cite{Britto:2004nc}), it is preferable to keep the momenta complex. In this case, $\bl$
is not the complex conjugate of $\lambda$.}

Equation \p{2'} allows us to determine $\lambda_{i}^{\a}$ and $\bl_{i}^{\da}$ in terms of the
(complex) momentum $(p_i)^{\da\a}$ up to a {\it local} (i.e., depending on the point $i$) complex
scale,
\begin{equation}\label{scale}
    \lambda_i^\a \ \to \ c_i\lambda_i^\a\,, \qqqquad \bl_i^\da \ \to \ c_i^{-1}\bl_i^\da\,.
\end{equation}
For real momenta $(p_i)^{\da\a}$ we have $|c_i|=1$, and the resulting  $U(1)$ phase can be
identified with the particle helicity at point $i$. The standard convention is that the spinors
$\lambda$ and $\bl$ carry helicities $-1/2$ and $+1/2$, respectively. Thus, the momentum
$(p_i)^{\da\a}$ has vanishing helicity. For complex momenta, we can still use the generalized notion
of `helicity', meaning the weight under the scale transformation \p{scale}.

Alternatively \cite{Witten:2003nn}, we can say that the spinor variable $\lambda_{i\,\a}$ appears in
the solution to the Weyl (or massless Dirac) equation for a chiral spin-1/2 particle:
\begin{equation}\label{plane}
   (p_i)^{\da\a} \psi_\a(p_i) = 0 \qquad \Longrightarrow \qquad \psi_\a(p_i) =  \lambda_{i\,\a}
   \Gamma(p_i)\,,
\end{equation}
provided the particle momentum satisfies the condition
\begin{equation}\label{parmomcon}
     (p_i)^{\da\a}\lambda_{i\,\a} = 0\,.
\end{equation}
The general solution to \p{parmomcon} for $(p_i)^{\da\a}$ is of the form \p{2'}, thus introducing
the antichiral spinor $\bl_{i}^{\da}$.

An important point is that the wave function $\psi_\a(p)$, belonging to a representation of the
Lorentz group (a Weyl spinor), cannot have helicity. Indeed, helicity is a label for the {\it
massless} representations of the Poincar\'e group, defined in a fixed  Lorentz frame where $p^\mu =
(p,0,0,p)$. The advantage of using the spinor variables $\lambda$ and $\bl$ is that we can make a
bridge between Lorentz and massless Poincar\'e representations. Thus, the helicity $+1/2$ of the
`particle' (Poincar\'e state) $\Gamma(p)$ is compensated by the helicity $-1/2$ of $\lambda_\a$,
while its spinor index makes the wave function $\psi_\a(p)$ transform as a Lorentz representation.
Similarly, we can relate an antichiral Weyl spinor field to a particle of helicity $-1/2$,
$\bar\psi_{\da}(p) = \bl_{\da} \bar\Gamma(p)$.

In the same way, we can introduce the spinor description of gluon states. On-shell gluons are
massless Poincar\'e states of helicity $\pm1$ described, correspondingly, by $G^{\pm}(p)$. Their
Lorentz covariant description makes use of the self-dual 
and
anti-self-dual
parts of the gluon field strength tensor,
$G_{\a\b}(p)$ and $\bar G_{\da\db}(p)$, respectively, satisfying the equations of motion
$p^{\da\a}G_{\a\b}(p) =
\bar G_{\da\db}(p)p^{\db\a} = 0$. Once again, the bridge between Poincar\'e and Lorentz
representations is made with the help of the spinor variables, $G_{\a\b}(p) = \lambda_\a \lambda_\b\
G^+(p)$ and $\bar G_{\da\db}(p) =  \bl_{\da} \bl_{\db}\ G^{-}(p)$.

Now, let us consider the simplest example of $n$-gluon MHV scattering amplitudes. By definition,
they involve only two gluons of, say negative helicity, while the other $n-2$ gluons have positive
helicity.~\footnote{For amplitudes with external particles different from gluons, their
classification (MHV, NMHV, $\ldots$) is based on the total helicity weight (the sum of helicities of
all particles). Namely, $n-$particle MHV amplitudes have the total helicity $n-4$, next-to-MHV
(NMHV) the helicity $n-6$, etc.} Different MHV gluon amplitudes are then defined by the positions of
the two negative-helicity gluons, e.g., $(--+\ldots+)$, $(-+-+\ldots+)$, etc. In the former case,
the negative-helicity gluons appear contiguously and not separated by positive-helicity gluons as in
the latter case. Such amplitudes are called `split-helicity amplitudes' \cite{Britto:2005dg}.

In the spinor formalism, the MHV tree-level amplitude with the  two negative-helicity gluons
occupying sites $i$ and $j$ is described by the following function of the spinor variables:
\begin{equation}\label{mhv}
{\cal A}_{n;0}\left(1^+\ldots i^-  \ldots  j^- \ldots n^+\right)  =
i(2\pi)^4\delta^{(4)}(\sum_{k=1}^n\, p_k)\ \frac{\vev{i\, j}^4}{\vev{1\, 2}\vev{2\, 3}\ldots\vev{n\,
1}}
\end{equation}
(here and in what follows ${\cal A}_{n;0}$ denotes a tree-level $n$-particle amplitude and $i^\pm$
stands for the gluon state $G^\pm(p_i)$). The delta function in \p{mhv} is the solution of the
momentum conservation condition \p{momcon}, or equivalently of the condition for  translation
invariance of the amplitude,
\begin{equation}\label{momconse}
{p}^{\da\a} \ {\cal A}_n(p) = 0 \qquad \Longrightarrow \qquad  {\cal A}_n(p) \propto
\delta^{(4)}\big(\sum_{i=1}^n\, p_i\big),
\end{equation}
where the total momentum
\begin{equation}\label{mathP}
    {p}^{\da\a} = \sum_{i=1}^n\ (p_i)^{\da\a}
\end{equation}
is the generator of translations in the momentum representation. The other factor in \p{mhv} is a rational and {holomorphic} function of the Lorentz invariant
contraction of spinor variables (see Appendix \ref{A} for our conventions for two-component
spinors),
\begin{equation}\label{twcont}
    \vev{i\, j} = -\vev{j\, i} = \lambda^\a_{i}\, \ep_{\a\b} \,\lambda^\b_{j} =  \lambda^\a_{i}\lambda_{j\,\a}\,.
\end{equation}
Since each spinor $\lambda$ carries helicity $(-1/2)$, this factor has helicity $(-1/2)$ at points
$i$ and $j$. By counting the degree of homogeneity in $\lambda$ on the right-hand side of \re{mhv},
we read off the helicities $(-1)$ at points $i$ and $j$, and $(+1)$ elsewhere.

We conclude this subsection by recalling that MHV tree-level amplitudes \re{mhv} have a conformal
symmetry \cite{Witten:2003nn}. This is the ordinary conformal symmetry $SO(2,4)$  of the
$\mathcal{N}=4$ SYM Lagrangian, but realized on the particle momenta (or, equivalently, on the
spinor variables $\lambda$ and $\bl$). In particular, the conformal boost generator takes the form
of a second-order differential operator
\begin{equation}\label{witt}
    k_{\a\da} = \sum_{i=1}^n\ \frac{\pa^2}{\pa \lambda^\a_i \pa \bl^{\da}_i}\,,
\end{equation}
leading to $k_{\a\da}{\cal A}_{n;0} = 0$.

\subsection{On-shell space and full space}\label{FS}

The scattering amplitudes have the following general form
\be\label{amplitude}
\mathcal{A}_n = i (2\pi)^4 \delta^{(4)}(\sum_{i=1}^n p_i) A_n(p_1,\ldots, p_n)\,.
\ee
The function $A_n$ depends on the momenta $p_i$ of the incoming on-shell particles which are
constrained in two ways. They are light-like vectors $p_i^2=0$ and satisfy the momentum conservation condition \p{momcon}.

As we saw in the previous subsection, it is natural to solve the on-shell constraints by introducing
spinor variables $\lambda_i$ and $\tilde\lambda_i$, Eq.~\re{2'}. For this reason we
call them coordinates of the `on-shell space'. However, we should remember that the on-shell coordinates still satisfy a constraint following from momentum conservation \p{momcon}:
\begin{equation}\label{momcon'}
    \sum_{i=1}^n\ \bl_{i}^{\da}\, \lambda_{i}^{\a} = 0\,.
\end{equation}
It means that not all of the variables $\lambda_{i}\,, \bl_{i}$ are independent. For instance, we
could eliminate any two spinors $\lambda_k$ and $\lambda_m$ (or two $\bl$'s) in terms of the
remaining $n-2$.

Alternatively, we might wish to first solve the momentum conservation condition. To this end we
introduce new `dual' coordinates $x_i$ (with $i=1,\ldots,n$)
\begin{align} \label{momconssol}
\sum_{i=1}^n (p_i)^{ \dot\alpha \alpha}=0  \implies (p_i)^{\dot\alpha \alpha} = (x_i)^{\dot\alpha
\alpha} - (x_{i+1})^{\dot\alpha\alpha}\,,
\end{align}
satisfying the cyclicity condition
\begin{equation}\label{cyc}
    x_{n+1} \equiv x_1\,.
\end{equation}
Of course, we still have to impose the on-shell conditions $p^2_i=0$, which imply that the dual coordinates are constrained,
\begin{equation}\label{llsep}
    (x_i - x_{i+1})^2 = 0\,.
\end{equation}

We must make it very clear that the
new variables $x_i$ have nothing to do with the coordinates in the original configuration space
(which are the Fourier conjugates of the particle momenta $p_{i}$). Indeed, as can be seen from
\p{momconssol}, the $x-$variables have the `wrong' dimension of mass. As we will see shortly, these
variables provide the natural framework for discussing the new `dual' conformal symmetry of the amplitude.

{At this stage we have proposed two sets of variables for describing the amplitude, the on-shell coordinates $\lambda_i,\tilde\lambda_i$ satisfying the constraint \p{momcon'}, and the dual coordinates $x_i$  satisfying the constraint \p{llsep}. We can now combine these two sets of variables into a single one, by defining the extended space with coordinates
$(\lambda_i,\tilde\lambda_i,x_i)$, which we call  the `full space'.}
From the compatibility of the solutions to
\p{2'} and \p{momconssol} it  follows that
\begin{equation}\label{masterconstraint}
    (x_i-x_{i+1})^{\da\a}  = \bl_{i}^{\da}\,\lambda_{i}^{\a} \,, \qquad x_{n+1} \equiv x_1\,.
\end{equation}
It is clear this identification yields both constraints  \p{momcon'} and  \p{llsep}.

We can think of the relation \p{masterconstraint} as defining a surface in the full space. Then we can interpret the
function $A_n(p_i)=A_n(\lambda_i,\tilde\lambda_i,x_i)$ appearing in the amplitude (\ref{amplitude}) as a
function defined on this surface.
Since $A_n$ is a function of the
particle momenta $p_i$, it can only depend on the dual coordinates through their differences
$x_i-x_j=x_{ij}$, thus implying a dual translation invariance,
\be
P_{\alpha \dot\alpha} \, A_n(\lambda_i,\tilde\lambda_i,x_i)  = 0 \label{dualtransinv}
\ee
with $P_{\alpha \dot\alpha}$ being the generator of translations
\begin{equation}\label{gendutr1}
    P_{\a\da} = \sum_{i=1}^n \frac{\pa}{\pa x^{\da\a}_i}\,.
\end{equation}
We would like to stress the fact that the generator $P_{\a\da}$ has nothing to do with the
conventional translation generator ${p}_{\a\da}$, Eq.~\p{mathP}. The latter acts as a shift of the
coordinates in the original configuration space of the particles, while \p{gendutr1} acts as a shift
of the {\it dual coordinates}.

To put it differently, we can say that the dual coordinates $x_i$ can be solved for from
\p{masterconstraint}, up to the freedom of choosing an arbitrary reference point, e.g., $x_1$:
\begin{equation}\label{xthrl1}
    (x_i)^{\dot\alpha\alpha} = (x_1)^{\dot\alpha\alpha} - \sum_{k=1}^{i-1}\
    \bl_{k}^{\dot\alpha}\lambda_{k}^{\alpha}\,.
\end{equation}
In other words, the definition of the dual coordinates \p{masterconstraint} is invariant under
shifts of the $x_i$'s by an arbitrary constant vector. Clearly, using the dual translation
invariance (\ref{dualtransinv}) and the constraint (\ref{masterconstraint}), we can always eliminate
$x_i$ and obtain $A_n$ as a function of $\lambda_i$ and $\tilde\lambda_i$ on the on-shell space. The
other possibility is to eliminate $\tilde\lambda_i$ with the help of \re{masterconstraint} and to
express $A_n$ as a function of $x_i$ and $\lambda_i$ only. This leads to a holomorphic description
of the amplitudes that we discuss in the following subsection.

\subsection{Dual space}\label{DS}

We can rewrite the constraint (\ref{masterconstraint}) without using the variables
$\tilde\lambda_i$,
\begin{equation}\label{twpr}
    (x_{i\ i+1})^{\da\a} \lambda_{i\,\a} = \lambda_i^\a\, (x_{i\ i+1})_{\a\da} =  0\,,
\end{equation}
where the shorthand notation $x_{i\ i+1} = x_i - x_{i+1}$ was used. This form of the
constraints is reminiscent of, and partially inspired by the conditions on the spin-1/2 particle
momentum \p{parmomcon}. Additional motivation for introducing the constraints in the form of
\re{twpr} will come from our discussion of dual supersymmetry in Sect.~\ref{DSS}.

The relations (\ref{twpr}) and  (\ref{masterconstraint}) are in fact equivalent, since the general
solution to \p{twpr} takes the form
\begin{equation}\label{takeform}
    (x_{i\ i+1})^{\da\a}  = \bl_{i}^{\da}\,\lambda_{i}^{\a}\,,
\end{equation}
thus introducing the {\it secondary} variables $\bl_i$. In other words, the $\bl$'s can be expressed
in terms of $x$ and $\lambda$ by projecting \p{takeform} with, e.g., $\lambda_{i+1}^\a$:
\begin{equation}\label{exprbarl1}
     \bl_{i}^{\da} =   \frac{(x_{i\ i+1})^{\da\a} \lambda_{i+1\,\a}}{\vev{i\ i+1}}\,.
\end{equation}
Once we have deduced \p{takeform} from the defining constraint \p{twpr}, we can make contact with
the momenta of the particles through the identification $\bl_{i}^{\da}\,\lambda_{i}^{\a} =
(p_{i})^{\da\a}$. Applying the relation \re{exprbarl1}, the function
$A_n(\lambda_i,\tilde\lambda_i,x_i)$ can now be regarded as a function of the variables $x_i$ and
$\lambda_i$.

We call the space with coordinates $(x_i,\lambda_i)$ satisfying the defining constraint \p{twpr} the `dual space'. Note that this space is  holomorphic -- we only need the (complex) variables $x_i,\lambda_i$, but not their complex conjugates. Later on,  in Sect.~\ref{DSS} we shall see that the holomorphic dual space has a natural extention to a chiral dual superspace.

In summary, we are proposing three different but equivalent descriptions of the scattering amplitudes in the on-shell space $(\lambda_i ,\tilde \lambda_i)$, in the full space $(x_i,
\lambda_i, \tilde \lambda_i)$ and in the dual space $(x_i, \lambda_i)$. The relations between these
equivalent descriptions are shown in the following diagram:

\vspace*{10mm}
\psfrag{Onshell}[cc][cc]{\parbox[t]{26mm}{ {On-shell space} \\[3mm]
\centerline{$(\lambda_i ,\tilde \lambda_i)$}}}
\psfrag{Dual}[cc][cc]{\parbox[t]{22mm}{ {Dual space} \\[3mm]
\centerline{$(x_i, \lambda_i)$}}}
\psfrag{Extended}[cc][cc]{\parbox[t]{18mm}{ {Full space} \\[3mm]
\centerline{$(x_i, \lambda_i, \tilde \lambda_i)$}}}
\psfrag{eq1}[cr][cr]{Eq.\,\re{xthrl1}}\psfrag{eq2}[cl][cl]{Eq.\,\re{exprbarl1}}
\psfrag{eq3}[cc][cc]{Eq.\,\re{masterconstraint}}
\centerline{{\epsfysize4cm \epsfbox{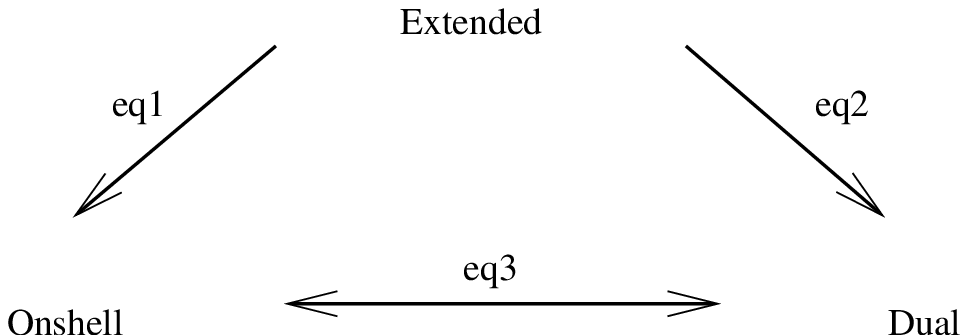}}}

\vspace{8mm}

\noindent As we will see in a moment, the dual space offers the most convenient framework for
understanding the new `dual conformal' symmetry of the scattering amplitudes.

\subsection{Dual conformal symmetry}\label{DCS}

As mentioned earlier, the main motivation for introducing the dual space was to exhibit a new,
unexpected conformal symmetry of the MHV amplitude. This is the conformal group $SO(2,4)$, which
acts {\it linearly} on the dual space coordinates, i.e. on the particle momenta (and not, we stress
again, on the coordinates of the particles in configuration space).

Our task in this section will be to learn how the dual conformal symmetry acts on the coordinates
$(x,\lambda)$ of the dual space. We shall assume that the dual coordinates $x^{\da\a}$ transform in
the standard way under the conformal group $SO(2,4)$, and then deduce the transformation properties
of $\lambda^\a$ by requiring that the defining relation \re{twpr} should remain covariant.

It is well known that the conformal group $SO(2,4)$ can be obtained from the Poincar\'e group by
adding the discrete operation of conformal inversion,
 \begin{equation}\label{inver}
   I[x_{\a\dot\beta}] = \frac{x_{\beta\da}}{x^2} \equiv (x^{-1})_{\beta\da}\,.
\end{equation}
Notice that the inversion changes the chirality of two-component spinor indices. Performing an
inversion, followed by an infinitesimal translation and then by another inversion, we obtain the
generators of special conformal transformations (boosts), $K^\mu = I P^\mu I$. Then, commuting
$P^\m$ with $K^\nu$ we get the rest of the conformal algebra $o(2,4)$, namely, Lorentz
transformations and dilatations. The operation of inversion is an involution, $I^2 = \mathbb{I}$,
which easily follows from the definition \p{inver}.

Let us now examine the effect of an inversion on the matrix $(x_i-x_{i+1})_{\a\dot\beta}$. Using
\p{inver} 
we obtain
\begin{equation}\label{weremthatt}
I [(x_{i\,j})_{\a\dot\beta}] = \left(x^{-1}_i - x^{-1}_{j} \right)_{\beta\da} =
-{(x^{-1}_{j})_{\b\db}}\ ( x_i- x_{j})^{\db\gamma}\ {(x^{-1}_i)_{\gamma\da}} \equiv -(x^{-1}_{i}
x_{ij} x^{-1}_{j})_{\b\da}\,.
\end{equation}
Specifying to the case $j=i+1$, this becomes
\begin{equation}\label{3}
  I [(x_{i\ i+1})_{\a\dot\beta}] = -{(x^{-1}_{i+1})_{\b\db}}\ ( x_{i\, i+1})^{\db\gamma}\
  {(x^{-1}_i)_{\gamma\da}}\,.
\end{equation}
Now, we can use \p{3} to deduce the inversion properties of the spinor variables $\lambda$, such
that the constraint \p{twpr} remains covariant,
\be
I\left[\lambda_i^\a\, (x_{i\ i+1})_{\a\db}\right] = -{(x^{-1}_{i+1})_{\b\db}}\ ( x_{i\,
i+1})^{\db\gamma}\ {(x^{-1}_i)_{\gamma\da}}I\left[\lambda_i^\a\right]= 0\,.
\ee
Comparing this identity with the first relation in \re{twpr}, we obtain that
${(x^{-1}_i)_{\gamma\da}}I\left[\lambda_i^\a\right]\sim \lambda_{i\,\gamma}$, or equivalently
\be\label{l-inv}
I\left[\lambda_i^\a\right] = \kappa_i\, (x_i)^{\da\b}\lambda_{i\,\b}\,,\qquad
I\left[\lambda_{i\,\a}\right] = \kappa_i\,\lambda_{i}^{\b} \,(x_i)_{\b\da}\,,
\ee
where $\kappa_i$ is an arbitrary ($x-$dependent) weight factor. Here the second relation follows
from the first one after we raise/lower spinor indices and take into account the standard rules for
inversion of the Levi-Civita tensors,\footnote{They can be obtained by considering, e.g., the
Lorentz invariant contraction of two spinors, $\lambda_{1 \a} \ep^{\a\b} \lambda_{2 \b}$. Inversion
changes the chirality of a spinor (see \p{l-inv}), therefore $\ep^{\a\b}$ and $\ep^{\db\da}$ have to
be swapped. }
\begin{equation}\label{LC}
    I[\ep^{\a\b}  ] = \ep^{\db\da}\,, \qquad I[\ep^{\da\db}] =  \ep^{\b\a}\,.
\end{equation}
It is straightforward to verify that the inversion defined in \p{l-inv} is an involution, $I^2 =
\mathbb{I}$, provided that $I[\kappa_i] = 1/\kappa_i$.

Using the transformation properties \p{l-inv}, we can easily show that the Lorentz invariant
contraction of two adjacent spinors $\vev{i\ i+1}$ is covariant under inversion:
\begin{eqnarray}\label{covvev1}
  I [\vev{i\ i+1}] &=& I[ \lambda_{i}^{\a}\ \lambda_{i+1\, \a}]
  = - \kappa_{i} \kappa_{i+1}\,\lambda_{i+1}^{\b} \,(x_{i+1}x_i)_{\b\a}\lambda_i^\a
  =  \kappa_{i} \kappa_{i+1}\vev{i\, i+1}x_{i+1}^2\,,
\end{eqnarray}
where in the last relation we took into account the constraint \re{twpr} to replace
$(x_{i+1}x_i)_{\b\a}\lambda_i^\a=(x_{i+1}x_{i+1})_{\b\a}\lambda_i^\a = x_{i+1}^2 \lambda_i^\b$.

To fix the value of the weight factor $\kappa_i$, it is convenient to examine how the antichiral spinors $\bl$ transformation.
We recall that they are not independent variables in the dual
space, being related to $x$ and $\lambda$ through \re{exprbarl1}. Applying
inversion to both sides of \re{exprbarl1} and taking into account \re{3}, \re{l-inv}
and \re{covvev1}, we find after some algebra that $\tilde\lambda$ also transform covariantly,
\be\label{lb-inv}
I[\tilde \lambda_i^\da] = \tilde\kappa_i\, {\tilde \lambda_{i\,\db} (x_i)^{\db\a}}  \,,\qquad
I[\tilde \lambda_{i\,\da}] = \tilde\kappa_i\, (x_i)_{\a\db}\tilde\lambda_{i}^\db\,,
\ee
with $\tilde\kappa_i = (\kappa_i x_{i}^2 x_{i+1}^2)^{-1}$.

Let us now compare the relations \re{l-inv} and \re{lb-inv}. If we wish to treat $\bl_i$ as the
complex conjugate of $\lambda_i$, the choice of the weight factor $\kappa_i$ is unambiguous (up to a
phase), $\kappa_i = \tilde \kappa_i = \lr{x^2_{i}x^2_{i+1}}^{-1/2}$. We prefer instead to follow the
holomorphic dual space description and treat $\bl$ as secondary, dependent variables. In this case,
we can take an arbitrary $\kappa_i$. A natural choice is $\kappa_i=1/x_i^2$ and $\tilde \kappa_i
=1/x_{i+1}^2$. Substituting these expressions into \re{l-inv} and \re{lb-inv}, we obtain
\begin{align}\label{l-lb-inv}
& I\left[\lambda_i^\a\right] = (x_i^{-1})^{\da\b}\lambda_{i\,\b}\,, & &
I\left[\lambda_{i\,\a}\right] = \lambda_{i}^{\b} \,(x_i^{-1})_{\b\da}\,,
\\[2mm] \notag
& I[\tilde \lambda_i^\da] =  {\tilde \lambda_{i\,\db} (x_{i+1}^{-1})^{\db\a}}  \,,& & I[\tilde
\lambda_{i\,\da}] = (x_{i+1}^{-1})_{\a\db}\tilde\lambda_{i+1}^\db\,,
\end{align}
where $(x^{-1})_{\a\da} = x_{\a\da}/x^2$ and we applied the constraint \re{twpr} in the second line
to replace $(x_i)_{\a\db}\tilde\lambda_{i}^\db=(x_{i+1})_{\a\db}\tilde\lambda_{i}^\db$.

We are now ready to discuss how to build covariants of the dual conformal transformations defined
above. Specifically, for the purpose of constructing the scattering amplitudes, we are interested in
Lorentz invariant functions of $x_i$, $\lambda_i$ and $\bl_i$. If we restrict ourselves to functions
of $x_i$ only, then the basic covariants are the `distances' between two points
$x^2_{ij}$:\footnote{Recall that in the dual space for scattering amplitudes $x^2_{i\ i+1}=p^2_i =
0$.}
\begin{equation}\label{inverx2}
    I[x^2_{ij}] = \frac{x^2_{ij}}{x^2_i x^2_j}\,.
\end{equation}
We have additional possibilities if we also include the spinor variables. Using the transformation
properties \p{l-lb-inv}, we can easily show that the Lorentz invariant contractions of two adjacent
(anti)chiral spinors $\vev{i\ i+1} \equiv \lambda_i^\a \lambda_{i+1\,\a}$ and $[i\ i+1] \equiv
\tilde\lambda_{i\,\da} \tilde\lambda_{i+1}^\da $ are covariant under inversion (see also
\re{covvev1}):
\begin{align} \label{covvev}
I \Big[\vev{i\ i+1}\Big] &=  (x^2_{i})^{-1}\ {\vev{i\ i+1}}\,,
\\ \notag
I\Big[ [i\ i+1] \Big] &= (x^2_{i+2})^{-1}\ {[i\ i+1]}\,.
\end{align}
Note, however, that all other Lorentz invariant contractions $\vev{ij}$ (or $[ij]$) with $j \neq
i+1$ are {\it not covariant} under the dual conformal symmetry. Besides these simplest examples,
there exists a large variety of dual conformally covariant and Lorentz invariant `strings' made of
$x_i$, $\lambda_i$ and $\bl_i$ that will be presented in Sect.~\ref{split}.


\subsection{Dual conformal transformation properties of the MHV tree-level gluon amplitudes}\label{Tpma}

In this section we examine the transformation properties of the MHV tree-level gluon amplitudes
\p{mhv} under the dual conformal transformations \re{inver} and \re{l-lb-inv}.

Let us first apply  \re{inver} and \re{l-lb-inv} to the MHV amplitude ${\cal A}^{\rm
MHV}_{n;0}\left( 1^- 2^-3^+ \ldots n^+\right)$, which is an example of a split-helicity amplitude.
According to \re{mhv}, this amplitude is given by a product of a momentum  delta function and a
rational holomorphic function of the spinor variables with $i=1$ and $j=2$. We start with the latter
and use \re{covvev} to obtain its conformal inversion,
\begin{equation}\label{invrat}
    I\left[ \frac{\vev{1\, 2}^4}{\vev{1\, 2}\vev{2\, 3}\ldots\vev{n\, 1}} \right] =
    \frac{x_2^2 x_3^2  \ldots x_n^2}{(x_1^2)^3  }\ \frac{\vev{1\, 2}^4}{\vev{1\, 2}\vev{2\, 3}\ldots\vev{n\,
    1}}\,.
\end{equation}
Note the appearance of conformal weights $(-3)$ at point $x_1$ and $(+1)$ at all remaining
points.\footnote{Here we use the same conventions for assigning conformal weights as in conformal
field theory. Namely, the two-point function of a primary scalar field with conformal weight $j$ has
the form  $\vev{\phi(x_1)\, \phi(x_2)} = 1/{x_{12}^{2j}}$. Under inversions it transforms as
$I[\vev{\phi(x_1)\, \phi(x_2)}] =  {(x^2_1 x^2_2)^j} \vev{\phi(x_1)\, \phi(x_2)} $, which explains
the weight assignments in \p{invrat}.  }

Let us now examine the conformal properties of the delta function in \p{mhv}. Its role is to impose
the momentum conservation condition \p{momcon}, or equivalently \p{momconssol}.  Any function,
multiplied by $\delta^{(4)}(\sum_{i=1}^n p_i)$ (for example, the factor discussed in \p{invrat}) is
thus defined on the constraint surface \re{masterconstraint} in the full space with the coordinates
$(x_i, \lambda_i, \bl_i)$. By construction,  the transformation properties \re{inver} and
\re{l-lb-inv} are consistent with the constraints \re{momconssol} and \p{masterconstraint},
therefore we may say that dual conformal transformations do not take us out of the constraint
surface. But we still have to answer the question how the delta function itself transforms.
Obviously, we cannot use the constraints \p{momconssol}, together with the cyclicity condition
$x_{n+1}=x_1$, because this will lead to the  vanishing of the argument of the delta function.

The answer to this question is found by realizing that the momentum conservation constraint is
solved by the substitution $p_i = x_i - x_{i+1}$ only if the cyclicity condition \p{cyc} is imposed.
Let us relax it for a moment, $x_{n+1} \neq x_1$, while still keeping the relations $(x_i -
x_{i+1})^{\da\a} = \bl_i^\da\lambda_i^\a$. In particular, we assume that $(x_n - x_{n+1})^{\da\a} =
\bl_n^\da\lambda_n^\a$ instead of the cyclic $(x_n - x_{1})^{\da\a} = \bl_n^\da\lambda_n^\a$. Then
$\sum_{i=1}^n p_i = x_1 - x_{n+1} \neq 0$, the delta function in \p{mhv} is replaced by $
\delta^{(4)}(x_1 - x_{n+1})$ and the amplitude takes the following form:
\be
{\cal A}^{\rm MHV}_{n;0}\left(1^-2^- 3^+ \ldots n^+ \right)  = i(2\pi)^4\delta^{(4)}(x_1 - x_{n+1})\
\frac{\vev{1\, 2}^4}{\vev{1\, 2}\vev{2\, 3}\ldots\vev{n\, 1}}\,.
\ee
The role of the delta function now is to impose the identification $x_{n+1} = x_1$ in the rational
factor in \p{mhv}, instead of the momentum conservation  \p{momcon}. Such a delta
function, defined in the dual space, is manifestly conformally covariant. It has conformal weight
four at point $1$ (needed to compensate that of the integration measure, $\int d^4 x_1\
\delta^{(4)}(x_1-x_{n+1}) = 1$). Thus, the tree-level MHV split-helicity amplitude transforms under
conformal inversions as follows,
\begin{equation}\label{invratcompl}
    I\left[{\cal A}^{\rm MHV}_{n;0}\left(1^-2^- 3^+ \ldots n^+\right) \right]
    = \lr{x_1^2 x_2^2   \ldots x_n^2}\ {\cal A}^{\rm MHV}_{n;0}\left(1^-2^- 3^+ \ldots n^+
    \right)\,,
\end{equation}
so we conclude that it has conformal weight $(+1)$ at all points. In a similar manner, it is
straightforward to verify that all tree-level MHV split-helicity amplitudes ${\cal A}^{\rm
MHV}_{n;0}\left(\ldots G^-_{i} G^-_{i+1} \ldots \right)$ are dual conformal covariant. Moreover, as
we will show in the next subsection, the same property holds for {\it  all split-helicity tree-level non-MHV
amplitudes} ${\cal A}_{n;0}\left(1^-\ldots q^- (q+1)^+ \ldots n^+ \right)$, i.e. those amplitudes in
which the negative-helicity gluons appear contiguously.

However, the tree-level non-split-helicity MHV amplitudes ${\cal A}^{\rm MHV}_{n;0}\left(\ldots i^-
\ldots j^- \ldots \right)$, Eq.~\re{mhv}, involve the spinor contractions $\vev{i\, j}$ which, as
noted above, are not covariant for $|i-j|>1$. Therefore the MHV gluon amplitudes  where the
negative-helicity gluons are not at adjacent points, are not dual conformal, at least not on their
own. At first sight, it might seem that dual conformal symmetry is an isolated
property of a very special class of gluon amplitudes, the split-helicity amplitudes. In fact, the full
understanding of the role of dual conformal symmetry is achieved when the gluon amplitudes are
combined together with the amplitudes involving other particles (gluinos, scalars) into a bigger and
unifying object, the $\cN=4$ superamplitude. We do this in Sect.~\ref{DSCPMHV} where we show that
only the complete MHV superamplitude (see \p{concorr} below) has well defined conformal properties.
There we also explain why its various components, i.e. the split-helicity and other gluon amplitudes
behave differently under dual conformal transformations.

\subsection{Split-helicity gluon amplitudes at tree level}\label{split}

Let us now show that dual conformality is a property of a much wider class of non-MHV amplitudes,
the split-helicity amplitudes. These are color-ordered $n-$gluon amplitudes with  the helicities
distributed as ${\cal A}_{n;0}\left(1^-\ldots q^-,(q+1)^+\ldots n^+\right)$.

The split-helicity amplitudes are special because they form a closed set under the BCF/BCFW
tree-level recursion relations \cite{Britto:2004ap,Britto:2005fq}. These relations have been solved
in \cite{Britto:2005dg} and the explicit expression for the tree level split-helicity amplitudes reads
\begin{equation}\label{zig-zag}
{\cal A}_{n;0}\left(1^-\ldots q^- (q+1)^+\ldots n^+\right) = i (2\pi)^4\delta^{(4)}\big(\sum_{i=1}^n
p_i \big)\sum_{k=0}^{{\rm min}(q-3,n-q-2)} \sum_{A_k,\,B_{k+1}} \frac{N_1^3 N_2 N_3}{D_1 D_2 D_3}.
\end{equation}
Here $A_k=(a_1,a_2,...,a_k)$ and $B_{k+1}=(b_1,b_2,...,b_{k+1})$ range over all subsets of the
indices $\{2,...,q-2 \}$ and $\{q+1,...,n-1\}$ of size $k$ and $k+1$ respectively. In terms of the dual variables, the quantities
$N_i$ are given by,
\begin{align}
N_1 &= \langle 1 | x_{1,\, b_1+1}x_{b_1+1,\, a_1+1}x_{a_1+1,\, b_2+1}\ldots x_{b_{k+1}+1,\,
q}|q\rangle\,,
\nn\\
N_2 &= \langle b_1+1, b_1 \rangle \ldots\langle b_{k+1}+1,
b_{k+1} \rangle\,, \nn\\
N_3 &=[a_1,a_1+1] \ldots [a_k,a_k +1]\,, \label{Ni}
\end{align}
and similarly for $D_i$,
\begin{align}
D_1 &= x_{2, b_1+1}^2 x_{b_1+1, a_1+1}^2 x_{a_1+1, b_2+1}^2
\ldots x_{b_{k+1}+1, q}^2\,, \nn\\
D_2 &= [2\, 3] [4 5] \ldots [q-2, q-1] \langle q , q+1 \rangle \langle q+1 , q+2 \rangle \ldots
\langle n
 1 \rangle \,, \nn\\
D_3 &= [2|x_{2, b_1+1}|b_1 +1 \rangle \langle b_1 |x_{b_1, a_1}|a_1] [a_1 +1|x_{a_1+1, b_2+1}|b_2 +1
\rangle \ldots \langle b_{k+1}
  |x_{b_{k+1}, q-1}| q-1]\,. \label{splithe}
\end{align}
The important property of the relations \re{Ni} and \re{splithe} is that the quantities $N_i$ and
$D_i$ are built from manifestly dual conformal covariant objects like $x_{ij}^2$, $\vev{i,i+1}$,
$[i,i+1]$ as well as strings of $x$'s `sandwiched' between $\lambda$'s and $\bl$'s. The shortest
string of the latter type is\footnote{Clearly,  this string vanishes for $j=i+1$.}
\begin{equation}\label{shch}
 \lambda^\a_i (x_{ij})_{\a\da} \bl^{\da}_i  \equiv \lan{i} x_{ij} |j] = \lan{i} x_{i+ 1\, j} |j]
 = \lan{i} x_{i+ 1\, j-1} |j]\,,
\end{equation}
where we have used the identities, e.g., $x_{ij} = x_{i\ i+1} + x_{i+1\, j}$ and $\lan{i} x_{i\ i+
1} = \vev{i i} [i| \equiv 0$. The first way of writing this string, $\lan{i} x_{ij} |j]$, clearly
shows that it is dual conformally covariant. Indeed,  from \p{weremthatt} and \p{l-lb-inv} we obtain
\begin{equation}\label{invshch'}
    I\Big[ \lan{i} x_{ij} |j] \Big] =  \lan{i}x^{-1}_i \cdot (x^{-1}_i x_{ij} x^{-1}_j)\cdot
    \frac{x_j}{x^2_{j+1}}|j]  = \frac{\lan{i} x_{ij} |j]}{x^2_{i}x^2_{j+1}}\,.
\end{equation}
Further, longer strings can be formed by multiplying together several $x-$matrices. For example,
\begin{equation}\label{invshch}
    I\Big[ \vev{i|x_{ij}x_{jk}|k} \Big] =  \frac{ \vev{i|x_{ij}x_{jk}|k}}{x^2_i x^2_j x^2_k}\,.
\end{equation}
It is straightforward to generalize this to strings built from an arbitrary number of $x$ insertions.
It is sufficient  that the two neighboring $x$'s have a common subscript to ensure that the entire
string transforms covariantly. This is exactly what we see in \re{Ni} and \p{splithe}.

Performing conformal inversion in \re{Ni} and \re{splithe} and combining the various
conformal weight factors, we find that the ratio of $N-$ and $D-$functions entering \re{zig-zag} has
conformal weight $(+1)$ at all points except for point $x_1$ which has weight $(-3)$. Similarly to
the MHV split-helicity amplitudes, the delta function $\delta^{(4)}\big(\sum_{i=1}^n p_i
\big)=\delta^{(4)}(x_1-x_{n+1})$ is conformally covariant, bringing in an additional conformal
weight $(+4)$ at point $x_1$, thus making the total weight equal to $(+1)$. Since the assignment of dual
conformal weights is independent of $A_k$ and $B_{k+1}$, every term in the sum \re{zig-zag} has the
same weight $(+1)$ at all points and, therefore, the whole expression for the split-helicity
amplitude \re{zig-zag} is manifestly dual conformal covariant,
\be
I\left[{\cal A}_{n;0}\left(1^-\ldots q^- (q+1)^+ \ldots n^+\right) \right] = \lr{x_1^2 x_2^2\ldots
x_n^2}\,{\cal A}_{n;0}\left(1^-\ldots q^- (q+1)^+ \ldots n^+\right)\,.
\ee

\subsection{Dual conformal boost generators in the full space}\label{sect27}

As discussed above, the form of the generators of infinitesimal dual conformal transformations can
be obtained through the relation $K=IPI$. An alternative, more geometrical approach is to consider the full
space with all the coordinates $x,\lambda,\tilde\lambda$.
In this space the amplitude has support only on a surface defined by the constraints \p{masterconstraint}.

To discuss the infinitesimal transformation properties of the amplitude under dual conformal symmetry we need to
construct generators which preserve the surface. This is achieved by complementing the conformal
generators acting on the $x$ coordinates,
\be
\sum_{i=1}^n x_{i}^{ \dot \alpha\beta} x_{i}^{\dot \beta\a}
\frac{\partial}{\partial x_i^{\dot\beta\beta }}\,,
\ee
by terms which act on  $\lambda$ and $\tilde\lambda$ in such a way that they commute with the
constraints modulo constraints. There is some ambiguity in this procedure, just as we saw that there
is an ambiguity in defining the  inversions of $\lambda$ and $\tilde\lambda$ in Sect.~\ref{DCS}. The choice which corresponds to the inversion defined there is
\be
K^{\dot\alpha\alpha} = \sum_{i=1}^n \biggl[ x_{i}^{ \dot \alpha\beta} x_{i}^{\dot \beta\a}
\frac{\partial}{\partial x_i^{\dot\beta\beta }}  + x_{i}^{ \dot \alpha\beta} \lambda_{i}^{ \alpha}
\frac{\partial}{\partial \lambda_i^{\beta}} +x_{i+1}^{\dot\beta \alpha} \tilde\lambda_{i}^{
\dot\alpha} \frac{\partial}{\partial \tilde\lambda_i^{\dot \beta}} \biggr]\,. \label{dualK}
\ee
This formula summarizes the infinitesimal dual conformal transformation of all variables in the full
space with coordinates $x_i,\lambda_i,\tilde\lambda_i$. {To derive the
action of the dual conformal generator in the on-shell space with coordinates
$\lambda_i,\tilde\lambda_i$ we can simply ignore the first term in (\ref{dualK}). Note however that
the action of $K^{\dot\alpha\alpha}$ on a function of $\lambda_i,\tilde\lambda_i$ will necessarily
introduce the dual coordinates $x_i$, so the on-shell space is not best suited for investigating the
dual conformal properties of amplitudes. To derive the action of the conformal generator in the dual
space with coordinates $x_i,\lambda_i$ we can ignore the third term in (\ref{dualK}). }

\subsection{Dual conformal properties of the complete MHV amplitude}\label{DCPFM}

The perturbative (loop) corrections to the tree-level amplitude \p{mhv} take the form
\begin{equation}\label{loopcorr}
    {\cal A}^{\rm MHV} = {\cal A}^{\rm MHV}_{n;0}\ M_n(x_i)
\end{equation}
where
\begin{equation}\label{wheref}
   M_n(x_i) = 1 + a  M_n^{(1)}(x_i) + a^2 M_n^{(2)}(x_i) + \ldots
\end{equation}
is a function of the dual coordinates $x_i$ (or, equivalently, of the momenta $p_i$)\footnote{The
reason why there is no explicit dependence on the spinor variables $\lambda, \bl$ in the function
$M$ is that the helicity weights of the amplitude are carried by the tree-level prefactor in
\p{loopcorr}. We know that $\lambda_i,\bl_i$ can be expressed in terms of $x_{i\ i+1}$ from
\p{takeform}, up to a helicity scale. Then any helicity neutral function $f(\lambda, \bl)$ can be
rewritten as a function of $x$.}, given by its perturbative expansion in terms of the coupling $a =
g^2N/8\pi^2$. Each term in this expansion is a combination of divergent loop momentum integrals. For
example, in the dimensional regularization scheme ($D=4-2\ep$ with $\ep>0$, regularization scale
$\mu$)  the one-loop correction is given by the function \cite{bddk94}
\begin{equation}\label{onelofu}
    M_n^{(1)}(x_i) = -\frac{1}{\ep^2} \sum_{i=1}^n
(-x_{i,i+2}^2\,\mu^{2})^{\epsilon} + F^{(1)}_n
\end{equation}
where
\be
F^{(1)}_n = \frac{1}{2} \sum_{i=1}^n g_{n,i}\,,\quad
g_{n,i} = - \sum_{r=2}^{\lfloor \tfrac{n}{2} \rfloor -1} \ln
\Bigl(\frac{x_{i,i+r}^2}{x_{i,i+r+1}^2}\Bigr) \ln
\Bigl(\frac{x_{i+1,i+r+1}^2}{x_{i,i+r+1}^2}\Bigr) + D_{n,i} + L_{n,i} +
\frac{3}{2} \zeta_2\,.
\ee
For $n$ even, $n=2m$, the functions $D_{n,i}$ and $L_{n,i}$ are
\begin{align}  \label{di2}
D_{n,i} &= - \sum_{r=2}^{m-2} {\rm Li}_2 \Bigl(1 - \frac{x_{i,i+r}^2
  x_{i-1,i+r+1}^2}{x_{i,i+r+1}^2 x_{i-1,i+r}^2} \Bigr) -
  \frac{1}{2}{\rm Li}_2\Bigl(1 - \frac{x_{i,i+m-1}^2
  x_{i-1,i+m}^2}{x_{i,i+m}^2 x_{i-1,i+m-1}^2} \Bigr)\ ,\\  \nn
L_{n,i} &= \frac{1}{4} \ln^2 \Big(\frac{x_{i,i+m}^2}{x_{i+1,i+m+1}^2}\Bigr)
\end{align}
{and for $n$ odd there are similar expressions}.

Clearly, the presence of divergences and consequently the need to use dimensional regularization
breaks dual conformal invariance. Remarkably, however, this breakdown {occurs} in a controlled way to all orders in the coupling, as we have shown in \cite{dhks1,dhks2}. To see this one splits $\ln M_n = \ln Z_n + \ln F_n$, where $\ln Z_n$
contains the infrared divergences (double and simple poles), while the finite part $\ln F_n$ is
subject to the anomalous conformal Ward identity
\begin{equation}\label{cwi}
 K^\mu  \ln {F}_n  =  \sum^n_{i=1} (2x_i^\nu x_i\cdot\pa_i - x_i^2 \pa_i^\nu) \ln {F}_n
  =  \frac{1}{2} \Gamma_{\rm cusp}(a) \sum_{i=1}^n  \ln \frac{x_{i,i+2}^2}{x_{i-1,i+1}^2} x^\nu_{i,i+1}\ .
\end{equation}
The operator on the left-hand side is the generator of dual conformal boosts $K^\mu$, obtained by applying \p{dualK} to a
function of the dual coordinates $x_i$ only. The anomaly term on the right-hand side
is determined by the cusp anomalous dimension $\Gamma_{\rm cusp}(a)$.

The general solution of \p{cwi} allows some freedom in the form of an arbitrary function of the
conformally invariant cross-ratios
\begin{equation}\label{crrat}
    u_{ijkl} = \frac{x_{ij}^2 x_{kl}^2}{x_{il}^2 x_{jk}^2}\,,
\end{equation}
if $n\geq 6$ (for $n=4,5$ there exist no cross-ratios, due to the light-like separation of adjacent points).

In summary, the MHV superamplitude consists of two factors, the tree-level prefactor ${\cal A}^{\rm
MHV}_{n;0}$ and the   perturbative corrections factor $M_n(x_i)$. The former is an exact dual
conformal covariant, while the latter has an anomalous dual conformal behavior controlled by
an all-order  Ward identity. We can conclude that the MHV superamplitude is compatible with dual
conformal symmetry, after taking the anomaly into account.

\section{$\cN=4$ supersymmetry and scattering amplitudes}


Apart from the two gluon states $G^{\pm}$ with helicities $\pm 1$, the $\mathcal{N}=4$ SYM theory
also describes eight fermion states (gluinos) $\Gamma_A$ and $\bar\Gamma^{A}$  with helicities $1/2$
and $-1/2$, respectively, and six scalars (helicity zero states) $S_{AB} = - S_{BA}$. Here
$A,B,C,D=1,\ldots,4$ are indices of the (anti)fundamental representation of the R symmetry group
$SU(4)$. These particles can scatter into each other in many different combinations, which results in a
large variety of amplitudes. For instance,  the MHV tree amplitude involving one negative-helicity
gluon and two gluinos reads:\footnote{It has total helicity $n-4$, therefore it is an MHV
amplitude.}
\begin{equation}\label{ggglu}
    {\cal A}_{n;0}(G^-(1)\Gamma_A(2)\bar\Gamma^B(3)G^+(4)\ldots G^+(n))
    = i(2\pi)^4\delta^B_A\ \delta^{(4)}(\sum_{i=1}^n\, p_i)\
    \frac{\vev{1\, 2}\vev{1\, 3}^3}{\vev{1\, 2}\vev{2\, 3}\ldots\vev{n\, 1}}\,.
\end{equation}
These various scattering amplitudes are related to each other through supersymmetric Ward
identities~\cite{Grisaru:1976vm,Grisaru:1977px}.

To discuss the symmetry properties of the scattering amplitudes, it would be desirable to find a way to
present all scattering amplitudes in the $\mathcal{N}=4$ theory as one simple and compact object
with manifest supersymmetry. In the simplest case of MHV tree-level amplitudes this has been
achieved some time ago by Nair \cite{Nair}\footnote{Here we use a modified version of Nair's
description proposed by Witten \cite{Witten:2003nn}.} who proposed to use a particular type of $\cN=4$ on-shell
superspace.

In this section we rederive the well-known expression for the MHV superamplitude by exploiting the
on-shell supersymmetry. Then we generalize the construction to an arbitrary $n$-particle
superamplitude. We start with a brief reminder of the structure of the $\cN=4$ supermultiplets of
massless states in Sect.~\ref{N4GS}. Then we reformulate these multiplets in on-shell (or
light-cone) superspace in Sect.~\ref{CLCN4SS}. This superspace is used  in Sect.~\ref{SAN4SYM} to
give a general description of all $n$-particle superamplitudes, including Nair's MHV amplitude, but
also all non-MHV amplitudes. These superamplitudes are expressed in terms of invariants of the
on-shell supersymmetry.

\subsection{$\cN=4$ gluon supermultiplet}\label{N4GS}

Here we recall (see, e.g., the textbook \cite{West:1990tg}) how one builds the massless representations (or supermultiplets) of the $\cN=4$ supersymmetry algebra
\begin{equation}\label{n4susyal}
    \{ q^A_\a, \bar{q}_{B\, \da}  \} = \delta^A_B\ p_{\a\da} \equiv \delta^A_B\ p^\mu
    (\sigma_\mu)_{\a\da}\,,
\end{equation}
where $\sigma_\mu = ({\mathbb I}, \vec{\sigma})$ and $\vec{\sigma}$ are the Pauli matrices. In the
massless case, $p_\mu^2=0$, we can choose the Lorentz frame in which $p^\mu = (p,0,0,p)$ and the
relation \p{n4susyal} becomes
\begin{equation}\label{2.35}
\{q^A_\alpha,\bar q_{B\, \dot\alpha}\} = \delta^A_B\ (1+\sigma_3)_{\alpha\dot\alpha}\ p \,,
\end{equation}
so the algebra \p{2.35} is reduced to the Clifford algebra
\begin{equation}\label{2.36}
\{q^A_1,\bar q_{B\, \dot 1}\} = 2\delta^A_B\ p
\end{equation}
with all the other anticommutators vanishing. In this frame the states (massless Poincar\'e
representations) are labeled by their helicity, the eigenvalue of the Lorentz generator $J_{12}$.
For chiral spinors it is $\frac{1}{2} (\sigma_{12})_\a{}^\b$, and the helicity of, e.g., $q^A_1$ is
$1/2$. For antichiral spinors $J_{12} = \frac{1}{2} (\tilde\sigma_{12})_{\da}{}^{\db}$, so that the
helicity of $\bar q_{A\, \dot 1}$ is $-1/2$.

Next, we define a vacuum state of helicity $h$ by the condition that it be annihilated by all those supersymmetry generators which anticommute among themselves (annihilation operators):
\begin{equation}
q^A_{1}\vert h\rangle =q^A_2\vert h\rangle =\bar q_{A\, \dot 2}\vert h\rangle = \lr{J_{12}-h}\vert
h\rangle = 0\,. \label{2.37}
\end{equation}
Then the massless supermultiplet of states is obtained by applying the four creation operators $\bar q_{A\, \dot 1}$
to the vacuum:
\begin{equation}
\begin{array}{ccc} {\phantom{mm}\rm State\phantom{mm}}&{\phantom{mm}\rm
Helicity\phantom{mm}}&\phantom{mm}\mbox{Multiplicity}\phantom{mm}\\
\begin{array}{r}\vert {h}\rangle \\ \bar q_{A\, \dot 1}\vert {h}\rangle \\
\bar q_{A\, \dot 1} \bar q_{B\, \dot 1}\vert {h}\rangle\\ \ep^{ABCD}\bar q_{A\, \dot 1} \bar q_{B\, \dot 1} \bar q_{C\, \dot 1}\vert {h}\rangle\\ \ep^{ABCD}\bar q_{A\, \dot 1} \bar q_{B\, \dot 1} \bar q_{C\, \dot 1} \bar q_{D\, \dot 1}\vert{h}\rangle \end{array}
&\begin{array}{l}{h}\\ {h} - 1/2\\ {h} -1\\ h-3/2\\ {h}
- 2 \end{array} &\begin{array}{c}1\\ 4\\ 6\\ 4\\ 1\end{array}
\end{array}
\label{2.38}
\end{equation}
In a physical theory the helicity should be $|h| \leq 2$, so in the case $\cN=4$ the allowed values
are $h=1,3/2,2$. We see that the multiplet obtained by choosing $h=1$ is self-conjugate under PCT,
since it contains all the helicities ranging from $+1$ to $-1$. This is the so-called $\cN=4$ gluon
supermultiplet, describing massless particles of helicities $\pm1$ (gluons), $\pm1/2$ (gluinos) and
$0$ (scalars).

\subsection{Covariant on-shell $\cN=4$  superspace}\label{CLCN4SS}

The construction of the preceding section has the drawback that it requires the choice of a special
frame, thus manifestly breaking Lorentz invariance. Having the spinor variables $\lambda_\alpha$ and
$\bl_{\dot\alpha}$ at our disposal, we can do better. We can reproduce the supermultiplet \p{2.38}
in a {\it manifestly Lorentz covariant way}.\footnote{The idea to use auxiliary commuting spinor
variables for a covariant description of light-cone supersymmetry has been introduced a long time
ago  under the name of `light-cone harmonic superspace' \cite{Sokatchev:1985tc}.}

Let us rewrite the supersymmetry algebra \p{n4susyal} using the representation \p{2'} of the on-shell momentum:
\begin{equation}\label{n4susyalcov}
    \{ q^A_\a, \bar{q}_{B\, \da}  \} = \delta^A_B\ \lambda_\a \bl_{\da}\,.
\end{equation}
The two-component spinor $q^A_\a$ has two Lorentz covariant projections, one `parallel' to
$\lambda_\a$, $q^A_{||\, \a} = \lambda_\a q^A$ (with $\lambda^\a q^A_{||\, \a} = 0$), the other
`orthogonal', $q^A_{\bot}=\lambda^\a  q^A_\a$. The same applies to $\bar{q}_{A\, \da}$. Multiplying
\p{n4susyalcov} by $\lambda^\a$ or by $\bl^{\da}$, we see that the projections $q^A_{\bot}$ and
$\bar q_{\bot\, A}$ anticommute with each other and with the rest of the generators. These are the
covariant analogs of the explicit light-cone projections $q^A_2$ and $\bar q_{A\, \dot 2}$ from
\p{2.37}. Then we substitute the projections $q^A_{||\, \a}$ and $\bar q_{||\, A\, \da}$ in
\p{n4susyalcov} and obtain
\begin{equation}\label{alg}
    \{ q^A, \bar q_B\} = \delta^A_B\,.
\end{equation}
Clearly, this is the covariant analog of the Clifford algebra \p{2.36}, with $q^A$  and $\bar q_A$
being the equivalents of the annihilation operator $q^A_{1}$ and  the creation operator $\bar q_{A\,
\dot 1}$, respectively.

It is well known that such algebras are most naturally realized in terms of anticommuting
(Grassmann) variables $\eta^A$:
\begin{equation}\label{grva}
    q^A = \eta^A\,, \qquad  \bar q_A = \frac{\pa}{\pa \eta^A}\,, \qquad \{\eta^A, \eta^B\} =0\,.
\end{equation}
Since the creation operator  $\bar q_{A\, \dot 1}$ has helicity $-1/2$, the variables $\eta^A$
should have  helicity $1/2$.

We can now use the generators \re{grva} to reproduce the content of the multiplet \p{2.38} in the
convenient and compact form of a super-wave function
\begin{eqnarray} \label{fitinto}
  \Phi(p,\eta) &=& G^{+}(p) + \eta^A \Gamma_A(p) + \frac{1}{2}\eta^A \eta^B S_{AB}(p)
  + \frac{1}{3!}\eta^A\eta^B\eta^C \ep_{ABCD} \bar\Gamma^{D}(p) \nn \\
  &&\  + \frac{1}{4!}\eta^A\eta^B\eta^C \eta^D \ep_{ABCD} G^{-}(p)\,.
\end{eqnarray}
The analog of the vacuum with helicity $h=1$ is the first term in \p{fitinto}, which can be identified
as $G^{+}(p) = \Phi(p,0)$. The next state in the multiplet is obtained by applying the creation
operator $\bar q_A$, i.e. $\Gamma_A(p) = \bar q_A \Phi(p,\eta)|_{\eta=0}$, etc. Notice that the
helicity of each component wave function in \p{fitinto} is balanced by that of the Grassmann
variables, so that the super-wave function $\Phi(p,\eta)$ carries overall helicity $(+1)$.

Let us recall the discussion of particle states and wave functions from Sect.~\ref{MHVGA}. There we
used the spinor variables to relate the Lorentz covariant wave function of, e.g., a gluon
$G_{\a\b}(p)$ to the corresponding state of helicity $+1$, $G_{\a\b}(p) = \lambda_\a \lambda_\b
G^{+}(p)$. Now we see all these states, i.e. wave functions `stripped' of their Lorentz structures,
gathered together in the super-wave function \p{fitinto}.

To summarize, with the help of the spinor variables we have been able to covariantly split the
supersymmetry generators into two halves (the covariant analogs of the light-cone projections). The
projections $q^A_{\bot}$ and $\bar q_{\bot\, A}$ play no role in the construction of the massless
supermultiplet, therefore we can set them to zero. Then, the on-shell supersymmetry generators are
realized on the light-cone super-wave functions \p{fitinto} as follows:
\begin{equation}\label{realq}
    q^A_\a = \lambda_\a \eta^A\,, \qquad \bar q_{A\, \da} = \bl_{\da} \frac{\pa}{\pa \eta^A}\,,
\end{equation}
so that under infinitesimal supersymmetry transformations we find
\begin{equation}\label{underinf}
\delta\Phi(p,\eta^A) = \lr{\epsilon^\alpha_A q_\alpha^A +\bar\epsilon^{A\dot\alpha} \bar q_{A\, \dot
\alpha}}\Phi(p,\eta^A) = \lr{\ep_A \eta^A +\bar\ep^A \frac{\pa}{\pa \eta^A}}\Phi(p,\eta^A)\,.
\end{equation}
Note that the super-wave function undergoes transformations with {\it covariantly projected}
parameters $\ep_A \equiv \epsilon^\alpha_A \lambda_\a$ and $\bar\ep^A \equiv
\bar\epsilon^{A\dot\alpha} \bl_{\da}$. In this way we obtain the supersymmetry transformations of
the component wave functions
\begin{equation}\label{susycomp}
    \delta G^{+} = \bar\ep^A \Gamma_A\,, \quad \delta \Gamma_A = \ep_A G^{+} - \bar\ep^B S_{BA}\,, \quad \mbox{etc.}
\end{equation}

Finally, we point out that our approach to the on-shell superspace is holomorphic.
We made our choice in favor of the Grassmann variables $\eta^A$, and did not use their conjugates
$\bar\eta_A$. Equivalently, we can say that here we favor a chiral description, since in \p{realq}
we chose to represent the chiral generator $q^A_\a$ as a multiplication operator, and not the
antichiral $\bar q_{A\, \da}$. This choice will subsequently determine our preference for a chiral
dual superspace in Sect.~\ref{DSS}. Of course, we could have equally well described the $\cN=4$
gluon supermultiplet by an anti-holomorphic super-wave function of overall helicity $(-1)$:
\begin{eqnarray} \label{antiholo}
  \bar\Phi(\bar\eta,p) &=& G^-(p) + \bar\eta_A \bar\Gamma^A(p) + \frac{1}{2}\bar\eta_A \bar\eta_B \bar S^{AB}(p)
  + \frac{1}{3!}\bar\eta_A \bar\eta_B \bar\eta_C  \epsilon^{ABCD} \Gamma_D(p) \nn \\
  &&\  + \frac{1}{4!}\bar\eta_A \bar\eta_B \bar\eta_C \bar\eta_D \epsilon^{ABCD} G^+(p)\,.
\end{eqnarray}
We see once again the special property of this multiplet of being PCT self-conjugate. In fact, this
is the reason why we can choose a purely holomorphic description of the multiplet and, subsequently,
a chiral dual superspace for the superamplitude. In a theory with $\cN=2$ or $\cN=1$ supersymmetry
the gluon multiplet is not self-conjugate, therefore we would need both a holomorphic {\it and} an
anti-holomorphic super-wave functions for the full theory.

The equivalence of the two descriptions of the $\cN=4$ gluon multiplet can also be shown by
establishing an explicit relation between \p{fitinto} and \p{antiholo}. It takes the form of a
Grassmann Fourier transform:
\begin{equation}\label{GrFtr}
    \bar\Phi(p,\bar\eta) = \int d^4\eta\, \e^{\bar\eta_A \eta^A} \,  \Phi(p,\eta)\,.
\end{equation}

\subsection{Superamplitudes in $\cN=4$ SYM}\label{SAN4SYM}

In this section we construct a superamplitude which gives a compact description of the scattering
amplitudes of all the particles in the  $\cN=4$ theory,
\begin{equation}\label{naamp}
    {\cal A}_n(\lambda,\bl,\eta) = {\cal A}(\Phi(1)\Phi(2)\ldots\Phi(n))\,,
\end{equation}
where $\Phi(i) = \Phi(p_i,\eta_i)$ (with $i=1,\ldots,n$) stands for the $\cN=4$ supermultiplet
\re{fitinto} and $(p_i)^{\da\a} = \bl_i^\da\lambda_i^\a$ is the on-shell momentum of the particles
in the supermultiplet. In general, ${\cal A}_n(\lambda_i,\bl_i,\eta_i)$ is an inhomogeneous
polynomial of degree $4n$ in the odd variables $\eta^A_i$. However, as we will see shortly,
invariance under on-shell supersymmetry restricts the minimal degree to be 8 and the maximal to be
$4n-8$. How can we construct such invariants?

We begin by remarking that the generator
\begin{equation}\label{ususupch}
    {q}^A_\a = \sum_{i=1}^n\, q^A_{i\, \a}\,,
\end{equation}
with each $q^A_{i\, \a}$ of the form \p{realq},  acts on the super-wave function by multiplication,
just like the translation generator \p{mathP} in momentum representation. Therefore, exactly as in
\p{momconse}, requiring the invariance of ${\cal A}_n(\lambda_i,\bl_i,\eta_i)$ we can
deduce\footnote{Consider the degenerate case where all $\lambda_i$ are `parallel', $\lambda_{i\, \a}
= c_i \lambda_\a$, for some coefficients $c_i$. Then we have $q^A_\alpha = \lambda_\alpha\sum
c_i\eta^A_i$ and the condition of $q$-invariance implies the presence of a factor of
$\delta^{(4)}(\sum c_i \eta_i)$ of Grassmann degree 4 in the amplitude. The only case where all the
$\lambda_i$ are parallel is the three-point $\overline{\rm MHV}$ amplitude (which requires
complexified momenta to exist), and hence this is the unique amplitude with Grassmann degree less
than 8.}
\begin{equation}\label{supmomconse}
  p_{\a\da} \ {\cal A}_n(\lambda,\bl,\eta) =   {q}^A_\a \ {\cal A}_n(\lambda,\bl,\eta) = 0\ \Rightarrow \
  {\cal A}_n(\lambda,\bl,\eta) = \delta^{(4)}(p_{\a\da})\ \delta^{(8)}(q^A_{\a})\ {\cal
  P}_n(\lambda,\bl,\eta)\,,
\end{equation}
where $\delta^{(8)}(q^A_{\a})=\prod_{A=1}^4 \prod_{\a=1,2}q^A_{\a}$ is a Grassmann delta function
and ${\cal P}_n$ is some polynomial in $\eta_i^A$. According to \re {supmomconse}, the superamplitude
\p{naamp} factorizes into $\delta^{(8)}(q^A_{\a})$ of Grassmann degree 8 and another polynomial
${\cal P}_n$. As we explain in a moment, the maximal degree of ${\cal P}_n$ in $\eta$ is not $4n-8$
but $4n-16$. Moreover, since each $\eta_i^A$ carries an $SU(4)$ index while ${\cal P}_n$ should be a
singlet, the polynomial ${\cal P}_n$ can be split into a sum of $SU(4)$ singlet {\it homogeneous}
polynomials of degree multiple of $4$,
\begin{equation}\label{concorr'}
    {\cal A}_n(\lambda,\bl,\eta) = i (2\pi)^4\ \delta^{(4)}(\sum_{i=1}^n\ \lambda_{i}^{\a}\, \bl_{i}^{\da})\
    \delta^{(8)} (\sum_{i=1}^n\ \lambda_{i}^{\a}\, \eta^A_i)\ \Big[
{\cal P}_n^{(0)} + {\cal P}_n^{(4)} + {\cal P}_n^{(8)} +  \ldots +{\cal P}_n^{(4n-16)} \Big]\,.
\end{equation}

We notice that ${\cal A}_n(\lambda,\bl,\eta)$  \re{concorr'} serves as a generating function for all
particle scattering amplitudes in the $\mathcal{N}=4$ SYM theory. In order to extract a particular
scattering amplitude out of the superamplitude \p{concorr'}, we need to expand it in terms of
$\eta^A_i$ and to collect terms of a given degree at each point, according to the content of the
super-wave function \p{fitinto}. For example, if at point $i$ we have a gluon $G^{+}$, we need
$(\eta_i)^0$; if the particle is a gluino $\Gamma_A$, we need $(\eta_i)^1$, etc. The highest degree
$(\eta_i)^4$ is obtained if a gluon $G^{-}$ occupies position $i$. In an amplitude with $m$
negative-helicity gluons, appearing at points $i_1, \ldots, i_m$, there will be a term of the form
$(\eta_{i_1})^4 \ldots (\eta_{i_m})^4$ of degree $4m$.

Now, the prefactor $\delta^{(8)}(q^A_{\a})$ in \p{concorr'}, whose presence is required by
supersymmetry, already has degree 8. This means that the gluon amplitudes extracted from
\p{concorr'} must have {\it at least two negative-helicity gluons}, which corresponds to MHV
amplitudes. They are described by the first term in \p{concorr'}, with the factor ${\cal P}_n^{(0)}$
of degree zero in $\eta$. We see here a very simple explanation of the well-known fact
\cite{Grisaru:1976vm,Grisaru:1977px} that
$\cN=4$ supersymmetry forbids gluon amplitudes with less than two negative-helicity gluons.

The last term in \p{concorr'}, ${\cal P}_n^{(4n-16)}$, multiplied by the Grassmann delta function,
$\delta^{(8)} (\sum_{i=1}^n\ \lambda_{i}^{\a}\, \eta^A_i)$, has degree $4(n-2)$. It contains a gluon
amplitude with $n-2$ negative-helicity gluons and two positive-helicity ones. This is an
$\overline{\rm MHV}$ amplitude which can be obtained from an MHV amplitude by PCT conjugation.
Having a term of higher degree in \p{concorr'} would imply the existence of amplitudes with less
than two positive-helicity gluons, which is forbidden by supersymmetry. This follows from the
equivalent antiholomorphic description of the superamplitude mentioned at the end of
Sect.~\ref{CLCN4SS}.\footnote{An alternative explanation is provided by $\bar q$ supersymmetry, see
below.} In a similar manner, the polynomial ${\cal P}_n^{(4k)}$ contains the $n-$particle non-MHV
scattering amplitudes with total helicity $n-2(k+2)$ including gluon amplitudes with $(k+2)$ gluons
of helicity $(-1)$ and the remaining $(n-k-2)$ gluons with helicity $(+1)$.

The simplest amplitude in \p{concorr'} is the MHV superamplitude
\begin{equation}\label{supmhv}
   {\cal A}^{\rm MHV}_n(\lambda,\bl,\eta) = i (2\pi)^4\ \delta^{(4)}(\sum_{i=1}^n\ \lambda_{i}^{\a}\, \bl_{i}^{\da})\
    \delta^{(8)} (\sum_{i=1}^n\ \lambda_{i}^{\a}\, \eta^A_i)\ {\cal P}_n^{(0)}\,.
\end{equation}
What can we say about the factor ${\cal P}_n^{(0)}$? It has Grassmann degree zero, so it is
independent of the odd variables $\eta$. The two delta functions in \p{supmhv} have no helicity,
while according to \p{naamp} the superamplitude must have helicity $+1$ at each point. This helicity
should be carried by ${\cal P}_n^{(0)}$. In the case of the tree-level MHV amplitude we can
determine this factor by comparing it, e.g., to the gluon amplitude \p{mhv}. As explained above, to
have negative-helicity gluons at points 1 and 2 we need to extract the term $(\eta_1)^4(\eta_2)^4$
from $\delta^{(8)} (\sum_{i=1}^n\ \lambda_{i}^{\a}\, \eta^A_i)$ (at the remaining points we need no
$\eta$'s). In doing so,  $\eta_{1}^A$ and $\eta_2^A$ form $SU(4)$ invariants, and the accompanying
spinor variables $\lambda_1$ and $\lambda_2$ contract into a Lorentz invariant. The result is
$\vev{1 2}^4 (\eta_1)^4 (\eta_2)^4$, which reproduces the numerator in \p{mhv}. Then the denominator
is obtained by setting
\begin{equation}\label{seta0}
    {\cal P}^{(0)}_{n;0} = \frac{1}{\vev{1\, 2}\vev{2\, 3}\ldots\vev{n\, 1}}\,,
\end{equation}
which has the required helicity $+1$ at each point. Thus, we have derived Nair's description of the
$n$-particle MHV tree-level superamplitude
\begin{equation}\label{concorr}
    {\cal A}^{\rm MHV}_{n;0} = \frac{i (2\pi)^4 \delta^{(4)}(\sum_{i=1}^n\ \lambda_{i}^{\a}\, \bl_{i}^{\da})\
    \delta^{(8)} (\sum_{i=1}^n\ \lambda_{i}^{\a}\, \eta^A_i)}{\vev{1\, 2}\vev{2\, 3}\ldots\vev{n\, 1}}\,.
\end{equation}
Extracting other partial amplitudes from \p{concorr} one follows the same procedure as above. For
instance, the mixed gluon/gluino amplitude \p{ggglu} is obtained by collecting the terms
$(\eta_1)^4(\eta_2)^1(\eta_3)^3$.

Let us come back to the supersymmetry generators \p{realq}. We have constructed the superamplitude
\re{concorr'} in such a way that the invariance under the first of them, ${q}^A_\a$, is manifest
(just like translation invariance). Let us now consider the second generator
\begin{equation}\label{secogen}
    \bar {q}_{A\, \da} =  \sum_{i=1}^n\ \bl_{i\, \da} \frac{\pa}{\pa \eta^A_i}\,.
\end{equation}
Acting on the argument of   $\delta^{(8)} (\sum_{j=1}^n\ \lambda_{i\,\a}\, \eta^A_i)$ in
\p{concorr'}, it gives
\begin{equation}\label{appltodel}
     \bar {q}_{A\, \da} (\sum_{i=1}^n\ \lambda_{i\,\a}\, \eta^B_i) =  \sum_{i=1}^n  \ \lambda_{i\,\a}\, \bl_{i\,\da}\,,
\end{equation}
which vanishes due to the momentum conservation delta function  $\delta^{(4)}(\sum_{i=1}^n\
\lambda_{i\,\a}\, \bl_{i\,\da})$ in \p{concorr'}. We conclude that the MHV superamplitude
\p{concorr} is {\it invariant under the full on-shell supersymmetry}. As to the other terms in
\p{concorr'}, the second supersymmetry condition,
\begin{equation}\label{invbarq}
    \bar{q}_{A\, \da}\ {\cal P}_n^{(4k)} = 0\,, \qquad (k=1,\ldots, n-4)\,,
\end{equation}
imposes restrictions on their $\eta$ dependence. For example, the absence of the two terms in
\p{concorr'}, ${\cal P}_n^{(4n-12)}={\cal P}_n^{(4n-8)}=0$, was explained above by the properties of
the $\overline{\rm MHV}$ amplitude. In fact, this is equivalent to requiring $\bar{q}-$invariance.
Indeed, the generator \re{secogen} acts on the $\eta$'s as follows:
\begin{equation}\label{susbq}
    \delta_{\bar q} \eta^{A}_i = \bar\rho_{\da}^{A}\bl^{\da}_{i}\,,
\end{equation}
with $\bar\rho_{\da}^{A}$ being an odd antichiral parameter. This parameter has two components,
which can be used to put to zero any two $\eta$'s. The remaining $(n-2)$ nonvanishing $\eta$'s form
a $\bar{q}-$invariant of maximal degree $4(n-2)$, of which 8 is already present in the compulsory
$\delta^{(8)} (\sum_{i=1}^n\lambda_{i\,\a}\eta^A_i)$. Thus, the maximal degree left for the factors
${\cal P}_n^{(4k)}$ in \re{concorr'} is $4(n-4)$, as stated above. Furthermore, as we will show in
Sect.~\ref{NMHVamplitude}, $\bar{q}-$supersymmetry imposes even stronger restrictions on the NMHV
amplitude (the factor ${\cal P}_n^{(4)}$ in \re{concorr'}).

As was mentioned at the end of Sect.~\ref{MHVGA}, the tree-level MHV amplitudes have a conventional
conformal symmetry~\cite{Witten:2003nn} and, in particular, they are annihilated by the the
conformal boost generator $k_{\a\da}$, Eq.~\p{witt}. Combined together, supersymmetry and
conformal symmetry lead to the $\cN=4$ {\it super}conformal symmetry of the superamplitudes
\p{concorr}. The corresponding special superconformal generators have the form:
\begin{equation}\label{specosu}
    s^\a_A = \sum_{i=1}^n \frac{\pa^2}{\pa \lambda_{i\, \a} \pa \eta^A_i}\,,
    \qqqquad \bar s^A_{\da} = \sum_{i=1}^n \eta^A_i\frac{\pa}{\pa \bl_{i}^{\da}}\,.
\end{equation}
To show that  $s^\a_A$ annihilates the amplitude \p{concorr} requires a short calculation
\cite{Witten:2003nn}. The invariance under $\bar s^A_{\da}$ is obvious, due to the fact that acting
on \re{concorr} the generator $\bar s^A_{\da}$ shifts the argument of the momentum delta function in
\re{concorr} by the amount proportional to the argument of the Grassmann delta function. {Together, these symmetries imply invariance of the superamplitude under $k_{\a\da}$.}

\section{Dual superconformal symmetry of the $\cN=4$\\  superamplitudes}\label{DSN4SA}

In this section we extend the notions of full space and dual space from Sect.~\ref{DCSGA} to chiral
superspaces (subsection \ref{DSS}), starting from the on-shell superspace of Sect.~\ref{CLCN4SS}.
With the help of these spaces we define the action of dual superconformal symmetry in
Sect.~\ref{DSCSY}. In Sect.~\ref{DSCPMHV} we  rewrite the tree-level MHV superamplitude as a dual
superconformal covariant function in dual superspace. Then, we discuss the dual conformal properties
of its various components and, in particular, explain why only the split-helicity amplitudes are
manifestly dual conformal.

\subsection{Full superspace}\label{FSS}

In the preceding section we saw that the superamplitudes have the following general form in the
on-shell superspace,
\be
 \mathcal{A}_n = i (2\pi)^4 \delta^{(4)}(\sum_{i=1}^n \lambda_i^\a \bl_i^\da) \delta^{(8)}(\sum_{i=1}^n \lambda_i^\a \eta_i^A)
 \mathcal{P}_n(\lambda_i,\bl_i,\eta_i).
\label{samplitude}
\ee
The function $\mathcal{P}_n$ depends on the  variables $\lambda_i$, $\bl_i$ and $\eta_i$ (with
$i=1,\ldots,n$) which are constrained by the two delta functions. As before, the spinor variables
$\lambda_i$ and $\bl_i$ have to verify the momentum conservation (see \p{momcon'}). In addition, the
variables $\lambda_i$ and $\eta_i$ have to satisfy the relation $\sum_{i=1}^n \lambda_i^\a
\eta_i^A=0$, which reflects the invariance of the superamplitude under $q$-supersymmetry. In very
much the same way as it was done in Sect.~\ref{FS}, both conditions can be trivially resolved by
introducing new dual variables. Namely, we introduce dual $x_i^{\da\a}$ coordinates to solve the
momentum conservation constraint and chiral dual $\theta_i^{A\,\a}$ coordinates to solve for the
supercharge conservation constraint,
\begin{align}\notag
\sum_{i=1}^n \lambda_{i}^{\a} \bl_{i}^{\da} =0 &\implies x_{i}^{\da\a} - x_{i+1}^{ \da\a} =
\bl_{\i}^{\da}\lambda_{i}^{\a}\, ,
\\ \label{thcon}
\sum_{i=1}^n \lambda_{i}^{\a} \eta_i^A = 0 &\implies \theta_{i}^{A\,\a} - \theta_{i+1}^{A\, a} =
\lambda_{i}^{\a} \eta_i^A\,,
\end{align}
and we impose the cyclicity conditions
\begin{equation}\label{cycodd}
    x_{n+1} \equiv x_1\,, \qquad \theta_{n+1} \equiv \theta_1\,.
\end{equation}
We will call the space with coordinates $(\lambda_i,\tilde\lambda_i,x_i,\eta_i,\theta_i)$ the `full
superspace'.

We can think of the relations (\ref{thcon}) as defining a surface in the full superspace. Then we
can interpret the function $\mathcal{P}_n$ appearing in the amplitude (\ref{samplitude}) as a
function on this surface. It is clear that $\mathcal{P}_n$ can only depend on the dual $x-$ and
$\theta-$coordinates through their differences $x_i-x_j=x_{ij}$ and $\theta_i- \theta_j =
\theta_{ij}$, thus implying dual (super)translation invariance,
\be
P_{\alpha \dot\alpha} \, \mathcal{P}_n = 0\,, \qquad\qquad Q_{A\,\a} \mathcal{P}_n = 0\,.
\label{dualpandq}
\ee
Again, we should stress that the generators of these symmetries,
\begin{equation}\label{gendutr}
    P_{\a\da} = \sum_{i=1}^n \frac{\pa}{\pa x^{\da\a}_i}\,, \qquad\qquad Q_{A\,\a}
    = \sum_{i=1}^n \frac{\pa}{\pa \theta_i^{A\,\a}}\,,
\end{equation}
are not related to the usual translation generator ${p}_{\a\da}$, Eq.~\p{mathP}, or supercharge
$q_{\a}^A$, Eq.~\p{ususupch}, (an obvious difference is the type of $SU(4)$ index of the dual
supercharge $Q_{\a A}$ as opposed to that of $q_\a^A$). As before, $P_{\a\da}$ generates shifts of
dual $x-$variables, while $Q_{A\a}$ generates shifts of $\theta$'s
\be\label{qsus}
\delta_Q \theta_i^{A\,\a} = \epsilon^{A\,\a}\,,
\ee
with $\epsilon^{A\, a}$ being a constant odd parameter (a chiral Weyl spinor).

The (super)translation invariance \p{dualpandq} can be equivalently interpreted as the possibility
to solve for the dual coordinates $x_i$ and $\theta_i$ from  (\ref{thcon}), up to the freedom of
choosing the arbitrary reference points, e.g., $x_1$ and $\theta_1$:
\begin{equation}\label{xthrl}
    (x_i)^{\da\a} = (x_1)^{\da\a} - \sum_{k=1}^{i-1}\ \bl_k^{\da} \lambda_k^\a \,, \qquad \qquad
    \q_i^{A\,\a} = \q_1^{A\,\a} - \sum_{k=1}^{i-1}\ \lambda_k^\a \,\eta_k^A\,.
\end{equation}
In other words, the definition of the dual coordinates \p{thcon} is invariant under shifts of  $x_i$
and $\theta_i$ by an arbitrary constant vector and spinor, respectively. Clearly, using the dual
translation and supersymmetry invariance (\ref{dualpandq}) and the constraints  (\ref{thcon}), we
can return to the on-shell superspace with just $\lambda_i,\tilde\lambda_i,\eta_i$ as coordinates.

\subsection{Dual superspace}\label{DSS}
Alternatively, we can give a holomorphic description of the superamplitudes by eliminating
$\tilde\lambda_i$ and $\eta_i$ instead of $x_i$ and $\theta_i$. As in the bosonic case (see
Sect.~\ref{DS}), we can rewrite the constraints  (\ref{thcon}) without using the variables
$\tilde\lambda_i$ and $\eta_i$,
\begin{equation}\label{stwpr}
     (x_{i\ i+1})^{\da\a} \lambda_{i\,\a} =  0\,, \qquad\qquad (\theta_{i\ i+1})^{A\,\a} \lambda_{i\,\a}  =0\,.
\end{equation}
These relations are equivalent to  (\ref{thcon}). Indeed, the general solution to \p{stwpr} takes
the form
\begin{equation}\label{stakeform}
    (x_{i\ i+1})^{\da\a}  = \bl_{i}^{\da}\,\lambda_{i}^{\a} \,, \qquad\qquad (\theta_{i\ i+1})^{A\,\a}
    = \lambda_{i}^{\a} \,\eta_i^A\,,
\end{equation}
thus introducing the {\it secondary} variables $\bl_i$ and $\eta_i$. Namely, $\bl_i$ and $\eta_i$
can be expressed in terms of the $x_i,\lambda_i$ and $\theta_i$ by projecting both relations in
\p{stakeform} by, e.g., $\lambda_{i+1\,\a}$:
\begin{equation}\label{exprbarl}
     \bl_{i}^{\da} = \frac{(x_{i\, i+1})^{\da\a}\lambda_{i+1\,\a}}{\vev{i\ i+1}}\,, \qquad\qquad
     \eta^A_i =  \frac{(\q_{i\ i+1})^{A\,\a}\lambda_{i+1\,\a}}{\vev{i\ i+1}}\,.
\end{equation}
With these relations taken into account, the function $\mathcal{P}_n$ can now be regarded as a
function of the variables $x_i,\lambda_i,\theta_i$.

We call the space with coordinates $(x_i,\lambda_i,\theta_i)$, satisfying the constraints \p{stwpr},
the `dual superspace'. It is important to realize that this is a {\it chiral} superspace, we only
use chiral spinors $\lambda_i^{\alpha }$ and $\theta_i^{A\,\alpha}$, but not their antichiral
complex conjugates. The relations between the on-shell, full and dual superspaces are summarized in
the following diagram:

\vspace*{10mm}
\psfrag{Onshell}[cc][cc]{\parbox[t]{36mm}{ {On-shell superspace} \\[3mm]
\centerline{$(\lambda_i^\alpha ,\tilde \lambda_i^{\dot \alpha },\eta_{i}^{A})$}}}
\psfrag{Dual}[cc][cc]{\parbox[t]{32mm}{ {Dual superspace} \\[3mm]
\centerline{$(x_i^{\alpha \dot \alpha }, \lambda_i^{\alpha },\theta_i^{A\,\alpha})$}}}
\psfrag{Extended}[cc][cc]{\parbox[t]{33mm}{ {Full superspace} \\[3mm]
\centerline{$(\lambda_i^\alpha,\tilde \lambda_i^{\dot \alpha},x_i^{\alpha \dot \alpha },\eta_i^A,
\theta_i^{A\,\alpha})$}}}
\psfrag{eq1}[cr][cr]{Eq.\,\re{xthrl}}\psfrag{eq2}[cl][cl]{Eq.\,\re{exprbarl}}
\psfrag{eq3}[cc][cc]{Eq.\,\re{thcon}}
\centerline{{\epsfysize4.5cm \epsfbox{diagram.eps}}} \vspace{10mm}

\subsection{Dual superconformal symmetry}\label{DSCSY}

In this subsection, we extend the previous analysis (see Sect.~\ref{DCS}) of dual conformal
properties of the bosonic coordinates $x,\lambda,\bl$ to the fermionic coordinates $\q,\eta$.
Starting from the known transformation properties of the `odd' dual coordinates $\q^{A\,\a}$ under
inversion, we derive those of the on-shell variables $\eta^A$. In this way we complete dual
conformal symmetry $SO(2,4)$ to the superconformal symmetry $SU(2,2|4)$.

\subsubsection{Dual Poincar\'e supersymmetry}

In the dual superspace with coordinates $(x_i,\lambda_i,\theta_i)$ we introduce the generators
\begin{equation}\label{dubarq}
  Q_{A\, \a} = \sum_{i=1}^n\ \frac{\pa}{\pa \q^{ A\, \a}_i}\,, \qqqquad
  \bar Q^A_{\da} = \sum_{i=1}^n\ \q^{A\, \a}_i \frac{\pa}{\pa x^{\da\a}_i}\,,
  \qqqquad  P_{\a\da} = \sum_{i=1}^n\ \frac{\pa}{\pa x^{\da\a}_i}\,,
\end{equation}
satisfying the $\cN=4$ Poincar\'e supersymmetry algebra
\begin{equation}\label{poincsusy}
    \{Q_{A\, \a}, \bar Q^B_{\da}  \} = \delta^B_A\ P_{\a\da}\,.
\end{equation}
The generator $\bar Q^A_{\da}$ has an induced action on the on-shell superspace variables $\eta$,
which follows from \p{exprbarl}:
\begin{equation}\label{indueta}
    \bar Q^A_{\da} = \sum_{i=1}^n \ \eta^A_i \frac{\pa}{\pa \bl^{\da}_i}\,.
\end{equation}
{Both forms of $\bar{Q}^A_{\da}$ can be obtained from its representation in the full superspace
(\ref{barQfss}) by restricting to the dual superspace or on-shell superspace, respectively. Note
that the action of $\bar{Q}^A_{\da}$ on the on-shell superspace (\ref{indueta}) is identical to the
ordinary superconformal generator $\bar s^A_{\da}$, Eq.~(\ref{specosu}), acting in the on-shell
superspace. We could say that half of the dual Poincar\'e supersymmetry, $\bar{Q}^A_{\da}$, is
induced by the ordinary superconformal symmetry $\bar{s}^A_{\da}$.} In Sect.~\ref{CRDSU} we will
extend this dual Poincar\'e supersymmetry to the full $\cN=4$ superconformal symmetry $SU(2,2|4)$.

Now we recall the discussion of the holomorphic approach to the (super)amplitudes from
Sect.~\ref{DS}. In the bosonic case we chose to describe the amplitude ${\cal A}(x,\lambda)$ in
terms of the dual space coordinates $(x_i,\lambda_i)$ and not to consider their complex conjugates.
In the supersymmetric case the analogous choice is that of the {\it chiral} dual superspace
$(x,\q,\lambda)$. What motivates this choice? In Sect.~\ref{DSS} we saw that from the chiral dual
superspace we can deduce the existence of the on-shell variables $\eta$. In Sect.~\ref{CLCN4SS} we
have shown that the complete, PCT self-conjugate on-shell gluon supermultiplet can be described in a
holomorphic way, in terms of $\eta^A$ only. It is precisely this special property of the $\cN=4$ SYM
theory which makes it possible to define purely chiral superamplitudes.

Thus, we can say that the choice of the chiral dual superspace   is determined by the holomorphic
description of the on-shell gluon multiplet. Further, the chiral Grassmann coordinates $\q^A_\a$
have twice the number of degrees of freedom of the on-shell variables $\eta^A$, which justifies the
fermionic defining constraint in \p{stakeform}. Without the auxiliary spinor variables, in order to
`halve' the chiral spinor $\q^A_\a$, we would have to  explicitly break Lorentz symmetry. As
discussed earlier in Sect.~\ref{CLCN4SS}, the role of the auxiliary spinor variables $\lambda$ is to
make these light-cone projections manifestly covariant.

Of course, we could make the equivalent choice of an  antichiral dual superspace, corresponding to
the antiholomorphic description of the gluon multiplet in terms of $\bar\eta$ (see \p{antiholo}).
The important point is that the specific nature of the $\cN=4$ gluon multiplet allows us to use
either the one or the other description, and does not oblige us to mix them.

\subsubsection{Chiral realization of the dual $SU(2,2|4)$}\label{CRDSU}

As in Sect.~\ref{DS}, we treat the coordinates of the chiral dual superspace $(x,\lambda,\theta)$ as
our `primary' variables. Let us first discuss their superconformal properties and, then, derive the
transformation rules for the `secondary' on-shell superspace variables $\eta$.

To begin with, we need to supplement the already known conformal inversion rules for $x$,
Eq.~\p{inver}, and for $\lambda$, Eq.~\p{l-lb-inv}, with the standard rule for the odd superspace
coordinates $\q$~\cite{West:1990tg},
\be\label{t-inv}
I\left[\q_{i}^{A\, \a}\right] = (x_i^{-1})^{\da\b}\q_{i\, \b}^A\,,  \qqqquad  I\left[\q_{i\,
\a}^A\right] =
 \q_{i}^{A\, \b} \,(x_i^{-1})_{\b\da}\,.
\ee
It is easy to see that the defining constraints \p{stakeform} transform covariantly under these
transformations, so the chiral dual superspace is closed under conformal inversion.

The combination of the dual supersymmetry transformation \p{qsus} with inversion implies another
continuous symmetry with generator  $\bar S^{\da}_{A} = I Q_{A\,\a} I$, in close analogy with the
conformal boosts $K^\mu = IP^\mu I$. Its action on the odd dual coordinates $\q$ is easy to work
out. After the inversion $\q_i$ becomes $\q_i x_i^{-1}$. Then, after a dual supersymmetry
transformation we get $(\q_i + \ep) x_i^{-1}$. Finally, the second inversion brings us to $\q_i +
\bar\rho x_i$ with $\bar\rho_{\da} = I[\ep_\a]$ (as usual, inversion changes the chirality of the
spinors, including the transformation parameters) leading to
\begin{equation}\label{ssus}
    \delta_{\bar S}  \q^{\a A}_i  = \bar\rho_{\da}^{A}x^{\da\a}_i\,.
\end{equation}
The generator of this transformation in the dual superspace only acts on $\theta_i$ and leaves the
other variables $x_i$ and $\lambda_i$ intact
\begin{equation}\label{ssusgener}
    \bar S^{\da}_{A} = \sum_{i=1}^n  \ x^{\da\a}_i \frac{\pa}{\pa \q^{A\,\a}_i}\,.
\end{equation}
Commuting $\bar S$ with the translation generator \p{gendutr}, we obtain
\begin{equation}\label{compbars}
    [\bar S^{\da}_{A}, P_{\b\db}] = \delta^{\da}_{\db} Q_{A\,\b}\,.
\end{equation}
Next, applying inversion to both sides of this commutator, using $\bar S = IQI$, $K=IPI$ and $I^2 =
{\mathbb I}$, we obtain
\begin{equation}\label{conbo}
    [Q_{\a A}, K_{\b\db}] = \ep_{\a\b} \bar S_{A\, \db}\,.
\end{equation}
We  identify the generators $P,K,Q,\bar S$ as part of the $\cN=4$ superconformal algebra
$su(2,2|4)$. The explicit form of the generators of this algebra and their commutation relations can
be found in Appendix~B.

The reason why we have $su(2,2|4)$ and not $psu(2,2|4)$ is that the algebra involves a central
charge. To see this, consider the anticommutators $\{Q,S\}$ and $\{\bar Q, \bar S \}$ from
Eq.~\re{comm-rel}. The Lorentz ($M$) and $SU(4)$ ($R$) generators annihilate the scalar and singlet
amplitude, while the  action of the dilatation operator $D$ and the central charge $C$ on the
tree-level superamplitude \re{concorr'} and \re{concorr} is given by
\be\label{C-act}
D\, {\cal A}_{n;0}(\lambda,\bl,\eta) = C \, {\cal A}_{n;0}(\lambda,\bl,\eta) = -n {\cal
A}_{n;0}(\lambda,\bl,\eta)\,.
\ee
Examining the explicit realizations of these two generators, Eqs.~\p{DD} and \p{CC}, respectively,
we see that $\lambda$ and $\bl$ have the same dilatation weight $(-1/2)$, while they have opposite
central charges, $(-1/2)$ for $\lambda$ and   $(+1/2)$ for $\bl$. This suggests to identify the
central charge with helicity.

\subsubsection{Induced action on the on-shell odd variables}

The fact that the $\eta$'s are determined by the $\q$'s and $\lambda$'s (recall \p{exprbarl})
implies that their conformal properties follow from those of $\q$ and $\lambda$. Let us begin by
making an inversion in the second equation in \p{stakeform}, {with the help of  \re{t-inv} and
\re{l-lb-inv}},
\begin{equation}\label{equ}
    {(\q^A_i x^{-1}_i)_{\da}} - {(\q^A_{i+1} x^{-1}_{i+1})_{\da}} = (\lambda_i x^{-1}_i)_{\da} \ I[\eta^A_i]\,.
\end{equation}
Then we multiply this equation by $( x_i)^{\da\a}$ from the right, replace $\q^A_{i+1}$ by $\q^A_i -
\lambda_i\eta^A_i$ using \re{stwpr} and apply \p{twpr} to arrive at
\begin{equation}\label{newequ}
    \lambda_{i}^{\a}\ {\frac{x^2_{i}}{x^2_{i+1}}}\ \left(\eta^A_i - {\q^A_i x^{-1}_i\bl_i} \right) = \lambda_{i}^{\a}\  I[\eta^A_i]\,.
\end{equation}
Since this equation should hold for any $\lambda_{i}^{\a}$, we deduce
\begin{equation}\label{inveta}
    I[\eta^A_i] = {\frac{x^2_{i}}{x^2_{i+1}}}\ \left(\eta^A_i - {\q^A_i x^{-1}_i\bl_i} \right)\,,
\end{equation}
{where the contraction of spinor indices is tacitly implied.} Repeating the inversion twice, we
obtain $I^2[\eta^A_i] = \eta^A_i$, as expected.

{It is important to realize} that, {contrary to $\theta-$variables}, the transformation of
$\eta$ in \re{inveta} is {\it not homogeneous} ($\eta$ transforms through itself and through $\q$).
In addition, the relation \re{inveta} explicitly involves the antichiral spinor variable $\bl$,
which takes us out of the holomorphic description. This shows that the $\eta-$variables  are not
well suited for the discussion of the dual conformal properties of the superamplitude. We will come
back to this important point in Sect.~\ref{DSCPMHV}.

The infinitesimal dual superconformal transformation \p{ssus} of $\q$ induces that of $\eta$,
\begin{equation}\label{sus}
    \delta_{\bar S} \eta^{A}_i = \bar\rho_{\da}^{A}\bl^{\da}_{i}
\end{equation}
and the corresponding generator $\bar{S}^{\da}_A$ in the on-shell superspace is~\footnote{The
representations of $\bar{S}^{\da}_A$ in the dual superspace (\ref{ssusgener}) and in the on-shell
superspace (\ref{ssusgener'}) can be obtained from the representation on the full superspace
(\ref{fssbarS}).}
\begin{equation}\label{ssusgener'}
    \bar S^{\da}_{A} = \sum_{i=1}^n \ \bl^{\da}_i \frac{\pa}{\pa \eta^A_i}
\end{equation}
We remark that the transformation \p{sus} is identical to \p{susbq}, and its generator
\p{ssusgener'} coincides with the supersymmetry generator $\bar q_{A\, \da}$ \p{realq} acting in the
on-shell superspace. We can reverse the argument and say that the light-cone supercharge $\bar
q_{A\, \da}$, initially acting on the on-shell superspace variables $\eta$,  induces the $\bar S$
transformations of the dual superspace coordinates $\q$ through the relation \p{thcon}.

\subsubsection{Transformation properties of the delta function in the superamplitude}

In Sect.~\ref{SAN4SYM} we have shown that the superamplitude \p{concorr'} contains a prefactor made
of two delta functions, bosonic and fermionic. The dual space interpretation of the bosonic delta
function was given in Sect.~\ref{Tpma}: we first broke the $n$-point cycle, $x_{n+1} \neq x_1$, and
then used the delta function  to impose back the identification $x_{n+1} = x_1$. In complete
analogy, we first  relax the cyclicity condition \p{cycodd} and then replace the product of two
delta functions in  \p{concorr'} by
\begin{equation}\label{equiodd}
   \delta^{(4)}(x_1-x_{n+1}) \delta^{(8)}(\q^{A\,a}_1-\q^{A\,\a}_{n+1})\,.
\end{equation}
This product imposes the condition \p{cycodd}. The advantage of this reformulation, from the point
of view of dual conformal symmetry, is that the  covariance of \p{equiodd} under inversion
\re{t-inv} is manifest, assuming that $\q^A_{n+1}$ transforms through $x_{n+1}$. As in the case of
the bosonic delta function, this creates some extra conformal weight at the preferred point 1. We
will come back to this point in Sect.~\ref{DSCPMHV}.

The invariance of the Grassmann delta function in \p{equiodd} under the dual $Q$ supersymmetry
\p{gendutr} is obvious. To show the invariance under the dual special conformal supersymmetry
\p{ssusgener} (which is equivalent to the light-cone supersymmetry \p{appltodel}), we again need the
help of the bosonic delta function.

Finally, an interesting question is what is the role of the other half of the `odd' $SU(2,2|4)$
generators  $\bar Q$, Eqs.~\p{dubarq} and \p{indueta}, and $S$, Eq.~\re{fssbarS}. Applying $\bar Q$
to the argument of bosonic delta function in \re{equiodd}, we obtain  $\bar Q^A (x_1-x_{n+1}) =
\q^A_1 - \q^A_{n+1}$, which vanishes due to the Grassmann delta function in \re{equiodd}. Hence, the
MHV tree-level superamplitude \p{concorr} is invariant under the full dual $\cN=4$ Poincar\'e
supersymmetry. Combining this with dual conformal symmetry, we can say that it is covariant under
the full dual $SU(2,2|4)$. However, as soon as we turn on the perturbative corrections, which
involve non-trivial dependence on $x$, the role of $\bar Q$ (and consequently of $S$) becomes less
clear. One point we can make is that, unlike $\bar S$,  $\bar Q$ cannot remain an exact symmetry of
the amplitude. Indeed, the two symmetries imply the compatibility condition (see \p{comm-rel})
\begin{equation}\label{comconditio}
    \bar Q \mathcal{A}_n = \bar S \mathcal{A}_n = 0 \qquad \Longrightarrow \qquad D\mathcal{A}_n = C
    \mathcal{A}_n\,.
\end{equation}
{This relation holds at tree level, but it does not survive loop corrections since the
dilatation symmetry becomes anomalous due to the presence of infrared divergences, while the central
charge $C$ still measures the helicity of the superamplitude and, therefore, is protected.} We hope
to come back to this issue in the future.

\subsection{Dual superconformal properties of the tree-level MHV superamplitude}\label{DSCPMHV}

The main result of the preceding subsections was the introduction of dual conformal symmetry
(inversion rules \re{l-lb-inv} and \re{t-inv}). We have shown that the denominator in the MHV
superamplitude \p{concorr} is covariant under these transformations. The superamplitude also
involves two delta functions whose origin is (super)translation invariance in the on-shell
superspace $(\lambda,\bl,\eta)$. As explained above, their dual conformal properties become manifest
if we first break the cyclicity of the amplitude  by introducing an extra point in dual superspace,
$x_{n+1}\neq x_1,\ \q_{n+1}\neq \q_1$, and then use the delta functions \re{equiodd} to identify the
end points. However, this creates extra conformal weight at the breaking point $(x_1,\theta_1)$,
which seems unnatural for a cyclicly symmetric amplitude.

Fortunately, in the special case of $\cN=4$ dual supersymmetry the product of two delta functions
\re{equiodd} has {\it vanishing conformal weight} at point $1 \equiv n+1$. Indeed, under inversion
the bosonic delta function transforms with a weight opposite to that of the space measure, $\int d^4
x\to \int d^4 x\ x^{-8}$, thus $\delta^{(4)}(x_1-x_{n+1})\to x_1^{8}\,\delta^{(4)}(x_1-x_{n+1})$. At
the same time, since $\q_1-\q_{n+1} \to x_1^{-1}\q_1-x_{n+1}^{-1}\q_{n+1}$ under inversion
\re{t-inv}, we have $\delta^{(8)}(\q_1-\q_{n+1})\ \rightarrow \ x_1^{-8}\delta^{(8)}(\q_1-\q_{n+1})$
due to $x_1=x_{n+1}$, so that the product \p{equiodd} remains invariant.

This shows that we could have chosen to break the cycle at any point $p$, without affecting the
conformal properties of the amplitude. To restore the cyclic symmetry we can sum over all such
choices. This leads the following manifestly dual {superconformal covariant} expression for the
tree-level MHV superamplitude
\begin{equation}\label{newold}
    {\cal A}^{\rm MHV}_{n;0}  = \frac{1}{n}\sum_{p=1}^n
\frac{\delta^{(4)}(x_p-x_{n+p})\ \delta^{(8)}(\q_p-\q_{n+p})}{\vev{1\, 2}\vev{2\, 3}\ldots\vev{n\,
1}}\,,
\end{equation}
where $x-$ and $\theta-$variables satisfy the defining relations \re{thcon} with $x_{n+p}\neq x_p$
and $\q_{n+p}\neq \q_p$. The superamplitude \re{newold} is obviously Lorentz and $SU(4)$ invariant.
It also has helicity weights $+1$ at each point, so that the total helicity equals $n$ in an
agreement with \re{C-act}. Moreover, it transforms covariantly under inversioncand has equal
conformal weights $+1$ at each point
\begin{equation}\label{trawiconwe}
    I[{\cal A}^{\rm MHV}_{n;0}] = \lr{ x^2_1 x^2_2 \ldots x^2_n} \ {\cal
A}^{\rm MHV}_{n;0}\,.
\end{equation}
However, the representation for the MHV superamplitude \re{newold} is not suitable for the practical
purpose of extracting various components of the superamplitude, e.g., gluon amplitudes like \p{mhv},
etc. To this end, it is necessary to go back to the original form  \p{concorr}, where we explicitly
see the on-shell superspace variables $\eta$.

In Sects.~\ref{Tpma} and \ref{split} we have shown that the special components of the
superamplitude, the split-helicity amplitudes, have covariant dual conformal transformations, while
the rest do not. What is the reason that the manifest dual conformal covariance of the
superamplitude \re{trawiconwe} is lost when going down to its components? The answer can be found in
the {\it inhomogeneous} transformation of the $\eta$'s, Eq.~\p{inveta}. Indeed, let us use $Q$
supersymmetry \re{qsus} to set, e.g., $\q_1=0$ (this choice is compatible with dual conformal
invariance, see \p{t-inv}). Then, using \p{xthrl} we can rewrite \p{inveta} as
\begin{equation}\label{etabeco}
    I[\eta_i] = {\frac{x^2_{i}}{x^2_{i+1}}}\left( \eta_i + \sum_{k=1}^{i-1} \eta_k \lan{k}x^{-1}_i|i]\right)\,.
\end{equation}
We see that, due to the presence of inhomogeneous term in this relation, the different $\eta$ terms
in the expansion of the superamplitude can mix with each other under inversion, yielding complicated
inhomogeneous transformations for their coefficients (partial scattering amplitudes).

Let us give an example which illustrates this effect.  Consider the MHV tree-level superamplitude
\p{newold} and  take the term $p=1$.  Its components originate form the expansion of the Grassmann
delta function in  \p{newold} or  \p{concorr}:
\begin{equation}\label{grabec}
  \delta^{(8)}(\q_1  - \q_{n+1}) =  \delta^{(8)}(\sum_{i=1}^n\ \lambda_i^\a\eta_i^A)
  = \sum_{1\leq i < j\leq n}\vev{i\, j}^4\ (\eta_{i})^4(\eta_j)^4  + \ldots\ ,
\end{equation}
where only purely gluon components are shown.
We already know the  behavior of this  delta function under inversion,
\begin{equation}\label{wefind}
    I\Big[\delta^{(8)}(\q_1  - \q_{n+1}) \Big] = x^{-8}_{1}\delta^{(2)}(\q_1  - \q_{n+1})
\end{equation}
(taking into account the bosonic $\delta^{(4)}(x_1-x_{n+1})$). What can we say about the conformal
properties of its components? In general they are not simple, because of the inhomogeneous term in
\p{inveta}. The exception are the split-helicity amplitudes, for which, e.g., the two
negative-helicity gluons appear at adjacent points  $i$ and $i+1$. Consider, e.g., the term
$\vev{n-1\ n}^4\ (\eta_{n-1})^4(\eta_n)^4$ in the right-hand side of \p{grabec}. From \p{etabeco} it
is clear that only this term is not affected by the inhomogeneous transformations, giving (recall
\p{covvev})
\begin{align}\notag
I[\vev{n-1\ n}^4\ (\eta_{n-1})^4(\eta_n)^4] & = \frac{\vev{n-1\ n}^4}{x^8_{n-1}}\
{\frac{x^8_{n-1}}{x^8_{1}}}\ (\eta_{n-1})^4(\eta_n)^4 + \ldots
\\ \label{givvving}
& = {x^{-8}_1}\ \vev{n-1\ n}^4\ (\eta_{n-1})^4(\eta_n)^4  + \ldots\,,
\end{align}
(here $\ldots$ denotes terms of different types), in accord with \p{wefind}. To show the covariance
of the other split helicity terms $(\eta_{i})^4(\eta_{i+1})^4$ in \p{grabec}, we need to make a
different choice for the `starting point' $\q_1=0$ of the cycle.

The same example shows what happens to other amplitudes, which are not of the split-helicity type.
Take, for instance, the term $(\eta_{n-2})^4(\eta_n)^4$ in \p{grabec}. It mixes under inversion with
similar term ${x^{-8}_1}\ \lan{n-2} x^{-1}_{n-1} x_{n-1\ n} \ran{n}^4\ (\eta_{n-2})^4(\eta_n)^4$
coming from inhomogenous transformation of  $(\eta_{n-1})^4(\eta_n)^4$.  This explains why the MHV
gluon amplitude $A_{n}(1^+\ldots (n-2)^- (n-1)^+ n^-)$  does not have a homogeneous dual conformal
transformation.

\subsection{Conventional and dual superconformal generators}

Just as in the purely bosonic case we can deduce the form of all generators of dual superconformal
transformations by working in the full superspace with coordinates
$x_i,\theta_i,\lambda_i,\tilde\lambda_i,\eta_i$. In this superspace the superamplitude is supported
on a surface defined by the constraints
\be\label{super-sur}
x_i^{\dot\alpha\alpha } - x_{i+1}^{\dot\alpha\alpha } -
\tilde\lambda_i^{\dot\alpha}\lambda_i^{\alpha} =0\,, \qquad\qquad \theta_i^{A\,\alpha} -
\theta_{i+1}^{A\,\alpha} - \lambda_i^{\alpha} \eta_i^A =0\,.
\ee
Following the same logic as we used for the bosonic constraints in Sect.~\ref{sect27}, we can extend
the dual conformal generator acting on $(x,\theta)$ to the surface \re{super-sur} in the full
superspace. The resulting generator,
\be
K^{ \dot\alpha\alpha} = \sum_{i=1}^n \biggl[ x_{i}^{\dot\beta\alpha} x_{i}^{\dot \alpha\beta}
\frac{\partial}{\partial x_i^{\dot\beta\beta }} + x_{i}^{\dot\alpha\beta} \theta_{i }^{B\,\alpha}
\frac{\partial}{\partial \theta_i^{\beta B}} + x_{i}^{\dot\alpha\beta} \lambda_{i}^{\alpha}
\frac{\partial}{\partial \lambda_i^{\beta}} +x_{i+1}^{\dot\beta\alpha} \tilde\lambda_{i}^
{\dot\alpha} \frac{\partial}{\partial \tilde\lambda_i^{\dot \beta}} + \tilde{\lambda}_{i}^{
\dot\alpha} \theta_{i+1}^{B\, \alpha} \frac{\partial}{\partial \eta_i^B} \biggr], \label{sdualK}
\ee
defines the dual conformal transformation of all variables,
$x_i,\theta_i,\lambda_i,\tilde\lambda_i,\eta_i$.

We can obtain the form of $K^{ \dot\alpha\alpha}$ in the on-shell superspace by ignoring the first
two terms in (\ref{sdualK}). As in the bosonic case, the conformal transformation introduces a
dependence on the variables $x_i,\theta_i$ which do not live in the on-shell superspace. To obtain
the generators acting in the dual superspace we can ignore the final two terms in (\ref{sdualK}). In
the same way we can find all generators of the dual superconformal algebra $u(2,2|4)$. These are
given in Appendix \ref{B}.

The following picture summarizes  the relationship between the two superconformal algebras, the
conventional one (acting in the configuration superspace of the particles)  and the dual one (acting
in the dual superspace and shown in capital letters):

\vspace{10mm} \psfrag{p}[cc][cc]{\parbox[t]{0mm}{ {\Large$p$}}} \psfrag{q}[cc][cc]{\parbox[t]{0mm}{
{\Large$q$}}} \psfrag{s}[cc][cc]{\parbox[t]{0mm}{ {\Large$s$}}} \psfrag{k}[cc][cc]{\parbox[t]{0mm}{
{\Large$k$}}} \psfrag{bq}[cc][cc]{\parbox[t]{30mm}{ {\Large\hspace{35pt}$\bar{q} =
\bar{S}$}}}
\psfrag{bs}[cc][cc]{\parbox[t]{30mm}{ {\Large\hspace{35pt}$\bar{s} =
\bar{Q}$}}}
\psfrag{P}[cc][cc]{\parbox[t]{0mm}{ {\Large$P$}}} \psfrag{K}[cc][cc]{\parbox[t]{0mm}{ {\Large$K$}}}
\psfrag{S}[cc][cc]{\parbox[t]{0mm}{ {\Large$S$}}} \psfrag{Q}[cc][cc]{\parbox[t]{0mm}{ {\Large$Q$}}}
\centerline{{\epsfysize4.5cm \epsfbox{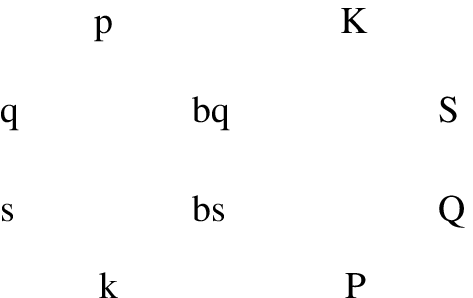}}} \vspace{10mm}

On the left-hand (conventional) side we have the generators $p$ and $q$ which are trivially realized
on the superamplitude (they vanish due to the delta functions). Further, the generators $k$ and $s$
are realized in terms of second-order differential operators (see \p{witt} and \p{specosu}), so the
implementation of these symmetries is not straightforward. On the dual side, $P$ and $Q$ act as
(super)translations leading to the elimination of one point, e.g., $(x_1,\q_1)$ in the dual
superspace. The generators $K$ and $S$ correspond to exact symmetries of the tree-level
superamplitude, but they become anomalous when the loop corrections are switched on. The overlap
between the two superalgebras is over the generators $\bar q = \bar S$ and $\bar s = \bar Q$. The
former remain exact symmetries of the full superamplitude, while the latter are again subject to
anomalies due to infrared divergences.

\section{The complete $n=6$ NMHV superamplitude}
\label{sect-super-NMHV6}

In this section, we shall construct the one-loop NMHV superamplitude for $n=6$ and compare it with
the one-loop expressions for six-gluon NMHV amplitudes computed in \cite{bddk94-2}. We would like to
mention that another approach to constructing the $n=6$ superamplitude  based on the unitary cuts
was proposed in Ref.~\cite{Huang}.

We recall that the superamplitude ${\cal A}^{\rm NMHV}_{6}$ is a generating function of the
scattering amplitudes (more precisely, planar color-ordered partial amplitudes) of scalars, gluinos
and gluons with the total helicity $n-6=0$. These amplitudes can be read as the coefficients in
expansion of ${\cal A}^{\rm NMHV}_{6}$ in powers of $\eta$'s. In particular, the six-gluon
amplitudes are accompanied by the $SU(4)$ singlets $(\eta_i)^4(\eta_j)^4(\eta_k)^4$ (with
$(\eta_i)^4\equiv \frac1{4!}\epsilon_{ABCD}\eta_i^A\eta_i^B\eta_i^C\eta_i^D$)
\be\label{gen-amp}
{\cal A}^{\rm NMHV}_{6} = \sum_{1\le i< j< k \le 6} (\eta_i)^4(\eta_j)^4(\eta_k)^4 \, {A}_{6}(
i^-j^- k^-)+\ldots
\ee
where ${A}^{\rm NMHV}_{6}( i^-j^- k^-)$ stands for  the NMHV six-gluon scattering amplitude with
$i^-$, $j^-$ and $k^-$ denoting gluons with helicity $(-1)$ and remaining gluons carry helicity
$(+1)$. Also, the ellipses represent the scattering amplitudes of scalars and gluinos.  The
scattering amplitudes ${A}_{6}( i^-j^- k^-)$ are invariant under cyclic shifts, $i\to i+1$ and
flips, $i\to 7-i$, of six gluons. As a result, for $n=6$ there are only three nontrivial NMHV
amplitudes~\cite{MP} that can be denoted according to the ordering of the gluon helicities as
$A^{+++---}$, $A^{++-+--}$ and $A^{+-+-+-}$, so that $A^{+++---}\equiv A(1^+2^+3^+4^-5^-6^-)$ and so
on.

\subsection{Tree level}

As we argued in Sect.~3.3, all six-particle amplitudes can be combined into a single superamplitude
$\mathcal{A}_{6}$ given by \re{concorr'}. The $n=6$ NMHV amplitude corresponds to the term involving
$\mathcal{P}_{6}^{(4)}$. At tree level, we take into account \re{concorr} and obtain
\begin{align}\label{n6-0}
{\cal A}^{\rm NMHV}_{6;0} = {\cal A}^{\rm MHV}_{6;0}\mathcal{P}_{6;0
}^{(4)}(\lambda_i,\tilde\lambda_i,\eta_i) \,,
\end{align}
where ${\cal A}^{\rm MHV}_{6;0}$ is the tree-level MHV amplitude
\be\label{superMHV}
{\cal A}^{\rm MHV}_{6;0} =  i(2\pi)^4\frac{\delta^{(4)}(\sum_{i=1}^6\, \lambda_i^\a \bl_i^\da)\
\delta^{(8)}(\sum_{i=1}^6\ \lambda_{i}^\alpha\, \eta_i^A)}{\vev{1\, 2}\vev{2\, 3}\ldots\vev{6\,
1}}\,,
\ee
and $\mathcal{P}_{6;0 }^{(4)}(\lambda_i,\tilde\lambda_i,\eta_i)$ is a homogenous polynomial in
$\eta$'s of degree $4$.

We expect that ${\cal A}^{\rm NMHV}_{6;0}$ should have the same transformation properties with
respect to dual superconformal transformations as ${\cal A}^{\rm MHV}_{6;0}$ and, therefore,
$\mathcal{P}_{6;0 }^{(4)}(\lambda_i,\tilde\lambda_i,\eta_i) $ should be superinvariant. We shall
construct such superinvariants for arbitrary $n$ in Sect.~\ref{sect62}. To simplify the
presentation, we give here the resulting expression for $\mathcal{P}_{6;0
}^{(4)}(\lambda_i,\tilde\lambda_i,\eta_i)$ and refer interested reader to Sect.~\ref{sect62} for
more details. The superinvariant $\mathcal{P}_{6;0 }^{(4)}(\lambda_i,\tilde\lambda_i,\eta_i)$ has
the factorized form
\be\label{P60}
{\cal P}^{(4)}_{6;0}(\lambda_i,\tilde\lambda_i,\eta_i) =  \frac16 \sum_{p,q,r=1}^6 c_{pqr}
\,\delta^{(4)}\lr{ \Xi_{pqr}}\,,
\ee
where sum runs over all possible $n=6$ superinvariants labeled by $p\neq q\neq r$. The Grassmann
valued $\Xi_{pqr}^A(\lambda_i,\tilde\lambda_i,\eta_i)$ are linear in $\eta$ and have the form
\be\label{Xi6}
  \Xi_{pqr}^A  = -\langle p| \left[x_{pq}  x_{qr} \sum_{i=p}^{r-1} |i\rangle \eta_i^A
  + x_{pr}  x_{rq}  \sum_{i=p}^{q-1} |i\rangle \eta_i^A \right],
\ee
with all indices subject to the periodicity condition $i\equiv i+6$. The coefficients
$c_{pqr}(\lambda_i,\tilde\lambda_i)$ do not depend on $\eta$'s and will be determined shortly by
matching \re{n6-0} and \re{P60} with the known result for six-gluon NMHV amplitude. It is
straightforward to verify $ \Xi_{pqr}^A$ transforms covariantly under dual superconformal
transformations (see \re{XiI} below). For ${\cal P}^{(4)}_{6;0}(\lambda_i,\tilde\lambda_i,\eta_i)$
to be invariant under these transformations, the coefficients $c_{pqr}$ should also transform
covariantly in such a way that the dual conformal weights of $c_{pqr}$ and
$\delta^{(4)}\lr{\Xi_{pqr}}$  should compensate each other. We shall verify this property by
explicit calculation in Sect.~\ref{sect62}.

Let us examine the triple sum on the right-hand side of \re{P60} and separate the contribution with
$p=1$. The remaining terms with $2\le p \le 6$ can be obtained from the one for $p=1$ by applying
cyclic shifts of the indices. As follows from the definition \re{Xi6}, $\Xi_{1pq}=\Xi_{1qp}$ and we
can impose the condition $p\le q$. In addition, making use of the on-shell condition \re{twpr},
$x_{i,i+1}\ket{i}=\bra{i} x_{i,i+1}=0$, one can verify that all $\Xi_{1pq}$ vanish except
$\Xi_{135}$, $\Xi_{136}$ and $\Xi_{146}$. Then, examining the explicit expressions \re{Xi6} for
these quantities we find that they are related to each other through cyclic shifts of the indices
\begin{align}\notag
\Xi_{136}  &=  \frac{\vev{12}}{\vev{23}}\Xi_{362} =  \frac{\vev{12}}{\vev{23}} \mathbb{P}^2\,
\Xi_{146}\,,
\\
\Xi_{135}  &=  \frac{\vev{12}}{\vev{23}}\Xi_{352} =  \frac{\vev{12}}{\vev{23}} \mathbb{P}^2\,
\Xi_{136}= \frac{\vev{12}}{\vev{23}} \frac{\vev{34}}{\vev{45}} \mathbb{P}^4\, \Xi_{146}\,,
\end{align}
where  $\mathbb{P}$ performs the cyclic shift of indices $i\to i+1$ modulo the periodicity condition
$i\equiv i+6$. These relations allow us to eliminate $\Xi_{136}$ and $\Xi_{135}$ as well as their
cyclic images from the triple sum  in \re{n6-0} and to rewrite ${\cal A}^{\rm NMHV}_{6;0} $ as
\be\label{n6-ansatz}
{\cal A}^{\rm NMHV}_{6;0} ={\cal A}^{\rm MHV}_{6;0}
%
  \left[ \widetilde c_{146}\, \delta^{(4)}\lr{\Xi_{146}}+ \text{(cyclic)}\right],
\ee
where the expression inside the square brackets is invariant under cyclic shift of indices and
 $\widetilde c_{146}$ is given by a linear combination of $c_{146}$ and cyclicly shifted $c_{135}$ and $c_{136}$
\be\label{c-tilde}
\widetilde c_{146} = \frac16 \left[ c_{146} + c_{514}\lr{\frac{\vev{56}}{\vev{61}}}^4 +
c_{351}\lr{\frac{\vev{56}}{\vev{61}} \frac{\vev{34}}{\vev{45}}}^4\right] \,.
\ee
The relation \re{n6-ansatz} defines the $n=6$ NMHV superamplitude at tree level and it involves only one
unknown function $\widetilde c_{146}$.

The function $\widetilde c_{146}$ can be determined by comparing scattering amplitudes of gluons,
gluinos and scalars generated by \re{n6-ansatz} with the known expressions. To perform the
comparison, we will need the expression for $\Xi_{146}$ in terms of $\lambda_i$, $\bar\lambda_i$ and
$\eta_i$. Using \re{Xi6} we find after some algebra
\be
\Xi_{146}^A = \vev{61}\vev{45}\left({\eta_4^A [56]+\eta_5^A[64]+\eta_6^A[45]}\right)\,.
\ee
Based on our analysis we expect that the expression for $\widetilde c_{146}$ has to ensure the dual
conformal invariance of the superamplitude \re{n6-ansatz}. In particular, to compensate the
conformal weight of $\Xi_{146}$ (see Eq.~\re{XiI} below) it should transform under inversions in the
dual space as
\begin{equation}
    I[\widetilde c_{146}] = \widetilde c_{146}\lr{x_1^2 x^2_4 x^2_6}^4\,,
\end{equation}
so that $I[\widetilde c_{146}\delta^{(4)}\lr{\Xi_{146}}] =\widetilde
c_{146}\delta^{(4)}\lr{\Xi_{146}}$.

\subsection{Generalization to all loops}

It is straightforward to generalize the relation \re{n6-ansatz} beyond tree level
\be\label{n6-ansatz1}
{\cal A}^{\rm NMHV}_{6} ={\cal A}^{\rm MHV}_{6}
  \left[ \widetilde c_{146}\, \delta^{(4)}\lr{\Xi_{146}} \left(1+a V_{146}\right)+ \text{(cyclic)}\right]+O(a^2)\,,
\ee
where ${\cal A}^{\rm NMHV}_{6}$ and ${\cal A}^{\rm MHV}_{6}$ stand for the all-loop superamplitudes,
$\widetilde c_{146}$ is  defined in \re{c-tilde} and $V_{146}$ is a scalar function of $x_i$. Note
that the form of $\Xi_{pqr}$ is fixed by dual superconformal symmetry and, therefore, $\Xi_{pqr}$ is
protected from perturbative corrections.

Writing down \re{n6-ansatz1} we have tacitly assumed that the expression inside square brackets in
the right-hand side of \re{n6-ansatz1}, to which we shall refer as the `ratio' of the
superamplitudes ${\cal A}^{\rm NMHV}_{6}$ and ${\cal A}^{\rm MHV}_{6}$, possesses the dual conformal
invariance beyond tree level. This property is extremely nontrivial given the fact that the dual
conformal invariance of the MHV amplitude ${\cal A}^{\rm MHV}_{6}$ is known to be broken already at
one-loop level. The reason for this is that loop corrections to the amplitude contain infrared
divergences which are regularized within dimensional regularization by evaluating the relevant
Feynman integral in $D=4-2\epsilon$ dimensions. This immediately breaks dual conformality of the
amplitude and induces an anomalous contribution to the conformal Ward identities. The phenomenon is
rather general and it applies to the amplitude ${\cal A}^{\rm NMHV}_{6}$ which also has infrared
divergences to any loop order. Note that the infrared divergences are not sensitive to the
helicities of external particles. They have the same universal form for MHV and NMHV amplitudes and,
therefore, they cancel in the ratio $R_6^{\rm NMHV}$ of the superamplitudes defined as
\be\label{R61}
{\cal A}^{\rm NMHV}_{6} = {\cal A}^{\rm MHV}_{6}\left[R_6^{\rm NMHV} + O(\epsilon) \right].
\ee
Contrary to the superamplitudes  ${\cal A}^{\rm NMHV}_{6}$ and ${\cal A}^{\rm MHV}_{6}$, the ratio
function $R_6^{\rm NMHV}$ is infrared finite and, therefore, it is well-defined in $D=4$ dimension.
This suggests that $R_6^{\rm NMHV}$ should possess dual superconformal invariance to all loops. If
so, then comparing \re{n6-ansatz1} and \re{R61}, we conclude that the ratio function
\be\label{R6}
R_6^{\rm NMHV} =   \widetilde c_{146}\left[1+ a V_{146}\right]\, \delta^{(4)}\lr{\Xi_{146}}+
\text{(cyclic)}+O(a^2)
\ee
should be invariant under dual superconformal transformations and, as a consequence, $R_6^{\rm
NMHV}$  has to satisfy a (nonanomalous) conformal Ward identity
\be\label{Ward-R6}
K^{a\dot a}R_6^{\rm NMHV} = D\, R_6^{\rm NMHV} =0
\ee
with the conformal boost operator $K^{a\dot a}$ and the dilatation operator $D$ defined in \re{KK}
and \re{DD}, respectively. As we will see in a moment, this is indeed the case, at least to one
loop.

To determine the function $\tilde c_{146}$ we shall apply \re{gen-amp} to extract from the $n=6$
NMHV amplitude \re{n6-ansatz1} the expressions for the six-gluon NMHV scattering amplitudes
${A}_{6}( i^-j^- k^-)$ and compare them with the known one-loop expressions \cite{bddk94-2}. To make
use of \re{n6-ansatz1} we have to specify the perturbative corrections to the MHV superamplitude
${\cal A}_6^{\rm MHV}$. The duality relation between the MHV amplitudes and light-like Wilson loops
allows us to write \footnote{Strictly speaking the duality between $\ln W_6$ and $\ln M_6^{(\rm
MHV)}$ holds up to unessential additive constant and involves a nontrivial redefinition of the
infrared/UV regulators. We refer interested reader to \cite{dhks4} for more details.}
\be\label{A=W}
{\cal A}_6^{(\rm MHV)}={\cal A}_{6;0}^{(\rm MHV)} M_6^{(\rm MHV)}\,, \qquad  \ln M_6^{(\rm MHV)} =
\ln W_6 + O(\epsilon,1/N^2)
\ee
where the tree level $n=6$ MHV superamplitude ${\cal A}_{6;0}^{\rm MHV}$ is given by \re{concorr}
and $W_6$ is the vacuum expectation value of the light-like Wilson loop evaluated over a hexagonal
contour with vertices located at the points $x_i^\mu$ (with $i=1,\ldots,6$) which are the dual
coordinates related to the gluon momenta $x_i-x_{i+1}=p_i$. To one-loop level, we find
\begin{align}\label{W6}
 \ln W_6  = &  a\,\bigg\{-\frac{1}{2\epsilon^2}\sum_{i=1}^6 \lr{-x_{i,i+2}^2\mu^2}^\epsilon+\frac{1}{2}\sum_{i=1}^6 \bigg[ -\ln\lr{\frac{x_{i,i+2}^2}{x_{i,i+3}^2}}\ln\lr{\frac{x_{i+1,i+3}^2}{x_{i,i+3}^2}}
\\ & \notag
+\frac14\ln^2\lr{\frac{x_{i,i+3}^2}{x_{i+1,i+4}^2}}-\frac12 {\rm
Li}_2\lr{1-\frac{x_{i,i+2}^2x_{i+3,i+5}^2}{x_{i,i+3}^2x_{i+2,i+5}^2}}\bigg]\bigg\}+ O(a^2)
\end{align}
where $a=g^2 N/(8\pi^2)$ and the periodicity condition $i\equiv i+6$ is imposed. Combining
the relations \re{n6-ansatz1}, \re{A=W} and \re{superMHV}, we obtain
\begin{align}\label{A6/W6}
{\cal A}^{\rm NMHV}_{6}/W_6  & =(2\pi)^4\delta^{(4)}(\sum_{i=1}^6p_i)    \frac{ \vev{61}^4\vev{45}^4
}{\vev{1\, 2}\vev{2\, 3}\ldots\vev{6\, 1}}\widetilde c_{146} \lr{1+a V_{146}}
\\ \notag
&\times \delta^{(8)}(\sum_{j=1}^6\ \lambda_{j}^\alpha\, \eta_j^A)\, \delta^{(4)}{ \big({\eta_4
[56]+\eta_5[64]+\eta_6[45]}\big)}+ \text{(cyclic)}\,.
\end{align}
To extract the six-gluon scattering amplitude  ${A}_{6}( i^-j^- k^-)$ from this relation, we make
use of \re{gen-amp}, expand the product of two delta functions on the right-hand side of \re{A6/W6}
and identify the coefficient in front of $\eta_i^4\eta_j^4\eta_k^4$. To simplify the calculation we
choose the amplitude  ${A}_{6}( 4^-5^- 6^-)\equiv A^{+++---}$ and concentrate on the terms
$\sim\eta_4^4\eta_5^4\eta_6^4$ only. Making use of the identity
\be
\delta^{(8)}(\sum_{j=1}^6\ \lambda_{j}^\alpha\, \eta_j^A) = \vev{ik}^4 \delta^{(4)} \bigg(\eta_i
+\sum_{j\neq k}\eta_j \frac{\vev{jk}}{\vev{ik}}\bigg) \delta^{(4)} \bigg(\eta_k +\sum_{j\neq
i}\eta_j \frac{\vev{ji}}{\vev{ki}}\bigg)
\ee
and choosing appropriately the indices $i$ and $k$, we find after some algebra
\begin{align}\label{A-hs}
A^{+++---}/W_6   =  \tilde C(a) &+\lr{\frac{[23]\vev{56} }{x_{25}^2}}^4 \, \mathbb{P}^{-2} \tilde
C(a) +\lr{\frac{ \bra{4}x_{41}|1]}{x_{25}^2}}^4 \, \mathbb{P} \,\tilde C(a) \,,
\\ \notag
& +\lr{\frac{[12]\vev{45} }{x_{36}^2}}^4\, \mathbb{P}^2 \tilde C(a) +
\lr{\frac{\bra{6}x_{63}|3]}{x_{36}^2}}^4\, \mathbb{P}^{-1} \tilde C(a)  \,,
\end{align}
where the following notation was introduced
\be\label{tilde-C}
\tilde C(a) = \widetilde c_{146}  \frac{\lr{ {\vev{61} \vev{45}}\,{x_{14}^2}}^4 }{\vev{1\,
2}\vev{2\, 3}\ldots\vev{6\, 1}}\left[1+ a V_{146} \right]
\ee
and $\mathbb{P}$ generates the cyclic shift of indices $i\to i+1$ so that  $\mathbb{P}^k\,\tilde
C(a)$ means that all indices in the expression for $\tilde C(a)$ should be shifted by $i\to i+k$.

\subsection{Six-gluon NMHV amplitudes to one loop}

To one-loop order, the six-gluon NMHV color-ordered amplitudes $A^{+++---}$, $A^{++-+--}$ and
$A^{+-+-+-}$ were computed in \cite{bddk94-2}. To perform a comparison with \re{A-hs} we will only need the
expression for the first  amplitude. It reads
\begin{align}\label{A-res}
& A^{+++---} =  A_{6;0} + g^2   A_{6;1}  + O(g^4)\,,
\end{align}
where the expansion coefficients are given by
\begin{align}
& A_{6;0} =  \frac12\left[B_1   + B_2  +B_3   \right]
\\ \notag
& A_{6;1}  = c_\Gamma N \left[B_1  F_6^{(1)} + B_2  F_6^{(2)} +B_3  F_6^{(3)}  \right].
\end{align}
Here $c_\Gamma = (4\pi)^{-2+\epsilon} {\Gamma(1+\epsilon)\Gamma^2(1-\epsilon)}/
{\Gamma(1-2\epsilon)}$ is a normalization factor  and $F_6^{(i)} $ stand for a combination of box
integrals evaluated within the dimensional regularization with $D=4-2\epsilon$
\begin{align}\notag
F_6^{(i)} =
&-\frac1{2\epsilon^2}\sum_{k=1}^6 \lr{\frac{\mu^2}{-x_{k,k+2}^2}}^\epsilon - \ln\lr{
\frac{x_{i,i+3}^2}{x_{ {i,i+2}}^2}}\ln\lr{\frac{x_{i,i+3}^2}{x_{i+1,i+3}^2}}
\\\notag
& - \ln\lr{\frac{x_{i,i+3}^2}{x_{i+3,i+5}^2}}\ln\lr{\frac{x_{i,i+3}^2}{x_{i+4,i+6}^2}} +
\ln\lr{\frac{x_{i,i+3}^2}{x_{i+2,i+4}^2}}\ln\lr{\frac{x_{i,i+3}^2}{x_{i+5,i+1}^2}}
\\\notag
& +\frac12 \ln\lr{\frac{x_{i,i+2}^2}{x_{i+3,i+5}^2}} \ln\lr{\frac{x_{i+1,i+3}^2}{x_{i+4,i+6}^2}}
+\frac12 \ln\lr{\frac{x_{i-1,i+1}^2}{x_{i,i+2}^2}} \ln\lr{\frac{x_{i+1,i+3}^2}{x_{i+2,i+4}^2}}
\\
& +\frac12 \ln\lr{\frac{x_{i+2,i+4}^2}{x_{i+3,i+5}^2}}
\ln\lr{\frac{x_{i+4,i+6}^2}{x_{i+5,i+1}^2}}+\frac{\pi^2}{3}\,,
\end{align}
where $x_{i,i+3}^2$ and $x_{i,i+2}^2$ are Mandelstam variables expressed in
terms of the dual coordinates. The explicit expressions for the functions $B_a$ are \cite{bddk94-2}%
\footnote{In these relations we used the expressions for $B_a$ from \cite{bddk94-2} in which we substituted
the gluon momenta by the dual coordinates $p_i=x_i-x_{i+1}$ and used the notations for contraction of
spinor indices specified in Appendix \ref{A}.}
\begin{align}\label{B1}
B_1=& i \frac{(x_{14}^2)^3}{\vev{12}\vev{23}[45][56] \bra{1}x_{14}|4]  \bra{3}x_{36}|6]}
\\ \notag
B_2= & \lr{\frac{[23]\vev{56} }{x_{25}^2}}^4 \, \mathbb{P}^{-2} B_1 +\lr{\frac{
\bra{4}x_{41}|1]}{x_{25}^2}}^4 \, \mathbb{P} \,B_1  \,,
\\ \notag
B_3=& \lr{\frac{[12]\vev{45} }{x_{36}^2}}^4\, \mathbb{P}^2 B_1 +
\lr{\frac{\bra{6}x_{63}|3]}{x_{36}^2}}^4\, \mathbb{P}^{-1} B_1
\end{align}
We observe a remarkable similarity between these expressions and those in the expansion of the
superamplitude \re{A-hs}.

To separate the infrared divergent and finite parts in $A^{+++---}$, we divide both sides of
\re{A-res} by the Wilson loop $W_6$ given by \re{W6}
\begin{align}
A^{+++---}/W_6 =  \frac12 B_1 \left(1 + a V_6^{(1)} \right)+\frac12 B_2 \left( 1 + a V_6^{(2)}
\right) +\frac12 B_3 \left( 1 + a V_6^{(3)} \right)+O(\epsilon)\,.
\end{align}
As was already mentioned, infrared divergences have a universal form for all scattering amplitudes
and, therefore, given the fact that the Wilson loop $W_6$ captures the divergences of the MHV
amplitudes, we expect that $A^{+++---}/W_6$ should be finite as $\epsilon\to 0$. Indeed, performing the calculation of $A^{+++---}/W_6$ to one loop, we find that the functions $V_6^{(i)}$ (with
$i=1,2,3$) do not contain infrared divergences and have the following form
\be
V_6^{(i)}  = - \ln u_i \ln u_{i+1} +\frac12\sum_{k=1}^3  \bigg[{ \ln u_k \ln u_{k+1} + {\rm
Li}_2(1-u_k) }\bigg]
\ee
where the periodicity condition $u_{i+3}=u_i$ is implied and $u_1$, $u_2$ and $u_3$ are conformal
cross-ratios in the  dual coordinates
\be
u_1=\frac{x_{13}^2x_{46}^2}{x_{14}^2x_{36}^2}\,,\qquad
u_2=\frac{x_{24}^2x_{15}^2}{x_{25}^2x_{14}^2}\,,\qquad
u_3=\frac{x_{35}^2x_{26}^2}{x_{36}^2x_{25}^2}\,.
\ee
Note that $V_6^{(i)}$ are invariant under cyclic shifts of the indices
\be\label{V-shift}
\mathbb{P}^3\,V_6^{(i)} = V_6^{(i+3)} = V_6^{(i)}
\ee
and, most importantly, $V_6^{(i)}$ are invariant under conformal transformations of $x_i$.

We observe here the same phenomenon as was already mentioned in Sect.~5.2. Namely, $A^{+++---}$ and
$W_6$ both contain divergences and their (dual) conformal invariance is broken in dimensional
regularization. Nevertheless, the ratio  $A^{+++---}/W_6$ is finite for $\epsilon\to 0$ and, as a
result, it is invariant under conformal transformations in dual $x-$variables.

\subsection{Conformal Ward identities for $n=6$ NMHV superamplitude}

Let us  compare the two expressions for the amplitude $A^{+++---}$ given by  \re{A-hs} and
\re{A-res}. We recall that the latter expression is  the result of the perturbative one-loop calculation
of Ref.~\cite{bddk94-2} while the former expression comes from the expansion of $n=6$ NMHV superamplitude
and involves yet unknown function $\tilde C(a)$ defined in \re{tilde-C}. This function depends in
turn on the function $\tilde c_{146}$ carrying the dependence on helicities of the external gluons
and the scalar function $V_{146}$.

Matching \re{A-hs} and \re{A-res} we take into account the identity \re{V-shift} and find that the
two expressions for the scattering amplitude coincide upon identification $\tilde C(a)= B_1 (1 + a
V_6^{(1)} )/2$, or equivalently
\be
\tilde C(a) = \frac12 B_1  \frac{\vev{1\, 2}\vev{2\, 3}\ldots\vev{6\, 1}}  {({x_{14}^2}{\vev{61}
\vev{45}})^4} \left(1 + a V_6^{(1)} \right)+ O(a^2)\,.
\ee
Then, we replace $B_1$ by its expression \re{B1}, compare the result with \re{tilde-C} and identify
the expressions for the three-level helicity function $\tilde c_{146}$
\begin{align}\label{tilde-c}
\tilde c_{146} = \frac{1}2 \frac{\vev{34}\vev{56}}{x_{14}^2\bra{1}\,x_{14}|4]
\bra{3}\,x_{36}|6]\lr{\vev{45}\vev{61}}^3 [45][56]}\,,
\end{align}
and for the one-loop scalar function $V_{146}=V_6^{(1)}$
\begin{align}\label{V146}
V_{146} = - \ln u_1 \ln u_{2} +\frac12\sum_{k=1}^3  \bigg[{ \ln u_k \ln u_{k+1} + {\rm Li}_2(1-u_k)
}\bigg]\,.
\end{align}
%
Substituting \re{tilde-c} and \re{V146} into \re{n6-ansatz1} and \re{R6} we obtain the one-loop
expression for the $n=6$ NMHV superamplitude and the corresponding ratio function, respectively.

By the construction, the superamplitude ${\cal A}^{\rm MHV}_{6}$ defined in \re{A6/W6} reproduces
the known one-loop expression for a particular six-gluon NMHV amplitude $A^{+++---}$. At the same
time, it also generate the two other gluon nMHV amplitudes $A^{++-+--}$ and $A^{+-+-+-}$ as well as
many other amplitudes containing gluinos and scalars. In particular, it follows from our analysis
that {\it all} tree-level NMHV amplitudes are described by the following compact expression
\begin{align}
{\cal A}^{\rm MHV}_{6;0} &=  i(2\pi)^4\frac{\delta^{(4)}(\sum_{i=1}^6\, \lambda_i \bl_i)\
\delta^{(8)}(\sum_{j=1}^6\ \lambda_{j}^\alpha\, \eta_j^A)}{\vev{1\, 2}\vev{2\, 3}\ldots\vev{6\, 1}}
\\ \notag
&\times
 \left[ \widetilde c_{146}\, \vev{61}^4\vev{45}^4\,\delta^{(4)} {\big({\eta_4 [56]
 +\eta_5[64]+\eta_6[45]}\big)}+ \text{(cyclic)}\right]
\end{align}
with $\widetilde c_{146}$ given by \re{tilde-c}. Moreover, adding the one-loop perturbative correction
to the ratio function \re{R6} simply amounts to inserting the additional factor $\lr{1+a V_{146}}$
involving \re{V146}
\be\label{R6-1}
R_{6}^{\rm NMHV} =   \widetilde c_{146}\vev{61}^4\vev{45}^4\, \left[1+ a V_{146}\right]\,
\delta^{(4)} {\big({\eta_4 [56]+\eta_5[64]+\eta_6[45]}\big)} + \text{(cyclic)}
\ee
As a nontrivial test of these relations, we have verified that, when expanded in powers of $\eta$'s,
the expressions for ${\cal A}^{\rm MHV}_{6;0}$ and ${\cal A}^{\rm MHV}_{6;1}$ correctly reproduce
all known expressions for tree-level and one-loop $n=6$ NMHV scattering amplitudes.

Next we would like to check the transformation properties of the $n=6$ NMHV superamplitude
\re{n6-ansatz1} under dual superconformal transformations. To this end, we examine the ratio
function \re{R6}. As before, the only nontrivial transformations are conformal inversions. Since
$V_{146}$ is conformal invariant, Eq.~\re{V146}, while $\Xi_{146}$ transforms covariantly  (see
Eq.~\re{XiI} below), we have to examine the action of conformal inversions on the function
$\widetilde c_{146}$, Eq.~\re{tilde-c}. To do this, it is convenient to obtain another, equivalent
representation for $\widetilde c_{146}$.

We recall that $\tilde c_{146}$ is given by a linear combination of three terms  \re{c-tilde}
involving the functions $c_{146}$, $c_{514}$ and $c_{351}$. Since the six-gluon NMHV superamplitude
only depends on $\tilde c_{146}$, the definition of the latter functions is ambiguous. We can make
use of this ambiguity to choose the following ansatz for $c_{pqr}$ (with $1\le p, q ,r \le 6$ and
$p\neq q\neq r$)
\begin{align}\label{c-pqr}
c_{pqr} = - \frac{\vev{q-1\, q}\vev{r-1\, r}}{x^2_{qr}\ \vev{p|x_{pr} x_{r\, q-1}| q-1} \
\vev{p|x_{pr} x_{r\, q}| q} \ \vev{p|x_{pq} x_{q\, r-1}| r-1} \ \vev{p|x_{pq} x_{q\, r}| r} }\,.
\end{align}
where we used the notation for the contraction of spinor indices explained in Appendix~A. It is
straightforward to verify that the three terms inside the square brackets on the right-hand side of
\re{c-tilde} produce the same contribution and reproduce the relation \re{tilde-c}
\be
\tilde c_{146} = \frac{1}2 c_{146}\,.
\ee
The relation \re{c-pqr} admits a natural generalization from $n=6$ to arbitrary $n> 6$.
As we will argue in the next section, the coefficient functions $c_{pqr}$ enter the expression
for the one-loop $n-$particle NMHV superamplitudes.

Another remarkable feature of \re{c-pqr} is that $c_{pqr}$ is built from exactly those conformal
covariant combinations of spinors that we already encountered before in Sect.~2.6. Making use of the
relations \re{invshch} and \re{covvev}, it is straightforward to verify that
\begin{equation}
    I[c_{pqr}] = c_{pqr}\lr{{x_1^2 x^2_4 x^2_6}}^4\,.
\end{equation}
It is remarkable that $c_{pqr}$ transforms covariantly under conformal inversions. Most importantly,
the corresponding conformal weight is exactly the one that is needed to compensate the conformal
weight of $\delta^{(4)}\lr{\Xi_{146}}$, Eq.~\re{XiI}. This means that the product $c_{146}\,
\delta^{(4)}\lr{\Xi_{146}}$ is invariant under inversion and therefore under the $SO(2,4)$ dual
conformal transformations. One can verify that it is also invariant under superconformal
transformations.

Thus, in agreement with our expectations,  the ratio function \re{R6} is superconformal invariant
and verifies the conformal Ward identities \re{Ward-R6}. This does not imply however that the
amplitude ${\cal A}^{\rm NMHV}_{6}$ is covariant under these transformations. On the contrary, its
conformal properties are broken by infrared divergences already at one loop but the corresponding
anomalous contribution cancels against a similar contribution from ${\cal A}^{\rm MHV}_{6}$ in such
a way that the corresponding ratio function  remains conformal.

\section{Next-to-MHV superamplitude}\label{NMHVamplitude}

In this section we propose the general form of the $n$--particle NMHV one-loop superamplitude. We
construct it from a particular set of three-point nilpotent dual superconformal invariants, which
encode the super-helicity structures. Each such invariant is accompanied by a finite and exactly
dual conformal invariant function made of one-loop momentum integrals. We briefly discuss the
twistor coplanarity properties of the super-helicity structures. We also propose a new, manifestly
Lorentz covariant form of the $n$--particle NMHV tree-level superamplitude.

\subsection{General structure of the NMHV superamplitude}

The NMHV gluon amplitudes are characterized by the presence of three negative-helicity  gluons. To
describe them we need the second term in the expansion \p{concorr'}, i.e. a superamplitude of
Grassmann degree of homogeneity 12. It contains terms of the type
$(\eta_i)^4(\eta_{j})^4(\eta_{k})^4$, whose coefficients are the amplitudes with gluons of helicity
$-1$ at points $i,j,k$ and $+1$ elsewhere. By the same counting, in the MHV case the required
Grassmann degree is 8, and it is indeed provided by the Grassmann delta function in \p{mhv}.

We can assume that the entire MHV amplitude \p{concorr} should appear as a prefactor in the NMHV
amplitude. The two delta functions are needed for conservation of the (super)momenta. Further, the
bosonic denominator in \p{concorr} supplies the necessary helicity and conformal weights $+1$ at
each point. Therefore, the generalization of the MHV superamplitude we are looking for should have
the form
\begin{equation}\label{genmhvam}
    {\cal A}^{\rm NMHV}_n = {\cal A}^{\rm MHV}_{n}\times [R^{\rm NMHV}_n + O(\ep)]
\end{equation}
where $R^{\rm NMHV}_n \propto (\eta)^{4}$ is a new factor of Grassmann degree $4$. Its perturbative
expansion starts with a tree-level part, after which come   the loop corrections. Since the MHV
prefactor carries the necessary helicity and conformal weights, we deduce that $R^{\rm NMHV}_n$ must
be a Lorentz scalar of vanishing helicity and be a {\it dual superconformal invariant} \footnote{At
least under the action of the dual supersymmetry generators $Q$ and $
\bar S$, see the discussion in Sect.~\ref{cnbhmvnftl}.}.

Let us try to determine the number of such superconformal invariants. First of all, invariance under
$Q$--supersymmetry \p{qsus} implies that they depend only on the $Q$--invariant variables $\eta$.
Further, using the two projections of the Grassmann parameter $\bar \rho_{\da}$ of $\bar
S$--supersymmetry \p{sus}, we can set to zero any two $\eta$'s. Then, we can solve for another pair
of $\eta$'s from the conservation condition \p{thcon}. Thus, in the end we find only $n-4$
independent $Q$-- and $\bar S$--supersymmetry invariant variables. We could use them as a basis for
constructing $SU(4)$ invariants  $\ep_{ABCD}\eta^A_k\eta^B_l\eta^C_m\eta^D_n$ of the required degree
four. The last step would be to try to make all this dual conformal invariant (not forgetting that
we need helicity weight zero). This is not so easy, given the inhomogeneous transformation law
\p{inveta}.

At first sight, the above procedure looks quite complicated. Fortunately, the relevant set of dual
superconformal invariants is suggested to us by comparison with the results on the $n-$gluon NMHV
amplitude of Bern et al \cite{Bern:2004bt}.\footnote{These invariants can be directly obtained by a
computation \cite{toappear} using a supersymmetrized version of the generalized four-particle cut
technique of \cite{Britto:2004nc}.} In the rest of this section we describe this construction and
formulate our proposal for the complete $n$-particle NMHV superamplitude. This proposal generalizes
the  detailed analysis of the simplest, $n=6$ NMHV superamplitude in Sect.~\ref{sect-super-NMHV6}.

\subsection{Three-point superconformal covariants}\label{sect62}

Our main building block for the new Grassmann factor $R^{\rm NMHV}_n$ will be a set of dual
superconformal invariants $R_{pqr}$ which we construct in Sect.~\ref{3ptscin}. These invariants are
in turn made of dual superconformal covariants $\Xi^A_{pqr}$, which are linear combinations of
$\theta$'s labeled by a triplet of points $p,q,r$ in dual superspace. They have the following
manifest properties: (i) invariance under $Q$-- and $\bar S$--supersymmetry; (ii) invariance under
translations ($P$); (iii) covariance under inversion ($I$) and $SU(4)$. In addition, we want
$\Xi^A_{pqr}$ to be Lorentz scalars (but they will carry helicity at point $p$). The key idea in
constructing such dual supercovariants is to use the two linear transformations with odd parameters
\p{qsus}, \p{ssus} in order to fix a coordinate frame in which two $\q$'s are set to zero. Such a
choice is consistent with conformal invariance, as follows from \p{t-inv}. Then we can easily
construct a conformal covariant from each of the remaining $\q$'s, which are $Q$-- and $\bar
S$--invariant in this frame. Finally, we can undo the supersymmetry transformation which lead to
this frame, thus obtaining the manifest supercovariant  $\Xi^A_{pqr}$.

Let us see how this works in detail.\footnote{To simplify the notation, we suppress the $SU(4)$
index $A$.} Choose any triplet of distinct points $p,q,r$, $p \neq q \neq r$. Using \p{qsus},
\p{ssus} we can shift away two of the $\q$'s of the triplet, e.g.,
\begin{equation}\label{susframe}
    \q_q=\q_r=0
\end{equation}
To this end we have to find parameters $\ep,\bar\rho$ such that
\begin{eqnarray}
  \q'_q &=& \q_q + \ep +x_q \bar\rho = 0 \nn\\
  \q'_r &=& \q_r + \ep +x_r \bar\rho = 0  \label{systwoeqs}
\end{eqnarray}
The solution of these {linear} equations is
\begin{equation}\label{solthese}
    \bar\rho = -x^{-1}_{qr} (\q_q-\q_r)\,, \qquad \ep = -x_q x^{-1}_{qr} \q_r + x_r x^{-1}_{qr} \q_q
\end{equation}
where we are assuming that $x^2_{qr}\neq 0$, i.e. $|q-r| \geq 2$.\footnote{Recall that for two
adjacent points in dual space, e.g., $q$ and $q+1$, the `distance' $x^2_{q\, q+1}=0$.}

We remark that the case $|q-r| = 2$ is exceptional. Take, for instance, $q=r+2$. Shifting away
$\q_r$ and $\q_{r+2}$ implies $\q_{r+1}=-\lambda_r \eta_r = \lambda_{r+1} \eta_{r+1}$.  Then the
linear independence of $\lambda_r$ and $\lambda_{r+1}$ yields $\q_{r+1}=0$, as well as $\eta_{r} =
\eta_{r+1} =0$. So, in this case we can shift away not two, but three neighboring $\q$'s.

In the special frame \p{susframe}, an obvious (but certainly not unique) conformal covariant and
Lorentz invariant, made of the remaining $\q_p$, is
 \begin{equation}\label{anobvi}
    \lambda^\a_p\, \q_{p\, \a} \equiv \vev{ p| \q_p}
\end{equation}

Let us now construct the supercovariant by `undoing' the supersymmetry transformations \p{systwoeqs}
which lead to \p{susframe}. This means to do another transformation with the {same parameters}. The
result is (to put the covariant in a more symmetric form, we multiply it by $x^2_{qr}$)
\begin{equation}\label{Xi}
    \Xi_{pqr} = \Xi_{prq} = \lan{p} x_{pq}  x_{qr} \ran{\q_r} + \lan{p}x_{pr}  x_{rq} \ran{\q_q} + x^2_{qr} \vev{p|\q_p}
\end{equation}
Note that this covariant depends on {\it three points} in dual superspace, $x_{p,q,r}$,
$\q_{p,q,r}$, as well as on the spinor variable $\lambda_p$. The latter carries helicity weight
$-1/2$ at point $p$.

This procedure automatically produces a dual conformal covariant. Indeed, using the transformation
rules \p{weremthatt}, \p{l-lb-inv}  and \p{t-inv}, the covariance under inversion is manifest,
\begin{equation}\label{XiI}
    I[\Xi_{pqr}] = \frac{\Xi_{pqr}}{x^2_p x^2_q x^2_r}
\end{equation}
The invariance under $Q$ follows from the identity $x_{pq}  x_{qr}  + x_{pr}  x_{rq}  + x^2_{qr}
\mathbb{I} = 0$, which also allows us to rewrite \p{Xi} as follows:
\begin{eqnarray}
  \Xi_{pqr} &=& \langle p| \left[x_{pq}  x_{qr} (\ran{\q_r}- \ran{\q_p}) + x_{pr}  x_{rq} (\ran{\q_q}- \ran{\q_p})\right] \label{Xi'}\\
   &=& -\langle p| \left[x_{pq}  x_{qr} \sum_{i=p+1}^{r-1} |i\rangle \eta_i + x_{pr}  x_{rq}  \sum_{i=p+1}^{q-1} |i\rangle \eta_i \right] \label{Xi''}
\end{eqnarray}
(where we have assumed that $p<\min(q,r)$). As explained earlier, these two symmetries yield
invariance under $\bar S$ (which is also very easy to check explicitly using \p{ssus}).

Note that besides the forbidden choice $|q-r| = 1$ (it is easy to see that \p{Xi} vanishes in this
case), there is another case where the covariant \p{Xi} is trivial, $q=p+1$ (or, equivalently,
$r=p+1$). The reason for this is simple. In the frame \p{systwoeqs}, if  for example $q=p+1$, we
have $\q_{p+1}=0$. But then $\q_p =\q_{p+1} + |p\rangle \eta_p = |p\rangle \eta_p$, so $\vev{ p|
\q_p} = 0$ and  $\Xi_{p,p+1,r}=0$.

Without loss of generality we can order the the points $p,q,r$ clockwise on the circle, $p < q < r\
\pmod n$. Then the supercovariants \p{Xi} exist for the following choices of the indices:
\begin{eqnarray}
  && 1 \leq p \leq n \nn \\
  && p+2 \leq q \leq p-3\ \pmod n \nn \\
  && q+2 \leq r \leq p-1\ \pmod n \label{restrlabels}
\end{eqnarray}
A convenient way to represent such supercovariants pictorially is to use the box diagrams from
\cite{Bern:2004bt}. There they were introduced to depict the coefficients of the three-mass box
integrals in the one-loop NMHV amplitude.

%
\psfrag{Xi}[cc][cc]{$\Xi_{pqr} \ =$}
\psfrag{p}[cc][cc]{$p$}\psfrag{q}[cc][cc]{$q$}\psfrag{r}[cc][cc]{$r$}
\centerline{{\epsfysize5cm \epsfbox{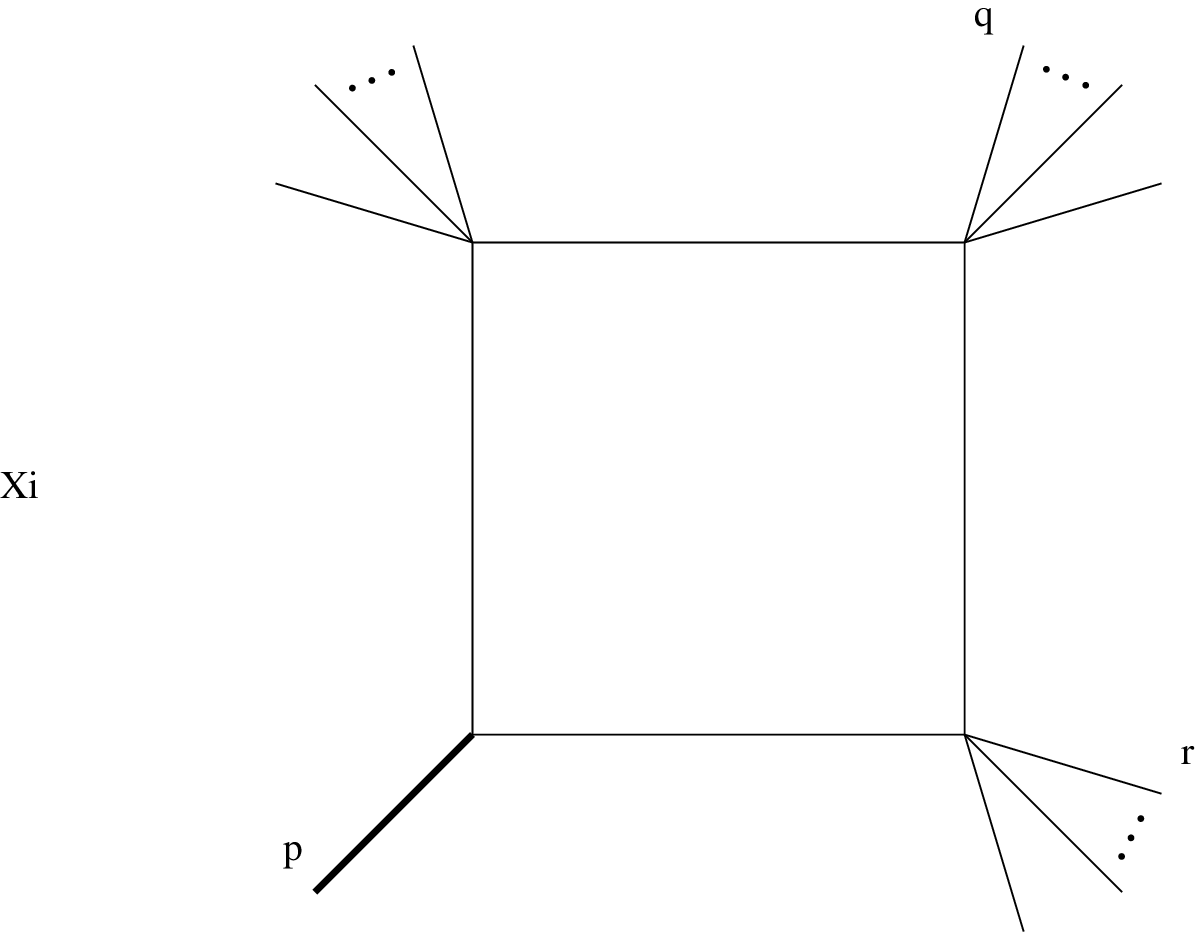}}}

\psfrag{p2}[cc][cc]{$p_2$}\psfrag{p3}[cc][cc]{$p_3$}\psfrag{pn}[cc][cc]{$p_n$}
\psfrag{x1}[cc][cc]{$x_1$}\psfrag{x2}[cc][cc]{$x_2$}\psfrag{x3}[cc][cc]{$x_3$}\psfrag{xn}[cc][cc]{$x_n$}
\psfrag{xn1}[cc][cc]{$x_{n-1}$}

It is easy to count the total number of supercovariants \p{Xi}, $n(n-3)(n-4)/2$. Clearly, it largely
exceeds the number $n-4$ of independent covariants that we found above. This suggests that there
exist many identities relating various $\Xi$'s. Here we present only one of them, which was needed
in Sect.~5. Consider the supercovariants $\Xi_{p, p+2,r}$ and $\Xi_{p+2,r,p+1}$ (the restriction $r
\geq p+4$ is assumed; the points are ordered clockwise on the circle). Let us fix the frame
$\q_{p+2}=\q_r=0$. Using \p{Xi'} and the fact that in this frame $\q_{p+1} = \ran{p+1}\eta_{p+1}$,
we can bring the two covariants to  the form
\begin{eqnarray}
  &&\Xi_{p, p+2,r} = x^2_{p+2,\, r} \vev{p|\q_p} = x^2_{p+2,\, r} \vev{p|\q_{p+1}}  = x^2_{p+2,\, r} \vev{p | p+1}\eta_{p+1} \nn \\
  &&\Xi_{p+2,r,p+1}  =  \vev{p+2|x_{p+2,\, r} x_{ r\,, p+1}|\q_{p+1}}  = -x^2_{p+2,\, r} \vev{p| p+1} \eta_{p+1}   \label{brtothef}
\end{eqnarray}
hence the relation
\begin{equation}\label{hencere}
    \Xi_{p, p+2,r} =  - \frac{\vev{p| p+1}}{\vev{p+2| p+1}}\ \Xi_{p+2,r,p+1}
\end{equation}

\subsection{Three-point superconformal invariants}\label{3ptscin}

Our next task is to construct the nilpotent factor $R^{\rm NMHV}_n$ \p{genmhvam} out of the
superconformal covariants \p{Xi}. To start with, we need to multiply four supercovariants (not
necessarily identical) together, in order to form an $SU(4)$ invariant of the required odd degree
four, $\ep_{ABCD} \Xi^A \Xi^B\Xi^C\Xi^D$. Then, remembering that  $R^{\rm NMHV}_n$ has to be a dual
conformal {invariant} without helicity (all the weights are carried by the MHV prefactor in
\p{genmhvam}), we will need to find an appropriate bosonic factor which compensates the conformal
and helicity weights of the $\Xi$'s. In principle, there are many ways how to do this, but one of
them is very special -- we wish to preserve the three-point nature of the supercovariant. Thus, we
need to take four copies of the same\footnote{Any other $\Xi_{p'q'r'}$ will bring in at least one
new point.} $\Xi_{pqr}$ and form the nilpotent $SU(4)$ invariant of degree four
\begin{equation}\label{formnilin}
    \delta^{(4)}(\Xi_{pqr}) \equiv \frac{1}{4!}\ep_{ABCD} \Xi^A_{pqr} \Xi^B_{pqr}\Xi^C_{pqr}\Xi^D_{pqr}
\end{equation}
According to \p{XiI}, it transforms with conformal weight $-4$ at each point,
\begin{equation}\label{del4I}
    I\left[\delta^{(4)}(\Xi_{pqr})\right] = \frac{\delta^{(4)}(\Xi_{pqr})}{x^8_p x^8_q x^8_r}
\end{equation}
and has helicity weight $-2$ at point $p$.

Next, we need to find a bosonic factor $c_{pqr}$ which compensates the above weights. Once again, we
wish to restrict ourselves to using only the three points $x_{p,q,r}$. Three-point string-type
conformal covariants and Lorentz invariants are shown in \p{invshch}. In our case, the following
four such strings are possible (recall \p{twpr}):
\begin{eqnarray}
  && \vev{p|x_{pr} x_{r\, q}| q-1} \equiv \vev{p|x_{pr} x_{r\, q-1}| q-1} \nn\\
  && \vev{p|x_{pr} x_{r\, q}| q} \nn\\
  && \vev{p|x_{pq} x_{q\, r-1}| r} \equiv \vev{p|x_{pq} x_{q\, r-1}| r-1} \nn\\
  && \vev{p|x_{pq} x_{q\, r}| r} \label{4string}
\end{eqnarray}
Multiplying them together, we obtain the necessary helicity weight $-2$ at point $p$. Finally, we
just need a couple of additional obvious factors, which adjust the conformal weights and cancel the
helicity weights at points $q-1, q,r-1,r$. The result is
\begin{equation}\label{defcurly}
    c_{pqr} = \frac{\vev{q-1\, q}\vev{r-1\, r}}{x^2_{qr}\ \vev{p|x_{pr} x_{r\, q}| q-1} \ \vev{p|x_{pr} x_{r\, q}| q} \ \vev{p|x_{pq} x_{q\, r}| r-1} \ \vev{p|x_{pq} x_{q\, r}| r} }
\end{equation}

Thus, we propose to build the nilpotent factor $R^{\rm NMHV}_n$ \p{genmhvam} as a linear combination
of the following three-point superconformal invariants:
\begin{equation}\label{npfac}
    R_{pqr} = c_{pqr}\ \delta^{(4)}(\Xi_{pqr})
\end{equation}
By construction, it is manifestly invariant under conformal inversion, as well as $Q$ and $\bar S$
supersymmetry. Remarkably, \p{npfac} turns out to be invariant under $\bar Q$ supersymmetry
\p{dubarq} as well (and consequently under $S=I\bar Q I$). The best way to see this is to use the
frame fixing procedure described above. We want to show that
\begin{equation}\label{shotha}
    \bar Q^A_{\da}\ R_{pqr} = 0
\end{equation}
We already know that $\bar S_A^{\da}\ R_{pqr} = 0$, so we derive the compatibility conditions (see
\p{comm-rel}) $(D-C)\ R_{pqr} = M^{\da}_{\db}\ R_{pqr} = R^A_B\ R_{pqr} = 0$. They are trivially
satisfied since   $R_{pqr}$ has neither conformal nor helicity weight, and is a Lorentz scalar and
$SU(4)$ singlet. The same argument shows that the left-hand side of Eq.~\p{shotha} is annihilated by
$\bar S$ (and also by $Q$, since $\{Q,\bar Q\}R_{pqr} = P\ R_{pqr} = 0$). This allows us to apply
the combined $Q$ and $\bar S$ transformation \p{systwoeqs} which leads to the fixed supersymmetry
frame \p{susframe}. Further, the generator $\bar Q$  \p{dubarq} only sees the three points
$x_{p,q,r}$ inside $R_{pqr}$ \p{npfac}, but its action on  $x_{q,r}$ is trivial in this fixed frame.
Moreover, in the same frame we have $\Xi_{pqr} =x^2_{qr}\, \vev{ p| \q_p}$ (see \p{anobvi}), so
\begin{eqnarray}
  \bar Q^A_{\da}\ R_{pqr} &=& \q^{A\, \a}_p\frac{\pa}{\pa x^{\da a}_p} \left( \frac{\vev{q-1\, q}\vev{r-1\, r}\ x^6_{qr}\ \delta^{(4)}(\vev{ p| \q_p})}{\vev{p|x_{pr} x_{r\, q}| q-1} \ \vev{p|x_{pr} x_{r\, q}| q} \ \vev{p|x_{pq} x_{q\, r}| r-1} \ \vev{p|x_{pq} x_{q\, r}| r}}\right) \nn \\
  &\propto& \vev{ p| \q_p} \ \delta^{(4)}(\vev{ p| \q_p}) = 0  \label{bqtop}
\end{eqnarray}
The  details of this proof explain why we insisted on the three-point nature of the superinvariant
$R_{pqr}$ -- it allowed us, after fixing the frame \p{susframe}, to drastically simplify the
structure of the invariant and of the generator $\bar Q$, and to profit from the odd delta function
$\delta^{(4)}(\Xi_{pqr})$ which annihilates the variation.

Summarizing the above argument, the minimal set of requirements that $R_{pqr}$ must be annihilated
by the generators of dual Poincar\'e supersymmetry $Q, \bar Q, P, M$, as well as being invariant
under dual conformal inversion and $SU(4)$, implies that it is a {\it dual superconformal invariant}
of the full $SU(2,2|4)$. Further, if we restrict ourselves to {\it three-point invariants}, then
\p{npfac} is the only solution. We point out once again that the three-point nature of the
invariants has to do with the 3-mass-box one-loop integrals in the gluon amplitude of
\cite{Bern:2004bt}. For NNMHV amplitudes, where also 4-mass boxes appear, the situation may change
and we may have to use four-point invariants. This question is still under investigation.

Another interesting property of the superinvariants \p{npfac} is their independence of $\bl$.
Indeed, both the odd covariants $\Xi_{pqr}$ \p{Xi} and the coefficients $c_{pqr}$ \p{defcurly} are
expressed in terms of the dual {\it chiral} superspace coordinates $x,\q,\lambda$. The dependence on
$\bl$ is only implicit, through the $x_{ij}$. This is another manifestation of the holomorphic
nature of the $\cN=4$ superamplitudes.

Finally, we note that the identity \p{hencere} between supercovariants can be upgraded to an
identity between the corresponding superinvariants,
\begin{equation}\label{identinv}
    R_{p, p+2,r} =  R_{p+2,r,p+1}
\end{equation}
It is most easily shown in a fixed frame of the type \p{susframe}.

\subsection{Twistor coplanarity of the superinvariants}

One of the main observations of Witten \cite{Witten:2003nn} was that the gluon amplitudes,
formulated in the on-shell space with coordinates $\lambda,\bl$, exhibit some unexpected simple
geometric properties after performing a `twistor transform' (a partial Fourier transform of the
variables $\bl$, but not of $\lambda$). One such property is `twistor coplanarity'. Formulated in
terms of the original amplitude ${\cal A}(\lambda,\bl)$ (before the twistor transform), it takes the
form of a second-order differential constraint:
\begin{equation}\label{twcop'}
    {\cal K}_{mnst}\ A(\lambda, \bl) = 0
\end{equation}
where the coplanarity operator is
  \begin{equation}\label{copoper}
    {\cal K}_{mnst} = \vev{mn}\kappa_{st} + \mbox{permutations}\,, \qquad \kappa_{st} = \pa_{\da\, s} \ep^{\da\db} \pa_{\da\, t}\,, \quad \pa_{\da\, s} \equiv \frac{\pa}{\pa\bl^{\da}_{s}}
\end{equation}
for any set of four points $m,n,s,t$. This property has been widely discussed in the literature. In
particular, in \cite{Bern:2004bt} it was shown that the 3-mass-box coefficients are annihilated by
the coplanarity operator.

Here we demonstrate that our dual superinvariants $R_{pqr}$ (which contain, among others, the
3-mass-box coefficients) are coplanar in the above sense, as a direct corollary of (i) $\bar Q$
supersymmetry  and (ii) a much simpler, supersymmetric version of \p{twcop'}. To do this, we regard
the  superinvariants as functions of the on-shell superspace variables, $R^{\rm
NMHV}_{pqr}(\lambda,\bl,\eta)$. Let us then define the `super-coplanarity' operator
\begin{equation}\label{supcpop}
    \Omega_{AB\, st} = \Omega_{BA\, st} = \pa_{A\, s}\pa_{B\, t} - \pa_{A\, t}\pa_{B\, s}\,, \qquad \pa_{A\, s} \equiv \frac{\pa}{\pa\eta^{A}_{s}}
\end{equation}
Now, let us apply this operator to $R_{pqr}$, assuming that both points $s,t$ fall within the range
of $\eta$'s in $\Xi_{pqr}$ \p{Xi''} (otherwise $\Omega_{AB\, st}$ annihilates $R_{pqr}$ trivially).
The derivatives in \p{supcpop} free two of the $SU(4)$ indices of $\ep_{ABCD}$, which contradicts
the symmetry of $\Omega$. Thus,
\begin{equation}\label{omannr}
    \Omega_{AB\, st} \ R_{pqr} = 0
\end{equation}

We have also seen that $R_{pqr}$ is annihilated by the dual supersymmetry generator $\bar Q$. Then,
(anti)commuting $\bar Q$ (in the form \p{indueta}) twice with $\Omega$, we find
\begin{equation}\label{combqom}
    \{\bar Q^D_{\db}, [\bar Q^C_{\da}, \Omega_{AB\, st}] \} \simeq \ep_{\da\db} \delta^C_{(A} \delta^D_{B)}\ \kappa_{st}
\end{equation}
We thus see that the coplanarity of the superinvariants,
\begin{equation}\label{kappannr}
    {\cal K}_{mnst}\  R_{pqr} = 0
\end{equation}
is an immediate consequence of super-coplanarity \p{omannr} and of the dual $\bar Q$ supersymmetry.
The former of these properties is valid at tree as well as at loop level (it is an obvious property
of the super-helicity structures),  while the latter is sensitive to the $x$--dependence and becomes
anomalous in the presence of infrared divergent loop corrections.

\subsection{The complete NMHV $n$-point one-loop superamplitude.\\ New form of the tree-level amplitude}\label{cnbhmvnftl}

We are now ready to formulate our proposal for the complete one-loop NMHV $n$-point superamplitude:
\begin{equation}\label{fullamp}
    {\cal A}^{\rm NMHV}_n  = {\cal A}_{n}^{\rm MHV}\left[{\sum_{p,q,r=1}^n\ R_{pqr}\ \big( 1 + a V_{pqr}(x_{ij})
+ O(\epsilon)\big) + O(a^2)}\right]
\end{equation}
Here the nilpotent superconformal invariants $R_{pqr}$ \p{npfac} encode the helicity structures for
all the component amplitudes. In particular, expanding \p{fullamp} in powers of $\eta$ and
collecting the terms $(\eta_{m_1})^4(\eta_{m_2})^4(\eta_{m_3})^4$, we obtain the helicity structures
of the gluon amplitudes. This straightforward procedure (illustrated in detail in
Sect.~\ref{sect-super-NMHV6} for the case $n=6$) exactly reproduces the full list of 3-mass-box
coefficients from \cite{Bern:2004bt}.

Further, the factors  $V_{pqr}(x_{ij})$ in \p{fullamp} contain the finite one-loop corrections (the
infrared divergent part is in the MHV prefactor in \p{fullamp}). They are given by appropriate
combinations of 3-, 2- and 1-mass-box one-loop momentum integrals. One of our main claims is that
$V_{pqr}(x_{ij})$ are {\it dual conformal invariant} functions of  the dual space coordinates. The
explicit $n=6$ example worked out in detail in Sect.~\ref{sect-super-NMHV6} confirms this
conjecture. Further evidence in favor of the proposal \p{fullamp} will be given in \cite{toappear}.

We recall that the manifest symmetries of each term in \p{fullamp} are: helicity covariance, Lorentz
invariance, as well as invariance under part of the dual superconformal algebra, namely $Q, \bar S,
P$. The rest of the dual $SU(2,2|4)$ could be obtained if we add $\bar Q$ (and hence $K$, $D$ and
$S$) to the list of generators annihilating the superamplitude. Above we showed that the nilpotent
structures $R_{pqr}$ do indeed have this extended symmetry. However, the loop corrections
$V^{(1)}_{pqr}(x_{ij})$ involve non-trivial dependence on the dual space coordinates, which is seen
by the generator $\bar Q$. Moreover, as mentioned in Sect.~2.8, the MHV prefactor in \p{genmhvam}
contains the dual conformal anomalous part of the amplitude, which inevitably must break the  $\bar
Q$ symmetry (since $\bar S$ is a manifest symmetry of the one-loop amplitude).

Collecting the $a^0$ terms in the perturbative expansion \p{fullamp}, we obtain the following new
form of the {\it tree-level NMHV amplitude}:
\begin{equation}\label{newtree}
    {\cal A}^{\rm NMHV}_{n;0}  = \frac{\delta^{(4)}(\sum_{i=1}^n\, \lambda_i \bl_i)\ \delta^{(8)}(\sum_{j=1}^n\ \lambda_{j}\, \eta_j)}{\vev{1\, 2}\vev{2\, 3}\ldots\vev{n\, 1}} \ \sum_{p,q,r}\ R_{pqr}
\end{equation}
Expanding \p{newtree} in $\eta$, we have checked that its gluon tree-level components coincide with
those given in \cite{Bern:2004bt}. The non-MHV tree-level amplitudes have been widely studied in the
literature. In particular, Cachazo, Svrcek and Witten \cite{CSW} have proposed a method for
constructing them by combining MHV vertices. The characteristic feature of this approach is that it
involves a constant fixed spinor (`reference spinor') which breaks manifest Lorentz
invariance.\footnote{In \cite{CSW} it was argued that the sum of all MHV$\times$MHV diagrams
effectively does not depend on the reference spinor. Later on Kosower \cite{Kosower:2004yz} was able
to rewrite the CSW NMHV tree amplitude in a form where the reference spinor manifestly drops out.}
In \cite{Khoze} Georgiou, Glover and Khoze adapted the CSW construction for NMHV amplitudes to
superspace. Their form of the tree amplitude bears some similarity to ours \p{newtree}, but the main
difference again is the use of a reference spinor and the loss of Lorentz invariance (see Appendix
\ref{Khoze} for details).

\section{Conclusions and outlook}

In this paper, we have argued that the dual conformal symmetry previously observed for MHV
amplitudes is part of a dual superconformal symmetry that governs the form of \textit{all} (NMHV,
NNMHV, ...) scattering amplitudes in $\mathcal{N}=4$ SYM.

It is important to realize that dual conformal symmetry is broken by the infrared regulator.
Therefore, in order to be able to make predictions about scattering amplitudes, one has to
control how this symmetry is broken. Insight into the mechanism of breakdown of dual conformal
symmetry has come from the study of a particular type of light-like Wilson loops, which are expected
to be dual to MHV gluon scattering amplitudes \cite{am1,dks,bht,dhks1,dhks2,dhks3,dhks4}. The Wilson loops satisfy anomalous
conformal Ward identities derived in \cite{dhks1,dhks2}, and so do the MHV amplitudes, through the
conjectured duality. More precisely, the finite part of the amplitude  satisfies the anomalous Ward
identity (see Sect.~\ref{DCPFM})
\begin{equation}
K^{\mu} {F}_{n}^{\rm MHV} = \frac{1}{2} \Gamma_{\rm cusp}(a) \sum_{i=1}^n  \ln
\frac{x_{i,i+2}^2}{x_{i-1,i+1}^2} x^\mu_{i,i+1} +O(\epsilon)\,.
\end{equation}

As we have shown in this paper, the dual conformal properties of non-MHV amplitudes can be made
manifest as well. In order to do this, one introduces the Grassmann variables $\eta_{i}$ to
describe the $\mathcal{N}=4$ supermultiplet, together with dual superspace coordinates
 \begin{equation}
   \lambda_{i\,\a}\, \eta^A_i =\q^A_{i\, \a} - \q^A_{i+1\, \a}\,,
\end{equation}
in close analogy with the dual bosonic coordinates defined by $\lambda_{i\,\a}
\tilde{\lambda}_{i\,\dot \a} = x_{i\,\a \dot{\a}} - x_{i+1\,\a \dot{\a}} $.
We conjecture that \textit{all} amplitudes satisfy the same anomalous dual conformal Ward
identities as the MHV amplitudes. In other words, we choose to write the full $n$-point
superamplitude as
\begin{equation}\label{conclusions-factorMHV}
{\cal A}_{n} =  {\cal A}_{n}^{\rm MHV} \left[ R_n(x_i, \q_i, \lambda_i)+O(\epsilon)\right]\,,
\end{equation}
where $R_n = 1+R_n^{(1)} +R_n^{(2)} + ... + R_n^{(n-4)}$ and $R_n^{(p)}$ contains the terms corresponding to the N${}^p$MHV amplitudes.
Then the infrared divergences and the conformal anomaly are contained in ${\cal A}_{n}^{\rm MHV}$, and therefore the (infrared finite)
function $R_n$ satisfies the conformal Ward identity
\begin{equation}\label{conclusions-CWI}
 {K}^{\mu} R_n(x_i, \q_i, \lambda_i)
 = 0\,,
\end{equation}
with the conformal boost operator $K^\mu$ defined appropriately on the superspace, see equation
(\ref{sdualK}) for its explicit expression. In order to check our conjecture
(\ref{conclusions-CWI}), we wrote various one-loop NMHV amplitudes that were available in the
literature \cite{bddk94-2,Bern:2004ky,Bern:2004bt,Risager:2005ke} in the superspace form using
$\eta_{i}$. We illustrated this by giving the explicit example of the $n=6$ NMHV superamplitude,
where $R^{(1)}_6$ could be written in a manifestly dual conformal form,  so that it obviously
vanishes under the action of $K^{\mu}$.

An equivalent way of stating our conjecture is the following. We can write the MHV factor in (\ref{conclusions-factorMHV}) as the product of its tree-level $\mathcal{A}_{n;0}^{\rm MHV}$ and loop-correction $M_n(x_i;\mu,\epsilon)$ parts  and then combine everything except the tree-level MHV factor into a single scalar function $\mathcal{M}_n$,

\be
\mathcal{A}_n = \mathcal{A}_{n;0}^{\rm MHV} \mathcal{M}_n(x_i,\theta_i,\lambda_i;\mu,\epsilon).
\ee
The function $\mathcal{M}_n$ is simply the product $M_n(x_i) \left[ R_n(x_i, \q_i, \lambda_i)+O(\epsilon)\right]$. Then one can take the logarithm of $\mathcal{M}_n$,
\be
{\rm ln}\mathcal{M}_n = [\text{IR divergences}] + F_n + O(\epsilon).
\ee
Finally the conjecture (\ref{conclusions-CWI}) can be restated as the fact that $F_n$ (which is nothing but $F_n^{\rm MHV} + \ln R_n$) satisfies the anomalous Ward identity,

\begin{equation}
K^{\mu} {F}_{n} = \frac{1}{2} \Gamma_{\rm cusp}(a) \sum_{i=1}^n  \ln
\frac{x_{i,i+2}^2}{x_{i-1,i+1}^2} x^\mu_{i,i+1} +O(\epsilon)\,.
\end{equation}

The reason why the dual conformal properties of the amplitudes become apparent only in superspace is
that under conformal boosts different component amplitudes transform into each other, and it is only
the superamplitude which has definite conformal properties. We pointed out that there is a sub-class
of amplitudes, the split-helicity amplitudes, which are exceptional. The do not 'mix' with other
amplitudes in the above sense and therefore are dual conformal on their own. This allows us to
suggest a two-loop test of our conjecture: for example, the ${\cal A}_{6}^{\rm NMHV}$ amplitude with
helicity assignment $(+++---)$ should be dual conformal on its own.

One of the results of this paper is a compact form for all NMHV one-loop amplitudes for arbitrary
external particles, namely
\be\label{conclusions-NMHV1loop}
{\cal A}^{\rm NMHV}_n  = {\cal A}_{n}^{\rm MHV}\left[{\sum_{p,q,r=1}^n\ R_{pqr}\ \big( 1 + a
V_{pqr}(x_{ij}) + O(\epsilon)\big) + O(a^2)}\right]
\ee
where the dual superinvariants $R_{pqr}$ were defined in (\ref{npfac}), and $V_{pqr}(x_{ij})$ is a
scalar dual conformal function. Beyond $n=6$ we obtained (\ref{conclusions-NMHV1loop}) from an
explicit calculation using the method of generalized unitarity cuts \cite{Britto:2004nc}. Using this
manifestly supersymmetric setup for the unitarity cuts, the form (\ref{conclusions-NMHV1loop}) of
the NMHV amplitudes emerges naturally. We will present the details of these results in a forthcoming
publication.

We stress that (\ref{conclusions-NMHV1loop}) also provides an explicit form of the tree amplitude,
\begin{equation}\label{conclusions-NMHVtree}
     {\cal A}^{\rm NMHV}_{n;0}  = \frac{\delta^{(4)}(\sum_{i=1}^n\, \lambda_i \bl_i)\ \delta^{(8)}(\sum_{j=1}^n\ \lambda_{j}\, \eta_j)}{\vev{1\, 2}\vev{2\, 3}\ldots\vev{n\, 1}} \ \sum_{p,q,r}\ R_{pqr}
\end{equation}
Such tree amplitudes appear for example when one computes one-loop amplitudes with a high number of
external legs using the (generalized) unitarity cut method. Indeed, using maximal cuts, loop
amplitudes can be constructed from the tree amplitudes, and therefore it is important to have
explicit expressions for the latter. A similar application may be the computation of amplitudes in
$\mathcal{N}=8$ SYM using the KLT relations \cite{KLT}. Notice that our formula for the tree-level
NMHV amplitudes does not depend on a reference spinor, as compared to \cite{CSW,Khoze,Freedman}. It
would be interesting to see how formulae obtained from recursion relations including a reference
spinor are equivalent to our formula.

It is natural to ask whether our amplitudes have a geometrical interpretation in (super-) twistor
space. For example, MHV tree-level amplitudes lie on lines in supertwistor space
\cite{Witten:2003nn}. The twistor space properties of NMHV gluon amplitudes were studied in
\cite{Bern:2004ky,Bern:2004bt}. We find it likely that our NMHV superamplitudes will allow for an
interpretation in supertwistor space. This point is currently under investigation.

\section*{Acknowledgements}

We would like to thank  F.~Alday, N.~Berkovits, Z.~Bern, L.~Dixon, H.~Elvang, D.~Freedman,
P.~Heslop, D.~Kosower, J.~Maldacena, G.~Travaglini, A.~Volovich for stimulating discussions. We are
grateful to the Center for Theoretical Studies, ETH Z\"urich for hospitality during the final stage
of this work. This research was supported in part by the French Agence Nationale de la Recherche
under grant ANR-06-BLAN-0142.

\section*{Appendices}

\appendix

\setcounter{section}{0} \setcounter{equation}{0}
\renewcommand{\theequation}{\Alph{section}.\arabic{equation}}

\section{Notations and conventions} \label{A}

In this paper we use the two-component spinor formalism.

The dotted and undotted spinor indices are raised and lowered as follows:
\begin{eqnarray}\label{2.57}
  &&\psi^\alpha = \epsilon^{\alpha\beta}\psi_\beta\,, \qquad \bar\chi^{\dot\alpha} =
  \epsilon^{\dot\alpha\dot\beta}\bar\chi_{\dot\beta}\;;\\
  &&\psi_\alpha = \epsilon_{\alpha\beta}\psi^\beta\,, \qquad \bar\chi_{\dot\alpha} =
  \epsilon_{\dot\alpha\dot\beta}\bar\chi^{\dot\beta}  \label{2.58}
\end{eqnarray}
where the antisymmetric $\epsilon$ symbols have the properties:
\begin{equation}\label{2.59}
  \epsilon_{12} = \epsilon_{\dot 1\dot 2} = -\epsilon^{12} = -\epsilon^{\dot 1\dot 2} =
  1\,, \qquad \epsilon_{\alpha\beta}\epsilon^{\beta\gamma} =
  \delta_\alpha^\gamma\,,  \qquad
\epsilon_{\dot\alpha\dot\beta}\epsilon^{\dot\beta\dot\gamma} =
  \delta_{\dot\alpha}^{\dot\gamma}
\end{equation}
The convention for the contraction of a pair of spinor indices is
\begin{equation}\label{2.60}
  \psi^\alpha\lambda_\alpha\equiv \vev{\psi\lambda}\,, \qquad
\bar\chi_{\dot\alpha}\bar\rho^{\dot\alpha}
  \equiv [\bar\chi\bar\rho]
\end{equation}
Two-component spinors satisfy the cyclic identity
\begin{equation}\label{cycide}
    \vev{\psi\lambda}\chi_\a + \vev{\lambda\chi}\psi_\a + \vev{\chi\psi}\lambda_\a = 0
\end{equation}
which simply means that the antisymmetrization over three two-component indices vanishes
identically.

The sigma matrices $\sigma^\mu$ are defined as follows:
\begin{equation}\label{2.61}
  (\sigma^\mu)_{\alpha\dot\alpha} = (1,\vec\sigma)_{\alpha\dot\alpha}\,, \qquad
  (\tilde\sigma^\mu)^{\dot\alpha\alpha} =
\epsilon^{\dot\alpha\dot\beta}\epsilon^{\alpha\beta}
  (\sigma^\mu)_{\beta\dot\beta} =  (1,-\vec\sigma)^{\dot\alpha\alpha}
\end{equation}
and have the basic properties:
\begin{eqnarray}
  &&\sigma^\mu \tilde\sigma^\nu = \eta^{\mu\nu} - i\sigma^{\mu\nu}\,,
  \qquad \tilde\sigma^\mu \sigma^\nu
  = \eta^{\mu\nu} - i\tilde\sigma^{\mu\nu}\,,\nonumber\\
  &&(\sigma^\mu)_{\alpha\dot\alpha}(\tilde\sigma_\mu)^{\dot\beta\beta} =
  2\delta^\beta_\alpha \delta^{\dot\beta}_{\dot\alpha}\,, \qquad
  (\sigma_\mu)_{\alpha\dot\alpha}(\tilde\sigma^\nu)^{\dot\alpha\alpha} =
  2\delta^\nu_\mu\,,\\ \label{2.62}
  && \sigma^{\mu\nu} = -\sigma^{\nu\mu}\,, \qquad \tilde\sigma^{\mu\nu} =
-\tilde\sigma^{\nu\mu}\,, \qquad (\sigma^{\mu\nu})_\alpha{}^\alpha =
(\tilde\sigma^{\mu\nu})_{\dot\alpha}{}^{\dot\alpha} = 0 \nonumber
\end{eqnarray}

A four-vector $x^\mu$ can be written as a two-component bispinor:
\begin{equation}\label{xsquared}
x_{\a\da} =x^{\mu}(\sigma_{\mu})_{\alpha\dot\alpha}\,, \qquad    x^{\dot\alpha\alpha} =
x^{\mu}\tilde\sigma_{\mu}^{\dot\alpha\alpha}\,, \qquad  x^{\mu} = {1\over 2}
x^{\dot\alpha\alpha}\sigma^{\mu}_{\alpha\dot\alpha}
\end{equation}
Its square $x^2$ can be expressed in various ways:
\begin{equation}\label{xsquared'}
x^2 = x^\mu x_\mu = {1\over 2} x^{\dot\alpha\alpha}x_{\alpha\dot\alpha}\,, \qquad x_{\a\da}\
x^{\da\b} = x^2 \delta^\b_\a\,, \qquad x^{\da\a}\ x_{\a\db} = x^2 \delta^{\da}_{\db}
\end{equation}
Its `inverse' $x^\mu/x^2$ takes the matrix form
\begin{equation}\label{id}
 (x^{-1})_{\a\da} = \frac{x_{\a\da}}{x^2}\,, \qquad    (x^{-1})_{\a\da}  x^{\da\b} =
\delta_\a^\b\,, \qquad x^{\da\a} (x^{-1})_{\a\db} = \delta^{\da}_{\db}
\end{equation}

We often have to deal with `strings' of commuting spinors $\lambda,\bl$ and vectors $x$, for which
we use the following short-hand notations, e.g.:
\begin{eqnarray}
  && \vev{p|x_{mn}x_{kl}|q} = \lambda^\a_p (x_{mn})_{\a\da} (x_{kl})^{\da\b} \lambda_{q\, \b}
= - \vev{q|x_{kl}x_{mn}|p} \label{lstring}\\
  && \lan{p} x_{mn} |q] = \lambda^\a_p (x_{mn})_{\a\da} \bl^{\da}_q\,, \qquad \mbox{etc.}
 \nn
\end{eqnarray}

\section{Conventional and dual superconformal generators} \label{B}

In this appendix we give the conventional and dual representations of the superconformal algebra. We
begin by listing the commutation relations of the algebra $u(2,2|4)$. The Lorentz generators
$\mathbb{M}_{\a \b}$, $\overline{\mathbb{M}}_{\da \db}$ and the $su(4)$ generators
$\mathbb{R}^{A}{}_{B}$ act canonically on the remaining generators carrying Lorentz or $su(4)$
indices. The dilatation $\mathbb{D}$ and hypercharge $\mathbb{B}$ act via
\be
[\mathbb{D},\mathbb{J}] = {\rm dim}(\mathbb{J}), \qquad [\mathbb{B},\mathbb{J}] = {\rm
hyp}(\mathbb{J}).
\ee
The non-zero dimensions and hypercharges of the various generators are
\begin{align} \notag
& {\rm dim}(\mathbb{P})=1, \qqquad {\rm dim}(\mathbb{Q}) = {\rm dim}(\overline{\mathbb{Q}}) =
\tfrac{1}{2},\qquad {\rm dim}(\mathbb{S}) = {\rm dim}(\overline{\mathbb{S}}) = -\tfrac{1}{2}
\\
&{\rm dim}(\mathbb{K})=-1,\qquad {\rm hyp}(\mathbb{Q}) = {\rm hyp}(\overline{\mathbb{S}}) =
\tfrac{1}{2}, \qquad~ {\rm hyp}(\overline{\mathbb{Q}}) = {\rm hyp}(\mathbb{S}) = - \tfrac{1}{2}.
\end{align}
The remaining non-trivial commutation relations are,
\begin{align} \notag
& \{\mathbb{Q}_{\a A},\overline{\mathbb{Q}}_{\da}^B\}  =  \delta_A^B \mathbb{P}_{\a \da},
   \qquad \{\mathbb{S}_{\a A},\overline{\mathbb{S}}_{\da}^B \} = \delta_A^B \mathbb{K}_{\a \da},
\\ \notag
& {}[\mathbb{P}_{\a \da},\mathbb{S}_{\b}^A] = \epsilon_{\da \db} \overline{\mathbb{Q}}_{\da}^A,
 \qqquad [\mathbb{K}_{\a \da},\mathbb{Q}_{\b A}] = \epsilon_{\a \b} \overline{\mathbb{S}}_\da^A,
\\ \notag
& {}[\mathbb{P}_{\a \da},\overline{\mathbb{S}}_{\db A}]  =  \epsilon_{\da \db} \mathbb{Q}_{\a A},
\qquad [\mathbb{K}_{\a \da}, \overline{\mathbb{Q}}_{\db A}]  =  \epsilon_{\da \db} \mathbb{S}_{\a
A},
\\ \notag
& [\mathbb{K}_{\a \da},\mathbb{P}^{\b \db}] = \delta_\a^\b \delta_\da^\db \mathbb{D} +
\mathbb{M}_{\a}{}^{\b}
 \delta_\da^\db + \overline{\mathbb{M}}_{\da}{}^{\db} \delta_\a^\b,
\\ \notag
& \{\mathbb{Q}_{\a A},\mathbb{S}_\b^B\} = \epsilon_{\a \b} \mathbb{R}^{B}{}_{A} + \mathbb{M}_{\a \b}
\delta_A^B + \epsilon_{\a \b}\delta_A^B (\mathbb{D}+\mathbb{C}),
\\ \label{comm-rel}
& \{\overline{\mathbb{Q}}_{\da}^{A},\overline{\mathbb{S}}_{\db B}\} = \epsilon_{\da \db}
\mathbb{R}^{A}{}_{B} + \overline{\mathbb{M}}_{\da \db} \delta_B^A + \epsilon_{\da \db}\delta_B^A
(\mathbb{D}-\mathbb{C}).
\end{align}
We now give the generators in both the conventional and dual representations of the superconformal
algebra. We will use the following shorthand notation:

\begin{align}
\partial_{i \alpha \dot{\alpha}} = \frac{\partial}{\partial
x_i^{\alpha \dot{\alpha}}}, \qquad \partial_{i \alpha A} = \frac{\partial}{\partial \theta_i^{\alpha
A}}, \qquad \partial_{i \alpha} = \frac{\partial}{\partial \lambda_i^{\alpha}}, \qquad
\partial_{i \dot{\alpha}} = \frac{\partial}{\partial
    \bar{\lambda}_i^{\dot{\alpha}}}, \qquad
\partial_{i A} = \frac{\partial}{\partial \eta_i^A}.
\end{align}
We first give the generators of the conventional superconformal symmetry, using lower case
characters to distinguish these generators from the dual superconformal generators which follow
afterwards.

$$
\begin{array}{cccccc}
p^{\alpha \dot{\alpha}} &=& \sum_i \lambda_i^{\alpha} \bar{\lambda}_i^{\dot{\alpha}}, &\qquad
k_{\alpha \dot{\alpha}} &=& \sum_i \partial_{i \alpha} \partial_{i \dot{\alpha}},\\
&&&&&\\
\overline{m}_{\dot{\alpha} \dot{\beta}} &=& \sum_i \bar{\lambda}_{i (\dot{\alpha}} \partial_{i
\dot{\beta} )}, &\qquad   m_{\alpha \beta} &=& \sum_i \lambda_{i (\alpha} \partial_{i \beta )}
,\\
&&&&&\\
d &=& \sum_i [\tfrac{1}{2}\lambda_i^{\alpha} \partial_{i \alpha} +\tfrac{1}{2}
\bar{\lambda}_i^{\dot{\alpha}} \partial_{i
    \dot{\alpha}} +2],& \qquad r^{A}{}_{B} &=& \sum_i [\eta_i^A \partial_{i B} - \tfrac{1}{4}\eta_i^C \partial_{i C}],\\
  &&&&&\\
q^{\alpha A} &=& \sum_i \lambda_i^{\alpha} \eta_i^A, &\qquad \bar{q}^{\dot\alpha}_A &=& \sum_i \bar\lambda_i^\da \partial_{i A}, \\
&&&&&\\
s^{\alpha}_A &=& \sum_i \partial_{i \alpha} \partial_{i A}, &\qquad
\bar{s}_{\dot\alpha}^A &=& \sum_i \eta_i^A \partial_{i \dot\alpha}.
\end{array}
$$

We can construct the generators of dual superconformal transformations by starting with the standard
chiral representation and extending the generators so that they commute with the constraints,
\be
(x_i-x_{i+1})_{\a\da}  - \lambda_{i\,\a}\, \bl_{i\,\da} = 0, \qquad (\theta_i -
\theta_{i+1})_\alpha^A - \lambda_{i \alpha} \eta_i^A = 0.
\ee
By construction they preserve the surface defined by these constraints, which is where the amplitude
has support. The generators are
\begin{align}
P_{\alpha \dot{\alpha}}&= \sum_i \partial_{i \alpha \dot{\alpha}},\\
Q_{\alpha A} &= \sum_i \partial_{i \alpha A}, \\
\overline{Q}_{\dot{\alpha}}^A &= \sum_i [\theta_i^{\alpha A}
  \partial_{i \alpha \dot{\alpha}} + \eta_i^A \partial_{i \dot{\alpha}}],\label{barQfss}\\
M_{\alpha \beta} &= \sum_i[x_{i ( \alpha}{}^{\dot{\alpha}}
  \partial_{i \beta ) \dot{\alpha}} + \theta_{i (\alpha}^A \partial_{i
  \beta) A} + \lambda_{i (\alpha} \partial_{i \beta)}],\\
\overline{M}_{\dot{\alpha} \dot{\beta}} &= \sum_i [x_{i
    (\dot{\alpha}}{}^{\alpha} \partial_{i \dot{\beta} ) \alpha} +
  \bar{\lambda}_{i(\dot{\alpha}} \partial_{i \dot{\beta})}],\\
R^{A}{}_{B} &= \sum_i [\theta_i^{\alpha A} \partial_{i \alpha B} +
  \eta_i^A \partial_{i B} - \tfrac{1}{4} \delta^A_B \theta_i^{\alpha
    C} \partial_{i \alpha C} - \tfrac{1}{4}\eta_i^C \partial_{i C}
],\\ \label{DD} D &= \sum_i [x_i^{\alpha \dot{\alpha}}\partial_{i \alpha \dot{\alpha}} +
  \tfrac{1}{2} \theta_i^{\alpha A} \partial_{i \alpha A} +
  \tfrac{1}{2} \lambda_i^{\alpha} \partial_{i \alpha} +\tfrac{1}{2}
  \bar{\lambda}_i^{\dot{\alpha}} \partial_{i \dot{\alpha}}],\\
  \label{CC}
C &=  \sum_i \tfrac{1}{2}[\lambda_i^{\alpha} \partial_{i \alpha} -
  \bar{\lambda}_i^{\dot{\alpha}} \partial_{i \dot{\alpha}} - \eta_i^A
  \partial_{i A}], \\
S_{\alpha}^A &= \sum_i [\theta_{i \alpha}^{B} \theta_i^{\beta A}
  \partial_{i \beta B} - x_{i \alpha}{}^{\dot{\beta}} \theta_i^{\beta
    A} \partial_{\beta \dot{\beta}} - \lambda_{i \alpha}
  \theta_{i}^{\gamma A} \partial_{i \gamma} - x_{i+1\,
    \alpha}{}^{\dot{\beta}} \eta_i^A \partial_{i \dot{\beta}} +
  \theta_{i+1\, \alpha}^B \eta_i^A \partial_{i B}],\\
\overline{S}_{\dot{\alpha} A} &= \sum_i [x_{i \dot{\alpha}}{}^{\beta}
  \partial_{i \beta A} + \bar{\lambda}_{i \dot{\alpha}}
  \partial_{iA}],\label{fssbarS}\\ \label{KK}
K_{\alpha \dot{\alpha}} &= \sum_i [x_{i \alpha}{}^{\dot{\beta}} x_{i
    \dot{\alpha}}{}^{\beta} \partial_{i \beta \dot{\beta}} + x_{i
    \dot{\alpha}}{}^{\beta} \theta_{i \alpha}^B \partial_{i \beta B} +
  x_{i \dot{\alpha}}{}^{\beta} \lambda_{i \alpha} \partial_{i \beta}
  + x_{i+1 \,\alpha}{}^{\dot{\beta}} \bar{\lambda}_{i \dot{\alpha}}
  \partial_{i \dot{\beta}} + \bar{\lambda}_{i \dot{\alpha}} \theta_{i+1\,
    \alpha}^B \partial_{i B}].
\end{align}
We also have the hypercharge $B$,
\begin{equation}
B=\sum_i \tfrac{1}{2}[\theta_i^{\alpha A} \partial_{i \alpha A} +
  \lambda_i^\alpha \partial_{i \alpha} -
  \bar{\lambda}_i^{\dot{\alpha}} \partial_{i \dot{\alpha}}]
\end{equation}
Note that if we restrict the dual generators $\bar{Q},\bar{S}$ to the on-shell superspace they
become identical to the conventional generators $\bar s, \bar q$.

\section{NMHV tree-level amplitudes with and without reference spinor}\label{Khoze}

Let us rewrite the tree amplitude \p{newtree}, in a way such that it resembles the GGK tree
amplitude \cite{Khoze} with the reference spinor of CSW \cite{CSW}. Define $\lan{I_q} =
\lan{1}x_{1r} x_{qr}$ and  $\lan{I_r} = \lan{1}x_{1q} x_{qr}$ (with $\lan{I_q} - \lan{I_r} =
x^2_{qr} \lan{1}$). Then \p{newtree} becomes (we have singled out the term $p=1$)
\begin{eqnarray}
  {\cal A}^{\rm NMHV}_{n;0} &=& \delta^{(4)}(\sum_{i=1}^n\, \lambda_i \bl_i)\ \delta^{(8)}(\sum_{j=1}^n\ \lambda_{j} \eta_j) \nn\\
  &\times&  \Big[ \sum_{q,r}  \frac{\delta^{(4)}(\sum_{k=q}^{r-1} \vev{I_r\,k} \eta_k  {+ \sum_{k=1}^{q-1} (\vev{I_r\,k}-\vev{I_q\,k}) \eta_k} )}{x^2_{qr}\  \vev{1\, 2}\ldots \vev{q-1 \, I_q} \vev{I_q \, q}\ldots  \vev{r-1 \, I_r} \vev{I_r \, r} \ldots \vev{n\, 1}} \nn\\
  &+& {\rm cycle} \Big] \label{fullampkhoze}
\end{eqnarray}

Now, compare this to Eqs.~(5.9), (5.10) from \cite{Khoze} for the same amplitude  (with the
identification $i \equiv q-1$, $j \equiv r-1$, $q^2_I \equiv x^2_{qr}$; the argument of GGK's
$\delta^{(4)}$ should be replaced by the complementary cluster $n_2$). GGK have a unique $\lan{I} =
[\xi_{\rm ref}| x_{qr}$ defined with the help of CSW's reference spinor $[\xi_{\rm ref}|$. Their
amplitude has the form:
\begin{eqnarray}
  {\cal A}^{\rm CSW-GGK}_{n;0} &=& \delta^{(4)}(\sum_{i=1}^n\, \lambda_i \bl_i)\ \delta^{(8)}(\sum_{j=1}^n\ \lambda_{j} \eta_j) \nn\\
  &\times&  \sum_{q,r}  \frac{\delta^{(4)}(\sum_{k=q}^{r-1} \vev{I\,k} \eta_k )}{x^2_{qr}\  \vev{1\, 2}\ldots \vev{q-1 \, I} \vev{I \, q}\ldots  \vev{r-1 \, I} \vev{I \, r} \ldots \vev{n\, 1}} \label{treekhoze}
\end{eqnarray}
The main difference is that in  \p{fullampkhoze} we are using two spinors, $[\xi_{q}| =
\lan{1}x_{1q}$ and $[\xi_{r}| = \lan{1}x_{1r}$ (being expressed in terms of the external particle
variables, they are not `reference' spinors), while in  \p{treekhoze} they have been merged into a
single {\it constant} spinor, $[\xi_{q}| = [\xi_{r}| \equiv [\xi_{\rm ref}|$, independent on the
external particle variables.

It is an interesting question to find out how the two expressions \p{fullampkhoze} and \p{treekhoze}
are equivalent.


 \end{document}